\documentclass[a4paper,11pt]{article}

\pdfoutput=1
\usepackage{jheppub}

\usepackage{amssymb}
\usepackage{latexsym}
\usepackage{graphicx}
\usepackage{amsmath}
\usepackage{array}

\allowdisplaybreaks[2]

\title{Unification of Galileon Dualities}
\author{Karol Kampf}
\author{and Ji\v{r}\'{\i} Novotn\'{y}}
\affiliation{Institute of Particle and Nuclear Physics, Faculty of
Mathematics and Physics,\\  Charles University, CZ-18000 Prague, Czech
Republic}

\emailAdd{karol.kampf@mff.cuni.cz}
\emailAdd{jiri.novotny@mff.cuni.cz}

\abstract{We study dualities of the general Galileon theory in $d$ dimensions 
in terms of coordinate transformations on the coset space corresponding to the 
spontaneously broken Galileon group.
The most general duality transformation is found to be determined uniquely up 
to four free parameters and under compositions these transformations form a 
group which can be identified with $GL(2,\mathbf{R})$.
This group  represents a unified framework for all the up to now known Galileon 
dualities. We discuss a representation of this group on the Galileon theory 
space and using concrete examples we illustrate its applicability both on the 
classical and quantum level.}

\begin{document}

\maketitle
\flushbottom

\section{Introduction}

The Galileons represent a particular class of models of real scalar field 
$\phi $ with derivative interactions and posses number of interesting 
properties. It emerges
in its simplest form as an effective theory of the Dvali-Gabadadze-Porrati
model \cite{Dvali:2000hr,Deffayet:2001pu} as well as of the de
Rham-Gabadadze-Tolley massive gravity theory \cite{deRham:2010kj} in the
decoupling limit. Generalization of the Galileon Lagrangian was proposed by
Nicolis, Rattazzi and Trincherini \cite{Nicolis:2008in} as the long distance
modifications of General relativity. In the seminal paper \cite%
{Nicolis:2008in} also the complete classification of possible terms of the
Galileon Lagrangian has been made and some of the physical consequences have
been studied in detail, i.a. it was demonstrated that such theories exhibit
the so-called Vainshtein mechanism \cite{Vainshtein:1972sx}. In fact the
general structures appearing in the Galileon Lagrangian have been already
discovered in the 70' as a building blocks of the Horndeski Lagrangian \cite%
{Horndeski:1974wa}, which is the most general Lagrangian built from no more
than the second order derivatives of the scalar field and leading to the
second-order Euler-Lagrange equations. Generalization of such Lagrangians to
curved backgrounds and arbitrary $p$-form fields has been studied in \cite%
{Deffayet:2009mn,Deffayet:2010zh}. From another point of view the Galileon
Lagrangian can be obtained as a special non-relativistic limit of the
Dirac-Born-Infeld Lagrangian describing the fluctuations of the $d$%
-dimensional brane in the $d+1$ dimensional space-time \cite{deRham:2010eu}.
For a pedagogical introduction into the Galileon physics as well as for the
complete list of literature see e.g. \cite{Khoury:2013tda}.

Putting aside very important cosmological aspects, the Galileon theory
itself has an amazing structure which has been studied intensively in the
literature (for pedagogical introductions into the technical aspects see
e.g. \cite{Curtright:2012gx,Deffayet:2013lga}). For instance on the quantum
level it exhibits the so-called non-renormalization theorem which prevents
the tree-level Galileon couplings from obtaining the quantum corrections
stemming from loops \cite{Luty:2003vm, Hinterbichler:2010xn, deRham:2012ew,
Brouzakis:2013lla}. Another interesting feature is the existence of
dualities, i.e. such transformations of fields and coordinates which
preserve the form of the Galileon Lagrangian, though it changes its
couplings. The duality transformations therefore interrelate different
Galileon theories on the contrary to the symmetry transformations which
leave the action invariant. The first such a duality has been recognized
already in the paper \cite{Nicolis:2008in} where it was shown that the
transformation $\phi \to \phi + \tfrac{1}{4}H^{2}x^{2}$ converts one form of
the Lagrangian into another one. The latter then describes the fluctuations
of the Galileon field about the de Sitter background solution. Another
example of duality was mentioned and studied in \cite{Curtright:2012gx} and
it corresponds to the dual Legendre transform of the field. The most
interesting duality has been discovered in \cite{Fasiello:2013woa} in the
context of massive gravity and bigravity and has been further studied in
\cite{deRham:2013hsa}.

In this paper we study these dualities from the unified point of view. We
make use of the fact that the general Galileon theory can be understood as a
low-energy effective theory describing the Goldstone bosons corresponding to
the spontaneously broken symmetry according to the pattern $GAL(d,1)\to
ISO(d-1,1)$ where $GAL(d,1)$ is the so-called Galileon group and its
Lagrangian can be identified with generalized Wess-Zumino-Witten terms \cite%
{Goon:2012dy}. This allows us to classify the most general duality
transformation and identify it as a non-linear coordinate transformations on
the coset space $GAL(d,1)/SO(d-1,1)$. As we will show such duality
transformations form a four-parametric group which can be identified with $%
GL(2,\mathbf{R})$ and which contains all the above mentioned dualities as
special cases. We will also study the representation of this duality group
on the Galileon theory space and give examples of physical applications of
the duality. Namely we discuss the duality of classical covariant phase
spaces and corresponding observables, the duality of fluctuations on the the
classical background, the dual realization of the symmetries, the duality of
the $S$ matrix and its applications on the tree and one-loop level. We also
classify the Galileon theories with respect to the duality generated with
specific subgroup of $GL(2,\mathbf{R})$ which leaves the $S$ matrix
invariant or under which the tree-level amplitudes trivially scale. We
illustrate most of the above topics by means of explicit examples.

This paper is organized as follows. First, in Section~\ref{motivations} we
introduce the Galileon symmetry and Lagrangian, discuss the Feynman rules
and as an illustration we calculate the tree-level amplitudes up to the
five-point one. In Section~\ref{sec3coset} we review the coset construction
of the Galileon Lagrangian. Section~\ref{sec4galileon} and~\ref{sec5gl2}
contain the main results of this work. In Section~\ref{sec4galileon} we
construct the most general duality transformations and in Section~\ref%
{sec5gl2} we discuss their group structure. Several applications then follow
in Section~\ref{sec6applications}. Some technical details and alternative
approaches are postponed in appendices.

\section{Introductory remarks on Galileon in flat space \label{motivations}}

In this section we fix our notation and introduce the classical Galileon
Lagrangian. Also some formulae which will be useful in the next sections are
presented. We also explicitly evaluate the Feynman rules and as a motivation
we calculate the tree-level scattering amplitudes up to five particles in
the in and out states.

\subsection{The Galileon Lagrangian}

The Galileon represents the most general theory of a real scalar field $\phi
$ in flat $d-$ dimensional space-time the action $S[\phi ]$ of which is
invariant with respect to the Galilean symmetry
\begin{equation}
\delta _{a,b}\phi =a+b\cdot x,  \label{galileon_symmetry}
\end{equation}%
where $a$ and $b_{\mu }$ are real parameters. Therefore the Galileon
Lagrangian $\mathcal{L}_{G}$ changes under this symmetry at most by a total
derivative%
\begin{equation}
\delta _{a,b}\mathcal{L}_{G}=\partial \cdot V_{a,b}.  \label{galileon}
\end{equation}%
At the quantum level the Lagrangian can be written in the general form%
\begin{equation}
\mathcal{L}_{G}=\mathcal{L}+\mathcal{L}_{CT}
\end{equation}%
where $\mathcal{L}$ is the leading (classical) part and $\mathcal{L}_{CT}$
corresponds to the higher order counterterms needed for a consistent
perturbative calculation of the quantum corrections. The latter part of the
Lagrangian will be discussed in more details in section \ref{CT}, here we
concentrate on the leading part $\mathcal{L}$. This can be determined
uniquely (up to $d+1$ arbitrary coupling constants) by a second requirement
demanding that the classical equations of motion corresponding to $\mathcal{L%
}$ contain at most second order derivatives of the field. As it has been
proven in the seminal paper \cite{Nicolis:2008in}, (see also \cite%
{Curtright:2012gx} and \cite{Deffayet:2013lga} for detailed pedagogical
introduction and many useful formulae), these conditions allow just $d+1$
possible terms in the Lagrangian
\begin{equation}
\mathcal{L}=\sum_{n=1}^{d+1}d_{n}\mathcal{L}_{n}=\sum_{n=1}^{d+1}d_{n}\phi
\mathcal{L}_{n-1}^{\mathrm{der}}  \label{galileon_Lagrangian}
\end{equation}%
where $d_{n}$ are real coupling constants and $\mathcal{L}_{n}^{\mathrm{der}%
} $ can be constructed from $d-$dimensional Levi-Civita tensor $\varepsilon
^{\mu _{1}\ldots \mu _{d}}$, the flat-space metric tensor $\eta _{\mu \nu }$
and the matrix of the second derivatives of the field $\partial \partial
\phi $ as follows\footnote{%
We use the convention $\eta _{\mu \nu }=\mathrm{diag}(1,-1,\ldots ,-1)$, $%
\varepsilon ^{0,1,\ldots ,d-1}=1$.}
\begin{equation}
\mathcal{L}_{n}^{\mathrm{der}}=\varepsilon ^{\mu _{1}\ldots \mu
_{d}}\varepsilon ^{\nu _{1}\ldots \nu _{d}}\prod\limits_{i=1}^{n}\partial
_{\mu _{i}}\partial _{\nu _{i}}\phi \prod\limits_{j=n+1}^{d}\eta _{\mu
_{j}\nu _{j}}=(-1)^{d-1}(d-n)!\det \left\{ \partial ^{\nu _{i}}\partial
_{\nu _{j}}\phi \right\} _{i,j=1}^{n}.  \label{basic_L}
\end{equation}%
In four dimensions we have explicitly\footnote{%
Here (and in what follows) we use condensed notation where the dot means
contraction of the adjacent Lorentz indices, e.g.%
\begin{equation*}
\partial \partial \phi \cdot \partial \partial \phi :\partial \partial \phi
=\partial _{\mu }\partial _{\sigma }\phi \cdot \partial ^{\sigma }\partial
_{\nu }\phi :\partial ^{\mu }\partial ^{\nu }\phi
\end{equation*}%
}%
\begin{eqnarray}
\mathcal{L}_{0}^{\mathrm{der}} &=&-4!  \notag \\
\mathcal{L}_{1}^{\mathrm{der}} &=&-6\square \phi  \notag \\
\mathcal{L}_{2}^{\mathrm{der}} &=&-2\left[ \left( \square \phi \right)
^{2}-\partial \partial \phi :\partial \partial \phi \right]  \notag \\
\mathcal{L}_{3}^{\mathrm{der}} &=&-\left[ \left( \square \phi \right)
^{3}+2\partial \partial \phi \cdot \partial \partial \phi :\partial \partial
\phi -3\square \phi \partial \partial \phi \cdot \partial \partial \phi %
\right]  \notag \\
\mathcal{L}_{4}^{\mathrm{der}} &=&-\left[ \left( \square \phi \right)
^{4}-6\left( \square \phi \right) ^{2}\partial \partial \phi :\partial
\partial \phi +8\square \phi \partial \partial \phi \cdot \partial \partial
\phi :\partial \partial \phi \right.  \notag \\
&&\left. -6\partial \partial \phi \cdot \partial \partial \phi \cdot
\partial \partial \phi :\partial \partial \phi +3\left( \partial \partial
\phi :\partial \partial \phi \right) ^{2}\right] .
\end{eqnarray}%
The equation of motion is then%
\begin{equation}
\frac{\delta S[\phi ]}{\delta \phi }=\sum_{n=1}^{d+1}nd_{n}\mathcal{L}%
_{n-1}^{\mathrm{der}}=0  \label{eom}
\end{equation}%
and involves just the second derivatives of the Galileon field.

Let us note that the operator basis $\mathcal{L}_{n}$ is not unique, we can
choose also another set which differs from (\ref{basic_L}) by a total
derivative and possible re-scaling. One of the many equivalent forms of the
Lagrangian which can be obtained from (\ref{basic_L}) by means of the
integration by parts and simple algebra is%
\begin{equation}
\widetilde{\mathcal{L}}=\sum_{n=1}^{d+1}c_{n}\widetilde{\mathcal{L}}%
_{n}=\sum_{n=1}^{d+1}c_{n}\left( \partial \phi \cdot \partial \phi \right)
\mathcal{L}_{n-2}^{\mathrm{der}}  \label{alternative_basic_L}
\end{equation}%
Let us mention useful formulae (for derivation see e.g. \cite%
{Deffayet:2013lga})%
\begin{equation}
\left( \partial \phi \cdot \partial \phi \right) \mathcal{L}_{n-1}^{\mathrm{%
der}}=-\frac{2(d-n+1)}{n+1}\left[ \phi \mathcal{L}_{n}^{\mathrm{der}%
}-\partial _{\mu }\left( H_{n}^{\mu }+\frac{n-1}{d-n+1}G_{n}^{\mu }\right) %
\right]  \label{lagrangian_relation}
\end{equation}%
where%
\begin{align}
H_{n}^{\mu }& =\phi \partial _{\nu _{1}}\phi \varepsilon ^{\mu \mu
_{2}\ldots \mu _{d}}\varepsilon ^{\nu _{1}\ldots \nu
_{d}}\prod\limits_{i=2}^{n}\partial _{\mu _{i}}\partial _{\nu _{i}}\phi
\prod\limits_{j=n+1}^{d}\eta _{\mu _{j}\nu _{j}} \\
G_{n}^{\mu }& =\frac{1}{2}\left( \partial \phi \cdot \partial \phi \right)
\partial ^{\mu }\phi \varepsilon ^{\mu _{2}\ldots \mu _{n}\alpha _{1}\ldots
\alpha _{d-n+1}}\varepsilon ^{\nu _{2}\ldots \nu _{n}\beta _{1}\ldots \beta
_{d-n+1}}\prod\limits_{j=1}^{d-n+1}\eta _{\alpha _{j}\beta
_{j}}\prod\limits_{k=2,k\neq i}^{n}\partial _{\mu _{k}}\partial _{\nu
_{k}}\phi .
\end{align}%
and thus after integration and omitting the surface terms we get
\begin{equation}
\int \mathrm{d}^{d}x\left( \partial \phi \cdot \partial \phi \right)
\mathcal{L}_{n-1}^{\mathrm{der}}=-\int \mathrm{d}^{d}x\frac{2(d-n+1)}{n+1}%
\phi \mathcal{L}_{n}^{\mathrm{der}}.  \label{prevod}
\end{equation}

\subsection{The Feynman rules and tree-level amplitudes}

For further convenience let us also write down explicitly the Feynman rule
for $n-$point vertex%
\begin{equation}
\mathcal{V}_{n}(1,2,\ldots ,n)=(-1)^{n}d_{n}(d-n+1)!(n-1)!\sum_{\sigma \in
Z_{n}}G(p_{\sigma (1)},p_{\sigma (2)},\ldots ,p_{\sigma (n-1)})\,,
\label{Feynman_rule}
\end{equation}%
where we have introduced the Gram determinant $G(p_{1},\ldots ,p_{n-1})$
\begin{equation}
G(p_{1},\ldots ,p_{n-1})=-\frac{1}{(d-n+1)!}\varepsilon ^{p_{1}\ldots
p_{n-1}\mu _{n}\ldots \mu _{d}}\varepsilon ^{p_{1},\ldots ,p_{n-1}\nu
_{n}\ldots \nu _{d}}\prod\limits_{j=n}^{d}\eta _{\mu _{j}\nu _{j}}
\end{equation}%
and where the sum is over the cyclic permutations only\footnote{%
Note that, the Gram determinant is independent on the ordering of the vector
arguments.}.

\begin{figure}[h]
\begin{center}
\includegraphics[scale=0.8]{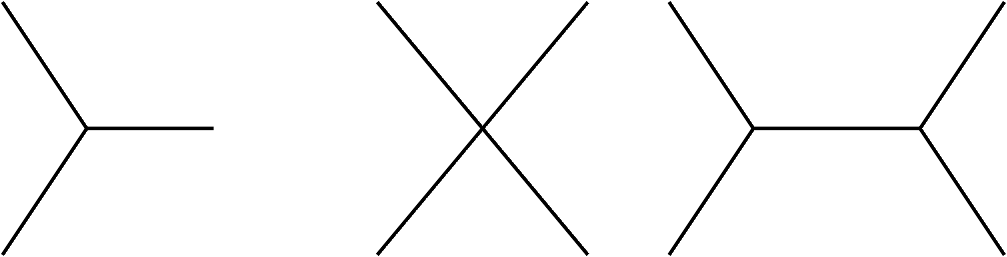}\\[0.2cm]
\includegraphics[scale=0.8]{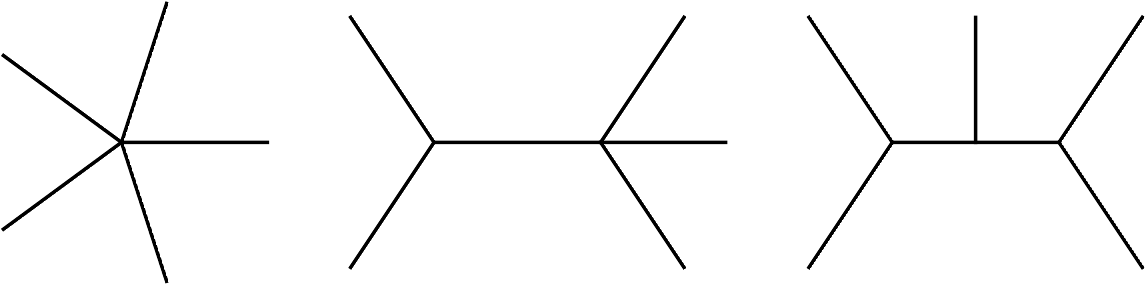}
\end{center}
\caption{The topologies of Feynman diagrams at the tree-level for the
three-point, four-point (first line) and five-point (second line) Galileon
scattering amplitudes.}
\label{fig:trees}
\end{figure}

Using this Feynman rules, one can in principle calculate any tree-level $n-$%
point amplitude in the pure Galileon theory. What it means for $n=3,4,5$ in
the language of Feynman diagrams is depicted in Fig.~\ref{fig:trees}. Note
that crossing is tacitly assumed for these graphs which finally leads to
four diagrams for 4-pt scattering and 26 for 5-pt scattering. However, due
to the complicated structure of the vertices the evaluation of the
individual contributions of the Feynman graphs is not an easy task. The most
economic way how to organize the rather lengthy and untransparent
calculation is the machinery of the Berends-Giele like recursion relations%
\footnote{%
For an application to a similar problem see e.g. \cite{naseamplitudy}.} \cite%
{Berends:1987me} which allows for an efficient computer algoritmization of
the problem.
In four dimension (in the theory without tadpole and with canonical kinetic
term, i.e. with $d_{1}=0$ and $d_{2}=1/12$) we get to the following results
\begin{eqnarray}
\mathcal{M}(1,2,3) &=&6d_{3}G(1,2)=\frac{3}{2}d_{3}p_{3}^{4}=0  \label{M3} \\
\mathcal{M}(1,2,3,4) &=&12(2d_{4}-9d_{3}^{2})G(1,2,3)  \label{M4} \\
\mathcal{M}(1,2,3,4,5) &=&-24\left( 72d_{3}^{3}-24d_{3}d_{4}+5d_{5}\right)
G(1,2,3,4)  \label{M5}
\end{eqnarray}%
(we were also capable to calculate the 6-pt diagrams which involves 235
Feynman diagrams). Without the deeper understanding of the structure of the
Galileon theory these results look suspiciously simple;\footnote{%
Note that, while the four- and five-point amplitudes are sums of Feynman
graphs including those with one and two propagators naively generating pole
terms (see Fig. \ref{fig:trees}), the resulting amplitude is represented by
a purely contact term.} in fact it was our main motivation for starting to
study this model more systematically. In what follows we shall i.a. show how
to understand these results and how they can be obtained almost without
calculation on a single sheet of paper.

\section{Coset construction of the Galileon action\label{sec3coset}}

The Galileon field can be also interpreted as a Goldstone boson
corresponding to the spontaneously broken Galileon symmetry \cite%
{Goon:2012dy}. Therefore, to obtain the most general Lagrangian for the
Galileon, the general theory of nonlinear realization \cite{Coleman:1969sm,
Callan:1969sn, Volkov:1973vd, Ogievetsky:1974} should be used. Because the
localized Galilean symmetries are non independent, the number of Goldstone
bosons is not equal to the number of the broken generators and an additional
constraint known as the inverse Higgs constraint \cite{Ivanov:1975zq} has to
be introduced. However, as discussed in \cite{Goon:2012dy}, only the
counterterm part $\mathcal{L}_{CT}$ can be obtained in this way. \ The
classical Galileon Lagrangian $\mathcal{L}$ is in fact invariant only up to
the total derivative and the corresponding action represents an analogue of
the Wess-Zumino-Witten term \cite{Wess:1971yu, Witten:1983tw,
D'Hoker:1994ti, D'Hoker:1995it} originally known from the effective low
energy theory of QCD. \ Such a term of the action can be reconstructed from
its variation and is thus expressed as $d+1$ dimensional integral. In this
section we give a brief review of the coset construction based on the
nonlinearly realized Galileon symmetry and of the interpretation of the
classical galileon Lagrangian as the generalized \ Wess-Zumino-Witten term.
Further details and generalizations can be found in the original paper \cite%
{Goon:2012dy}.

\subsection{Nonlinear realization of the Galilean symmetry}

The Galilean symmetry is a prominent example of the so called non-uniform
symmetry, i.e. a symmetry which does not commute with the space-time
translations \cite{Watanabe:2011ec, Brauner:2014aha}. Indeed, denoting the
infinitesimal translations and Galilean transformations of the Galileon
field $\delta _{c}\phi $ and $\delta _{a,b}\phi $ respectively,
\begin{eqnarray}
\delta _{c}\phi &=&c\cdot \partial \phi \\
\delta _{a,b}\phi &=&a+b\cdot x,
\end{eqnarray}%
we get
\begin{equation}
\left[ \delta _{c},\delta _{a,b}\right] \phi =c\cdot b=\delta _{c\cdot
b,0}\phi
\end{equation}%
Let us add to this transformations also the Lorentz rotations and boosts $%
\delta _{\omega }$%
\begin{equation}
\delta _{\omega }\phi =\frac{1}{2}\omega ^{\mu \nu }(x_{\mu }\partial _{\nu
}-x_{\nu }\partial _{\mu })\phi
\end{equation}%
we get then%
\begin{equation}
\left[ \delta _{\omega },\delta _{a,b}\right] \phi =-b\cdot \omega \cdot
x=\delta _{0,-b\cdot \omega }\phi =\delta _{0,\omega \cdot b}\phi
\end{equation}%
Therefore the infinitesimal transformations $\ \delta _{c}$, $\delta
_{\omega }$, $\delta _{a,b}$ form a closed algebra with generators $P_{a}$, $%
J_{ab}=-J_{ba}$, $A$ and $B_{a}$ respectively. In terms of these generators
\begin{eqnarray}
\delta _{c} &=&-\mathrm{i}c^{a}P_{a} \\
\delta _{\omega } &=&-\frac{\mathrm{i}}{2}\omega ^{ab}J_{ab} \\
\delta _{a,b} &=&-\mathrm{i}aA-\mathrm{i}b^{a}B_{a}
\end{eqnarray}%
and the commutator algebra can be rewritten in the form of the Galileon
algebra $\mathfrak{gal}(d,1)$
\begin{eqnarray}
\left[ P_{a},P_{b}\right] &=&[P_{a},A]=[B_{a},A]=\left[ J_{ab},A\right] =0
\notag \\
\lbrack P_{a},B_{b}] &=&\mathrm{i}\eta _{ab}A  \notag \\
\lbrack J_{ab},P_{c}] &=&\mathrm{i}\left( \eta _{bc}P_{a}-\eta
_{ac}P_{b}\right)  \notag \\
\lbrack J_{ab},B_{c}] &=&\mathrm{i}\left( \eta _{bc}B_{a}-\eta
_{ac}B_{b}\right)  \notag \\
\lbrack J_{ab},J_{cd}] &=&\mathrm{i}\left( \eta _{bc}J_{ad}+\eta
_{ad}J_{bc}-\eta _{ac}J_{bd}-\eta _{bd}J_{ac}\right)
\end{eqnarray}%
which corresponds to the Galileon group $GAL(d,1)$ (see \cite{Goon:2012dy}).

Within the Galileon theory this group is realized non-linearly on the fields
$\phi $ and space-time coordinates $x^{\mu }$. Indeed, for the generators $%
P_{a}$, $A$ a $B_{a}$ we have%
\begin{eqnarray}
-\mathrm{i}P_{a}x^{\mu } &=&\delta^{\mu }_{a}  \notag \\
-\mathrm{i}A\phi &=&1  \notag \\
-\mathrm{i}B_{a}\phi &=&x_{a}.
\end{eqnarray}%
The generators $A$ a $B_{a}$ are spontaneously broken, the order parameter
can be identified with
\begin{equation*}
\left\langle 0|\delta _{a,b}\phi |0\right\rangle =a+b\cdot x.
\end{equation*}%
This corresponds to the symmetry breaking pattern $GAL(d,1)\rightarrow
ISO(d-1,1)$. Let us note that the above transformations are not completely
independent in the sense of refs. \cite{Low:2001bw, Brauner:2014aha}.
Indeed, their localized forms with space-time dependent parameters $a(x)$
and $b_{\mu }(x)$ yield the same local transformation
\begin{equation}
\delta _{a(x),b(x)}\phi =a(x)+b(x)\cdot x
\end{equation}%
which corresponds to the local shift of the Galileon field $\phi $. More
precisely, writing $a_{b}(x)=x\cdot b(x)$, we can identify
\begin{equation}
\delta _{a_{b}(x),0}=\delta _{0,b(x)}.
\end{equation}%
Physically this means that the local fluctuations of the order parameter
which correspond to the Goldstone modes are not independent. As a result the
particle spectrum does not contain the same number of Goldstone bosons as is
the number of the broken generators (i.e. $d+1$) but just one zero mass mode
which can be identified with the Galileon field $\phi $. (see \cite%
{Brauner:2014aha, McArthur:2010zm} for recent discussion of this issue).

Construction of the low energy effective Lagrangian describing the dynamics
of the Goldstone bosons corresponding to the spontaneous breakdown of the
non-uniform symmetries is a generalization of the coset construction of
Callan, Coleman, Wess and Zumino \cite{Coleman:1969sm, Callan:1969sn} and
has been formulated by Volkov \cite{Volkov:1973vd} and Ogievetsky \cite%
{Ogievetsky:1974}. Applied to the Galileon case, where the only linearly
realized generators of the Galileon group are the Lorentz rotations and
boosts $J_{ab}$, \ the coset space is $GAL(d,1)/SO(d-1,1)$ the elements of
which are the left cosets $\{gSO(d-1,1)\}$ where $g\in GAL(d,1)$. The
coordinates on this coset space can be chosen in a standard way by means of
a unique choice of the representant $U$ of each left coset. Such a
representant can be written in terms of the coset coordinates $x^{a}$, $\phi
$ and $L^{a}$ as%
\begin{equation}
U\equiv U(x,\phi ,L)=\exp (\mathrm{i}x^{a}P_{a})\exp \left( \mathrm{i}\phi A+%
\mathrm{i}L^{a}B_{a}\right)  \label{coset_parametrization}
\end{equation}%
The general element of the galileon group $g\in GAL(d,1)$
\begin{equation}
g=\exp \left( \frac{\mathrm{i}}{2}\omega ^{ab}J_{ab}\right) \exp \left(
\mathrm{i}c^{a}P_{a}\right) \exp \left( \mathrm{i}aA+\mathrm{i}%
b^{a}B_{a}\right)  \label{group_element}
\end{equation}%
acts on the cosets by means of the left multiplication and consequently the
coset coordinates transform according to
\begin{equation}
U^{\prime }\equiv U(x^{\prime },\phi ^{\prime },L^{\prime })=gUh^{-1}
\end{equation}%
where $h\equiv h(g,x,\phi ,L)\in SO(d-1,1)$ is the compensator arranging $%
U^{\prime }$ to be of the form (\ref{coset_parametrization}). As usual, the
stability group, which is the Lorentz group $SO(d-1,1)$ here, is realized
linearly ($\phi $ transformed as a scalar and $x$ and $L$ are vectors), and
the general element (\ref{group_element}) of the Galileon group $g\in
GAL(d,1)$ acts on $U$ as follows%
\begin{equation}
gU(x,\phi ,L)=\exp (\mathrm{i}x^{a\prime }P_{a})\exp \left( \mathrm{i}\phi
^{\prime }A+\mathrm{i}L^{a\prime }B_{a}\right) \exp \left( \frac{\mathrm{i}}{%
2}\omega ^{ab}J_{ab}\right)
\end{equation}%
where%
\begin{equation*}
x^{a\prime }=\Lambda (\omega )_{b}^{a}(x^{b}+c^{b}),~\phi ^{\prime }=\phi
+a+b\cdot x,~~L^{a\prime }=\Lambda (\omega )_{b}^{a}\left( L^{b}+b^{b}\right)
\end{equation*}%
and where%
\begin{equation}
\Lambda (\omega )=\exp \left( \frac{1}{2}\omega ^{ab}M_{ab}\right)
,~~~\left( M_{ab}\right) _{d}^{c}=\delta _{a}^{c}\eta _{bd}-\delta
_{b}^{c}\eta _{ad}.  \notag
\end{equation}%
As a result, for the general element of the Galileon group $g\in GAL(d,1)$
we have the following compensator
\begin{equation}
h(g,x,\phi ,L)=\exp \left( \frac{\mathrm{i}}{2}\omega ^{ab}J_{ab}\right) .
\label{compensator}
\end{equation}%
Note that, the compensator does not depend on the coset coordinates $\left(
x,\phi ,L\right) $ and therefore treating $\phi $ and $L^{a}$ as space-time
dependent fields, the compensator has no explicit or implicit $x$
dependence. This simplifies the application of the general recipe \cite%
{Volkov:1973vd, Ogievetsky:1974} significantly, because the requirement of
the invariance with respect to the local stability group can be replaced by
much simpler requirement of global invariance.

\subsection{Construction of the invariant Lagrangian}

The basic object for the construction of the effective Lagrangian is the
Maurer-Cartan form, which can be expressed in the coordinates $x^{a}$, $\phi
$ and $L^{a}$ as%
\begin{eqnarray}
\frac{1}{\mathrm{i}}U^{-1}\mathrm{d}U &=&\exp \left( -\mathrm{i}\phi A-%
\mathrm{i}L^{b}B_{b}\right) \exp (-\mathrm{i}x^{d}P_{d})\mathrm{d}\left(
\exp (\mathrm{i}x^{c}P_{c})\exp \left( \mathrm{i}\phi A+\mathrm{i}%
L^{a}B_{a}\right) \right)  \notag \\
&=&\exp \left( -\mathrm{i}L^{b}B_{b}\right) \left( \mathrm{d}x^{c}P_{c}+%
\mathrm{d}\phi A+\mathrm{d}L^{d}B_{d}\right) \exp \left( \mathrm{i}%
L^{a}B_{a}\right)
\end{eqnarray}%
where in the second line we have used the fact that $A$ commutes with all
the other generators. Using further
\begin{equation}
\exp \left( -\mathrm{i}L^{b}B_{b}\right) P_{c}\exp \left( \mathrm{i}%
L^{a}B_{a}\right) =P_{c}-L^{b}\eta _{bc}A
\end{equation}
we get finally
\begin{eqnarray}
\frac{1}{\mathrm{i}}U^{-1}\mathrm{d}U &=&\mathrm{d}x^{c}P_{c}+\left( \mathrm{%
d}\phi -L^{b}\eta _{bc}\mathrm{d}x^{c}\right) A+\mathrm{d}L^{d}B_{d}  \notag
\\
&\equiv &\omega _{P}^{c}P_{c}+\omega _{A}A+\omega _{B}^{d}B_{d}
\label{Maurer_Cartan}
\end{eqnarray}%
The form $\omega _{P}^{c}$ is particularly simple. In the general case we
get $\omega _{P}^{a}=e_{\mu }^{a}(x)\mathrm{d}x^{\mu }$ and $e_{\mu }^{a}$
plays a role of $d-$bein, intertwining the abstract group indices $a,\ldots $
with space-time indices $\mu ,\ldots $ and the flat metric $\eta _{ab}$ with
the effective space-time metric $g_{\mu \nu }$ according to%
\begin{equation}
g_{\mu \nu }=\eta _{ab}e_{\mu }^{a}e_{\nu }^{b}
\end{equation}%
In our case $e_{\mu }^{a}=\delta _{\mu }^{a}$ , the space-time metric is
therefore flat%
\begin{equation}
g_{\mu \nu }=\eta _{\mu \nu }
\end{equation}%
and the abstract group indices are identical with the space-time ones. This
also ensures that the volume element $\mathrm{d}^{d}x$ is invariant with
respect to the non-linearly realized Galileon group. Note also that there is
no term of the form $\omega _{J}^{ab}J_{ab}$ on the right hand side of (\ref%
{Maurer_Cartan}). This implies that the usual group covariant derivative is
in our case identical with ordinary partial derivatives $\partial _{\alpha }$%
. The forms $\omega _{P}^{c}$, $\omega _{A}$ and $\omega _{B}^{d}$ transform
under a general element of the Galileon (\ref{group_element}) group $g\in
GAL(d,1)$ (cf. (\ref{compensator})) according to%
\begin{eqnarray}
\omega _{P}^{\prime a} &=&\Lambda (\omega )_{b}^{a}\omega _{P}^{b}\,  \notag
\\
\omega _{B}^{\prime a} &=&\Lambda (\omega )_{b}^{a}\omega _{B}^{b}  \notag \\
\omega _{A}^{\prime } &=&\omega _{A}.
\end{eqnarray}%
These forms span three irreducible representations of the stability group $%
SO(d-1,1)$ (namely two vectors and one scalar) and can be therefore used
separately as the basic building blocks for the construction of the
effective Lagrangian. The general recipe requires to use this building block
and their (covariant) derivatives to construct all the possible terms which
are invariant with respect to local stability group. As we have mentioned
above, in our case we make do with ordinary partial derivatives and the last
requirement can be rephrased as the global $SO(d-1,1)$ invariance when we
identify the abstract group and space-time indices with help of the trivial $%
d$-bein $\delta _{\mu }^{a}$. Therefore, writing%
\begin{equation*}
~~\mathrm{d}\phi (x)=\partial _{\mu }\phi (x)\mathrm{d}x^{\mu },~~\mathrm{d}%
L^{\nu }(x)=\partial _{\mu }L^{\nu }(x)\mathrm{d}x^{\mu },
\end{equation*}%
the most general invariant term of the Lagrangian is the Lorentz invariant
combinations of the fields $\partial _{\mu }L^{\nu }$ and $D_{\mu }\phi $,
where
\begin{equation}
D_{\mu }\phi \equiv \partial _{\mu }\phi -L_{\mu }
\end{equation}%
and their derivatives.

Apparently we have ended up with $d+1$ Goldstone fields $\phi $ and $L_{\mu
} $ however this is not the final answer. In fact these fields are not
independent. The standard possibility how to eliminate the unwanted degrees
of freedom is to require an additional constraint \cite{Goon:2012dy,
Brauner:2014aha}
\begin{equation}
\omega _{A}=0\Longleftrightarrow L_{\mu }=\partial _{\mu }\phi ,  \label{IHC}
\end{equation}%
which is invariant with respect to the group $GAL(d,1)$ and which is known
as the inverse Higgs constraint (IHC) \cite{Ivanov:1975zq}. Then the only
remaining nontrivial building blocks are $\partial _{\mu }\partial _{\nu
}\phi $ and its derivatives\footnote{%
Another possibility how to treat the problem of additional degrees of
freedom is based on the field redefinition
\begin{equation*}
L^{\mu }=\psi ^{\mu }+\partial ^{\mu }\phi
\end{equation*}%
where $\psi ^{\mu }$ are new fields. Then%
\begin{equation*}
\mathcal{L}(\partial _{\mu }L^{\nu },D_{\mu }\phi )=\mathcal{L}(\partial
_{\mu }L^{\nu },\partial _{\mu }\phi -L_{\mu })=\mathcal{L}(\partial _{\mu
}\psi ^{\nu }+\partial _{\mu }\partial ^{\nu }\phi ,\psi ^{\mu })
\end{equation*}%
The invariant term $M^{2}D_{\mu }\phi D^{\mu }\phi $, which was responsible
for the kinetic term of the field $\phi $ in the original Lagrangian goes
within the new parametrization in terms of $\phi $ and $\psi ^{\mu }$ into
the mass term of the field $\psi ^{\mu }$. This field then does not
correspond more to the Goldstone boson and can be integrated out from the
effective Lagrangian, provided we are interested in the dynamics of the
field $\phi $ only. We end up again with the just one nontrivial building
block $\partial _{\mu }\partial _{\nu }\phi $. See \cite{Brauner:2014aha}
for detailed discussion of this aspect of the spontaneously broken
non-uniform symmetries.} and the general Lagrangian is%
\begin{equation}
\mathcal{L}_{\mathrm{inv}}=\mathcal{L}_{\mathrm{inv}}(\partial _{\mu
}\partial _{\nu }\phi ,\partial _{\lambda }\partial _{\mu }\partial _{\nu
}\phi ,\ldots ).
\end{equation}

\subsection{Generalized Wess-Zumino-Witten terms}

The Galileon Lagrangian represents a different type of possible terms
contributing to the invariant action, namely those which are not strictly
invariant on the Lagrangian level, but are invariant only up to a total
derivative. Such terms can be identified as the generalized
Wess-Zumino-Witten (WZW) terms \cite{Wess:1971yu, Witten:1983tw}, as was
proved and discussed in detail in \cite{D'Hoker:1994ti, D'Hoker:1995it}.
From the point of view of the coset construction, the WZW terms originate in
the integrals of the closed invariant $\left( d+1\right) $-forms\footnote{%
More precisely we integrate the pull-back of the form $\omega _{d+1}$ with
respect to the map $B_{d+1}\rightarrow GAL(d,1)/SO(d-1,1)$ which maps $%
(x^{\mu },x^{d})\rightarrow (\delta _{\mu }^{a}x^{\mu },\phi (x^{\mu
},x^{d}),L^{\mu }(x^{\mu },x^{d}))$.} $\omega _{d+1}$ on $GAL(d,1)/SO(d-1,1)$
(these correspond to the variation of the action) over the $d+1$ dimensional
ball $B_{d+1}$ the boundary of which is the compactified space-time $%
S_{d}=\partial B_{d+1}$%
\begin{equation}
S^{WZW}=\int_{B_{d+1}}\omega _{d+1}  \label{S_WZW}
\end{equation}%
In order to prevent these contributions to the action to degenerate into the
strictly invariant Lagrangian terms discussed above it is necessary that the
form $\omega _{d+1}$ is not an exterior derivative of the invariant $d$-form
on $GAL(d,1)/SO(d-1,1)$. This means that $\omega _{d+1}$ has to be a
nontrivial element of the cohomology $H^{d+1}\left( GAL(d,1)/SO(d-1,1),%
\mathbf{R}\right) $ (see \cite{deAzcarraga:1997gn} and also \cite%
{deAzcarraga:1998uy} for a recent review on this topic). In the case of
Galileon such forms can be constructed out of the covariant $1$-forms $%
\omega _{P}^{\mu }$, $\omega _{A}$ a $\omega _{B}^{\mu }$ with indices
contracted appropriately to get Lorentz invariant combinations. As was shown
in \cite{Goon:2012dy}, there are $d+1$ such $\omega _{d+1}$, namely%
\begin{eqnarray}
\omega _{d+1}^{(n)} &=&\varepsilon _{\mu _{1}\ldots \mu _{d}}\omega
_{A}\wedge \omega _{B}^{\mu _{1}}\wedge \ldots \wedge \omega _{B}^{\mu
_{n-1}}\wedge \omega _{P}^{\mu _{n}}\wedge \ldots \wedge \omega _{P}^{\mu
_{d}}  \notag \\
&=&\varepsilon _{\mu _{1}\ldots \mu _{d}}\left( \mathrm{d}\phi -L_{\mu }%
\mathrm{d}x^{\mu }\right) \wedge \mathrm{d}L^{\mu _{1}}\wedge \ldots \wedge
\mathrm{d}L^{\mu _{n-1}}\wedge \mathrm{d}x^{\mu _{n}}\wedge \ldots \wedge
\mathrm{d}x^{\mu _{d}},  \label{omega_d+1}
\end{eqnarray}%
where $n=1,2,\ldots ,d+1$. These forms are closed%
\begin{equation}
\omega _{d+1}^{(n)}=\mathrm{d}\beta _{d}^{(n)}
\end{equation}%
where\footnote{%
The opposite sign of the second term in comparison with \cite{Goon:2012dy}
stems from different convention for the metric tensor $\eta _{\mu \nu }$.}
\cite{Goon:2012dy}
\begin{eqnarray}
\beta _{d}^{(n)} &=&\varepsilon _{\mu _{1}\ldots \mu _{d}}\phi ~\mathrm{d}%
L^{\mu _{1}}\wedge \ldots \wedge \mathrm{d}L^{\mu _{n-1}}\wedge \mathrm{d}%
x^{\mu _{n}}\wedge \ldots \wedge \mathrm{d}x^{\mu _{d}}  \notag \\
&&+\frac{n-1}{2(d-n+2)!}\varepsilon _{\mu _{1}\ldots \mu _{d}}L^{2}\mathrm{d}%
L^{\mu _{1}}\wedge \ldots \wedge \mathrm{d}L^{\mu _{n-2}}\wedge \mathrm{d}%
x^{\mu _{n-1}}\wedge \ldots \wedge \mathrm{d}x^{\mu _{d}}
\end{eqnarray}%
and therefore%
\begin{equation}
\int_{B_{d+1}}\omega _{d+1}^{(n)}=\int_{\partial B_{d+1}}\beta
_{d}^{(n)}=\int_{S_{d}}\beta _{d}^{(n)}
\end{equation}%
Note that the $d$-forms $\beta _{d}^{(n)}$are not invariant, and therefore $%
\omega _{d+1}^{(n)}$ are nontrivial elements of $H^{d+1}\left(
GAL(d,1)/SO(d-1,1),\mathbf{R}\right) $. Imposing now the IHC constraint (\ref%
{IHC}), we can finally identify%
\begin{equation}
\int_{S_{d}}\beta _{d}^{(n)}=\frac{1}{n}\int_{S_{d}}\mathrm{d}^{d}x\mathcal{L%
}_{n}  \label{identifikaceLn}
\end{equation}%
The classical Galileon action is therefore a linear combination of the
generalized Wess-Zumino-Witten terms corresponding to the $d+1$ elements $%
\omega _{d+1}^{(n)}=\mathrm{d}\beta _{d}^{(n)}$ of the cohomology $%
H^{d+1}\left( GAL(d,1)/SO(d-1,1),\mathbf{R}\right) $ built with help of the
forms $\omega _{P}^{\mu }$, $\omega _{A}$ a $\omega _{B}^{\mu }$.

\section{Galileon duality as a coset coordinate transformation\label%
{sec4galileon}}

The canonical coordinates $\left( x,\phi ,L\right) $ on the coset space $%
GAL(d,1)/SO(d-1,1)$ which we have defined according to (\ref%
{coset_parametrization}) are not the only possible ones. We can freely use
any other set of coordinates connected with them by a general coordinate
transformation of the form
\begin{eqnarray}
x^{\mu } &=&\xi ^{\mu }(x^{\prime },L^{\prime },\phi ^{\prime })  \notag \\
L^{\mu } &=&\Lambda ^{\mu }(x^{\prime },L^{\prime },\phi ^{\prime })  \notag
\\
\phi &=&f(x^{\prime },L^{\prime },\phi ^{\prime }).  \label{coordinat_change}
\end{eqnarray}%
Not all such new coordinates are of any use, e.g. those transformations (\ref%
{coordinat_change}) which are not covariant with respect to the $SO(d-1,1)$
symmetry will hide this symmetry in the effective Lagrangian. Even if the
covariance is respected, in the general case the resulting Lagrangian might
be difficult to recognize as a Galileon theory. In this section we shall
make a classification of those coordinate changes which preserve the general
form of the Galileon action as a linear combination of the $d+1$ terms
discussed in the previous sections (though we allow for change of the
couplings). Such a transformation of the coset coordinates can be then
interpreted as a Galileon duality.

It is obvious from (\ref{S_WZW}) and (\ref{omega_d+1}) that, provided the
forms $\omega _{P}^{\mu }$, $\omega _{A}$ a $\omega _{B}^{\mu }$ can be
expressed in the primed coordinates as a (covariant) linear combination
(with constant coefficients) of the primed forms $\omega _{P}^{\prime \mu }$%
, $\omega _{A}^{\prime }$ and $\omega _{B}^{\prime \mu }$ where%
\begin{eqnarray}
\omega _{P}^{\prime \mu } &=&\mathrm{d}x^{\prime \mu }  \notag \\
\omega _{A}^{\prime } &=&\mathrm{d}\phi ^{\prime }-L_{\mu }^{\prime }\mathrm{%
d}x^{\prime \mu }  \notag \\
\omega _{B}^{\prime \mu } &=&\mathrm{d}L^{\prime \mu },
\end{eqnarray}%
the coordinate transformation corresponds to a duality transformation of the
Galileon action. Indeed, provided\footnote{%
Note that, the constants $\alpha _{IJ}$ cannot be decorated with any Lorentz
index because the only invariant tensors at our disposal are $\eta _{\mu \nu
}$ and $\varepsilon _{\mu _{1}\ldots \mu _{d}}$.}%
\begin{eqnarray}
\omega _{B}^{\mu } &=&\alpha _{BB}\omega _{B}^{\prime \mu }+\alpha
_{BP}\omega _{P}^{\prime \mu }  \notag \\
\omega _{P}^{\mu } &=&\alpha _{PB}\omega _{B}^{\prime \mu }+\alpha
_{PP}\omega _{P}^{\prime \mu }  \notag \\
\omega _{A} &=&\alpha _{AA}\omega _{A}^{\prime }  \label{omega_omega_prime}
\end{eqnarray}%
we have%
\begin{eqnarray}
\omega _{d+1}^{(n)} &=&\varepsilon _{\mu _{1}\ldots \mu _{d}}\omega
_{A}\wedge \omega _{B}^{\mu _{1}}\wedge \ldots \wedge \omega _{B}^{\mu
_{n-1}}\wedge \omega _{P}^{\mu _{n}}\wedge \ldots \wedge \omega _{P}^{\mu
_{d}}  \notag \\
&=&\alpha _{AA}\sum_{k=0}^{d-n+1}\sum_{l=0}^{n-1}\left(
\begin{array}{c}
d-n+1 \\
k%
\end{array}%
\right) \left(
\begin{array}{c}
n-1 \\
l%
\end{array}%
\right) \alpha _{PB}^{k}\alpha _{BB}^{l}\alpha _{PP}^{d-n+1-k}\alpha
_{BP}^{n-1-l}\omega _{d+1}^{\prime (l+k+1)}  \notag \\
&&
\end{eqnarray}%
and, after imposing the IHC constraint\footnote{%
Note that, the formula $\omega _{A}=\alpha _{AA}\omega _{A}^{\prime }$
ensures a compatibility of the IHC constraint with the coordinate
transformation.} (\ref{IHC}), the corresponding term in the action satisfies%
\begin{equation}
\int_{S_{d}}\beta _{d}^{(n)}=\alpha
_{AA}\sum_{k=0}^{d-n+1}\sum_{l=0}^{n-1}\left(
\begin{array}{c}
d-n+1 \\
k%
\end{array}%
\right) \left(
\begin{array}{c}
n-1 \\
l%
\end{array}%
\right) \alpha _{PB}^{k}\alpha _{BB}^{l}\alpha _{PP}^{d-n+1-k}\alpha
_{BP}^{n-1-l}\int_{S_{d}^{\prime }}\beta _{d+1}^{\prime (l+k+1)}.
\end{equation}%
This means that the coordinate transformation maps linear combination of the
$d+1$ basic building block of the Galileon action onto different linear
combination of the same building blocks and the two apparently different
Galileon theories are in fact dual to each other.

The conditions (\ref{omega_omega_prime}) constraint the form of the duality
transformation (\ref{coordinat_change}) strongly. We have in the primed
coordinates (here and in what follows the superscript at the symbol of
partial derivative indicates the corresponding primed variable, e.g. $%
\partial ^{(\phi )}\equiv \partial /\partial \phi ^{\prime }$)%
\begin{eqnarray}
\omega _{P}^{\mu } &=&\partial _{\nu }^{(L)}\xi ^{\mu }\omega _{B}^{\prime
\nu }+\left( \partial _{\nu }^{(x)}\xi ^{\mu }+L_{\nu }^{\prime }\partial
^{\left( \phi \right) }\xi ^{\mu }\right) \omega _{P}^{\prime \nu }+\partial
^{\left( \phi \right) }\xi ^{\mu }\omega _{A}^{\prime }  \notag \\
\omega _{B}^{\mu } &=&\partial _{\nu }^{(L)}\Lambda ^{\mu }\omega
_{B}^{\prime \nu }+\left( \partial _{\nu }^{(x)}\Lambda ^{\mu }+L_{\nu
}^{\prime }\partial ^{\left( \phi \right) }\Lambda ^{\mu }\right) \omega
_{P}^{\prime \nu }+\partial ^{\left( \phi \right) }\Lambda ^{\mu }\omega
_{A}^{\prime }  \notag \\
\omega _{A} &=&\left( \partial ^{\phi }f-\Lambda _{\mu }\partial ^{\left(
\phi \right) }\xi ^{\mu }\right) \omega _{A}^{\prime }+\left( \partial _{\mu
}^{(L)}f-\Lambda _{\nu }\partial _{\mu }^{(L)}\xi ^{\nu }\right) \omega
_{B}^{\prime \mu }  \notag \\
&&+\left[ \partial _{\nu }^{(x)}f+L_{\nu }^{\prime }\partial ^{\left( \phi
\right) }f-\Lambda _{\mu }\left( \partial _{\nu }^{(x)}\xi ^{\mu }+L_{\nu
}^{\prime }\partial ^{\left( \phi \right) }\xi ^{\mu }\right) \right] \omega
_{P}^{\prime \nu }
\end{eqnarray}%
and comparing the coefficients at $\omega _{P}^{\prime \nu }$, $\omega
_{A}^{\prime }$ and $\omega _{B}^{\prime \nu }$ in the expressions for $%
\omega _{P}^{\mu }$ and $\omega _{B}^{\mu }$ with the corresponding right
hand sides of (\ref{omega_omega_prime}) we get the following set of
differential equations for $\xi ^{\mu }$ and $\Lambda ^{\mu }$%
\begin{eqnarray}
\partial ^{(\phi )}\xi ^{\mu } &=&0,~~~~\partial _{\nu }^{(L)}\xi ^{\mu
}=\delta _{\nu }^{\mu }\alpha _{PB},~~~~\partial _{\nu }^{(x)}\xi ^{\mu
}+L_{\nu }^{\prime }\partial ^{(\phi )}\xi ^{\mu }=\delta _{\nu }^{\mu
}\alpha _{PP},  \notag \\
\partial ^{(\phi )}\Lambda ^{\mu } &=&0,~~~\partial _{\nu }^{(L)}\Lambda
^{\mu }=\delta _{\nu }^{\mu }\alpha _{BB},~~~~\partial _{\nu }^{(x)}\Lambda
^{\mu }+L_{\nu }^{\prime }\partial ^{(\phi )}\Lambda ^{\mu }=\delta _{\nu
}^{\mu }\alpha _{BP}.
\end{eqnarray}%
Integration of these equations is trivial, we get (up to the additive
constants\footnote{%
We have set these additive constants equal to zero. The reason is that, if
nonzero, they corresponds to the additional combination of the space-time
translation and Galileon transformation. Both these additional contributions
are exact symmetries of the Galileon theory and does not bring about
anything new.})
\begin{eqnarray}
\xi ^{\mu } &=&\alpha _{PB}L^{\prime \mu }+\alpha _{PP}x^{\prime \mu }
\notag \\
\Lambda ^{\mu } &=&\alpha _{BB}L^{\prime \mu }+\alpha _{BP}x^{\prime \mu }.
\label{f_g}
\end{eqnarray}%
Comparison of coefficients in both expressions for $\omega _{A}$ gives,
after using the explicit form (\ref{f_g}) of $\xi ^{\mu }$ and $\Lambda
^{\mu }$, the following differential equations for $f$
\begin{equation*}
\partial ^{(\phi )}f=\alpha _{AA},~~~\partial _{\mu }^{(L)}f=\alpha
_{PB}\left( \alpha _{BB}L_{\mu }^{\prime }+\alpha _{BP}x_{\mu }^{\prime
}\right) ,~~\partial _{\nu }^{(x)}f+L_{\nu }^{\prime }\partial ^{(\phi
)}f=\alpha _{PP}\left( \alpha _{BB}L_{\nu }^{\prime }+\alpha _{BP}x_{\nu
}^{\prime }\right) ~
\end{equation*}%
From the first equation it follows%
\begin{equation}
f=\alpha _{AA}\phi ^{\prime }+F(x^{\prime },L^{\prime })
\end{equation}%
where the function $F$ of two variables satisfies%
\begin{eqnarray}
\partial _{\mu }^{(L)}F &=&\alpha _{PB}\left( \alpha _{BB}L_{\mu }^{\prime
}+\alpha _{BP}x_{\mu }^{\prime }\right)  \notag \\
\partial _{\nu }^{(x)}F &=&\alpha _{PP}\left( \alpha _{BB}L_{\nu }^{\prime
}+\alpha _{BP}x_{\nu }^{\prime }\right) -\alpha _{AA}L_{\nu }^{\prime }.
\label{h_equations}
\end{eqnarray}%
Integration of these equations is possible only if the integrability
conditions are satisfied
\begin{equation}
\partial _{\nu }^{(x)}\partial _{\mu }^{(L)}F=\partial _{\mu }^{(L)}\partial
_{\nu }^{(x)}F
\end{equation}%
This constraints the possible values of the constants $\alpha _{IJ}$
\begin{equation}
\alpha _{PB}\alpha _{BP}=\alpha _{PP}\alpha _{BB}-\alpha _{AA}
\end{equation}%
which means%
\begin{equation}
\alpha _{AA}=\det \left( \mathbf{\alpha }\right) \equiv \det \left(
\begin{array}{cc}
\alpha _{PP} & \alpha _{PB} \\
\alpha _{BP} & \alpha _{BB}%
\end{array}%
\right) .  \label{integrability_alpha}
\end{equation}%
Imposing this constraint, the equations (\ref{h_equations}) transforms into
the form%
\begin{eqnarray}
\partial _{\mu }^{(L)}F &=&\alpha _{PB}\alpha _{BB}L_{\mu }^{\prime }+\alpha
_{PB}\alpha _{BP}x_{\mu }^{\prime }  \notag \\
\partial _{\nu }^{(x)}F &=&\alpha _{BP}\alpha _{PB}L_{\nu }^{\prime }+\alpha
_{PP}\alpha _{BP}x_{\nu }^{\prime }.
\end{eqnarray}%
which can be easily integrated (again up to the additive constant
corresponding to trivial shift of $\phi $)
\begin{eqnarray}
F &=&\int_{0}^{\left( x^{\prime },L^{\prime }\right) }\left( \mathrm{d}%
x^{\prime }\cdot \partial ^{(x)}F+\mathrm{d}L^{\prime }\cdot \partial
^{(L)}F\right)  \notag \\
&=&\frac{1}{2}\left( \alpha _{PB}\alpha _{BB}L^{\prime 2}+2\alpha
_{PB}\alpha _{BP}x^{\prime }\cdot L^{\prime }+\alpha _{PP}\alpha
_{BP}x^{\prime 2}\right)
\end{eqnarray}%
As a result we get the most general formulae\footnote{%
Up to the remnants of the omitted additive constants, as discussed above.}
for the duality transformation of the coset coordinates in the form
\begin{eqnarray}
x^{\mu } &=&\alpha _{PP}x^{\prime \mu }+\alpha _{PB}L^{\prime \mu
},~~~L^{\mu }=\alpha _{BB}L^{\prime \mu }+\alpha _{BP}x^{\prime \mu }  \notag
\\
\phi &=&\det \left( \mathbf{\alpha }\right) \phi ^{\prime }+\frac{1}{2}%
\left( \alpha _{PB}\alpha _{BB}L^{\prime 2}+2\alpha _{PB}\alpha
_{BP}x^{\prime }\cdot L^{\prime }+\alpha _{PP}\alpha _{BP}x^{\prime
2}\right) .  \label{u_dualita}
\end{eqnarray}%
Under this transformation the basic building blocs of the Galileon
Lagrangian transform as
\begin{eqnarray}
\omega _{B}^{\mu } &=&\alpha _{BB}\omega _{B}^{\prime \mu }+\alpha
_{BP}\omega _{P}^{\prime \mu }  \notag \\
\omega _{P}^{\mu } &=&\alpha _{PB}\omega _{B}^{\prime \mu }+\alpha
_{PP}\omega _{P}^{\prime \mu }  \notag \\
\omega _{A} &=&\det \left( \mathbf{\alpha }\right) \omega _{A}^{\prime }.
\label{omega_duality}
\end{eqnarray}%
These transformations are parametrized by four constants arranged in the real
$2\times 2$ matrix%
\begin{equation}
\mathbf{\alpha }=\left(
\begin{array}{cc}
\alpha _{PP} & \alpha _{PB} \\
\alpha _{BP} & \alpha _{BB}%
\end{array}%
\right) .
\end{equation}%
Imposing the IHC constraint (\ref{IHC}) we get finally
\begin{eqnarray}
x &=&\alpha _{PP}x^{\prime }+\alpha _{PB}\partial ^{\prime }\phi ^{\prime }
\notag \\
\phi &=&\det \left( \mathbf{\alpha }\right) \phi ^{\prime }+\frac{1}{2}%
\left( \alpha _{PB}\alpha _{BB}\partial ^{\prime }\phi ^{\prime }\cdot
\partial ^{\prime }\phi ^{\prime }+2\alpha _{PB}\alpha _{BP}x^{\prime }\cdot
\partial ^{\prime }\phi ^{\prime }+\alpha _{PP}\alpha _{BP}x^{\prime
2}\right)  \notag \\
\partial \phi &=&\alpha _{BB}\partial ^{\prime }\phi ^{\prime }+\alpha
_{BP}x^{\prime }.  \label{generalized_duality_alpha}
\end{eqnarray}%
Let us note that the last formula of (\ref{generalized_duality_alpha}) (the
transformation of $\partial \phi $) is compatible with the first two as a
result of the compatibility of the IHC constraint with the coordinate
transformation mentioned above. We can also prove this easily by explicit
calculation (see Appendix B).

Let us finally write down the explicit formula for the duality in terms of
the Galileon action. It is expressed by the identity%
\begin{equation}
S[\phi ]=S_{\mathbf{\alpha }}[\phi ^{\prime }]
\label{dual_action_definition}
\end{equation}%
where%
\begin{eqnarray}
S[\phi ] &=&\int \mathrm{d}^{d}x\sum_{n=1}^{d+1}d_{n}\mathcal{L}_{n}
\label{S_alpha} \\
S_{\mathbf{\alpha }}[\phi ] &=&\int \mathrm{d}^{d}x\sum_{n=1}^{d+1}d_{n}(%
\mathbf{\alpha })\mathcal{L}_{n}
\end{eqnarray}%
and the couplings of the two dual action are interrelated as%
\begin{equation}
d_{n}(\mathbf{\alpha })=\sum_{m=1}^{d+1}A_{nm}(\mathbf{\alpha })d_{m}
\end{equation}%
where the matrix $A_{nm}(\mathbf{\alpha })$ has the following form%
\begin{equation}
A_{nm}(\mathbf{\alpha })=\det \left( \mathbf{\alpha }\right) \frac{m}{n}%
\sum_{k=0}^{d-m+1}\sum_{l=0}^{m-1}\left(
\begin{array}{c}
d-m+1 \\
k%
\end{array}%
\right) \left(
\begin{array}{c}
m-1 \\
l%
\end{array}%
\right) \alpha _{PB}^{k}\alpha _{BB}^{l}\alpha _{PP}^{d-m+1-k}\alpha
_{BP}^{m-1-l}\delta _{n,l+k+1}.  \label{theory_space_representation}
\end{equation}

\section{$GL(2,\mathbf{R})$ group of the Galileon dualities\label{sec5gl2}}

The duality transformations introduced in the previous section has natural $%
GL(2,\mathbf{R})$ group structure under compositions. This is immediately
seen from their action on the $1$-forms $\omega _{A}$, $\omega _{P}^{\mu }$
and $\omega _{B}^{\mu }$ (cf. (\ref{omega_duality})) and on the coset
coordinates $x^{\mu }$ and $L^{\mu }$. The duality transformation is in
one-to-one correspondence with the matrix%
\begin{equation}
\mathbf{\alpha }=\left(
\begin{array}{cc}
\alpha _{PP} & \alpha _{PB} \\
\alpha _{BP} & \alpha _{BB}%
\end{array}%
\right)  \label{matrix_notation1}
\end{equation}%
and composition of two duality transformations corresponding to the matrices
$\mathbf{\alpha }$ and $\mathbf{\beta }$ is again a duality transformation
described by matrix $\mathbf{\alpha }\cdot \mathbf{\beta }$. The condition $%
\det \mathbf{\alpha }\neq 0$ ensures regularity\footnote{%
The Jacobian of the transformation (\ref{u_dualita}) is $(\det (\mathbf{%
\alpha }))^{2}$} of the transformation of the coordinates on the coset space
(\ref{u_dualita}).

A little bit less obvious is the group property for the duality
transformation of $\phi $. To demonstrate it let us rewrite (\ref{u_dualita}%
) in the form
\begin{eqnarray}
X &=&\mathbf{\alpha }\cdot X^{\prime }  \label{matrix_notation2} \\
\phi &=&\det \mathbf{\alpha ~}\phi ^{\prime }+\frac{1}{2}X^{\prime T}\cdot
\widehat{\mathbf{\alpha }}\cdot \mathbf{\alpha }\cdot X^{\prime }
\end{eqnarray}%
where
\begin{equation}
X=\left(
\begin{array}{c}
x \\
L%
\end{array}%
\right) ,~~~\widehat{\mathbf{\alpha }}=\left(
\begin{array}{cc}
\alpha _{BP} & 0 \\
0 & \alpha _{PB}%
\end{array}%
\right)  \label{matrix_notation3}
\end{equation}%
Then a composition of two dualities means
\begin{eqnarray}
X &=&\mathbf{\alpha }\cdot X^{\prime }=\mathbf{\alpha }\cdot \left( \mathbf{%
\beta }\cdot X^{\prime \prime }\right) =\left( \mathbf{\alpha }\cdot \mathbf{%
\beta }\right) \cdot X^{\prime \prime }  \notag \\
\phi &=&\det \mathbf{\alpha ~}\phi ^{\prime }+\frac{1}{2}X^{\prime T}\cdot
\widehat{\mathbf{\alpha }}\cdot \mathbf{\alpha }\cdot X^{\prime }  \notag \\
&=&\det \left( \mathbf{\alpha \cdot \beta }\right) \phi ^{\prime \prime }+%
\frac{1}{2}X^{\prime \prime T}\cdot \left( \left( \widehat{\mathbf{\beta }}%
\cdot \mathbf{\beta }\right) \det \mathbf{\alpha +\beta }^{T}\cdot \widehat{%
\mathbf{\alpha }}\cdot \mathbf{\alpha }\cdot \mathbf{\beta }\right) \cdot
X^{\prime \prime }
\end{eqnarray}%
However, as can be proved by direct calculation,
\begin{equation}
\left( \widehat{\mathbf{\beta }}\cdot \mathbf{\beta }\right) \det \mathbf{%
\alpha +\beta }^{T}\cdot \widehat{\mathbf{\alpha }}\cdot \mathbf{\alpha }%
\cdot \mathbf{\beta =}\left( \widehat{\mathbf{\alpha \cdot \beta }}\right)
\cdot \left( \mathbf{\alpha \cdot \beta }\right)
\end{equation}%
and therefore%
\begin{equation}
\phi =\det \left( \mathbf{\alpha \cdot \beta }\right) \phi ^{\prime \prime }+%
\frac{1}{2}X^{\prime \prime T}\cdot \left( \widehat{\mathbf{\alpha \cdot
\beta }}\right) \cdot \left( \mathbf{\alpha \cdot \beta }\right) \cdot
X^{\prime \prime }
\end{equation}%
as expected.

On the space $D_{d+1}$ of the Galileon theories, which can be treated as a $%
d+1$ dimensional real space with elements
\begin{equation}
\mathbf{d=}\left(
\begin{array}{c}
d_{1} \\
d_{2} \\
\vdots \\
d_{d+1}%
\end{array}%
\right) ,  \label{theory_space}
\end{equation}%
corresponding to $d+1$-tuples of the couplings $d_{n}$ , we have a linear
representation of the duality group $GL(2,\mathbf{R})$ by the matrices $%
A_{nm}(\mathbf{\alpha })$ explicitly given by (\ref%
{theory_space_representation}).

\subsection{Special cases}

Let us now discuss some important special cases. The duality transformations
corresponding to the one-parameter subgroup of matrices%
\begin{equation}
\mathbf{\alpha }_{\mathrm{dS}}(\zeta )=\left(
\begin{array}{cc}
1 & 0 \\
2\zeta & 1%
\end{array}%
\right)
\end{equation}%
which satisfy%
\begin{equation}
\mathbf{\alpha }_{\mathrm{dS}}(\zeta )\cdot \mathbf{\alpha }_{\mathrm{dS}%
}(\zeta ^{\prime })=\mathbf{\alpha }_{\mathrm{dS}}(\zeta ^{\prime })\cdot
\mathbf{\alpha }_{\mathrm{dS}}(\zeta )=\mathbf{\alpha }_{\mathrm{dS}}(\zeta
+\zeta ^{\prime }),
\end{equation}%
result in the following explicit transformation%
\begin{equation}
x=x^{\prime },~~~\phi =\phi ^{\prime }+\zeta x^{\prime 2},~~~\partial \phi
=\partial ^{\prime }\phi ^{\prime }+2\zeta x^{\prime }.
\end{equation}%
Fixing the parameter $\zeta =H^{2}/4$ the dual theory can be interpreted as
an expansion of the original Galileon field about the de Sitter solution
\begin{equation}
\phi _{\mathrm{dS}}=\frac{1}{4}H^{2}x^{2}
\end{equation}%
The fact, that the fluctuations $\phi ^{\prime }$ about such a background
are described by a dual Galileon Lagrangian has been established already in
the seminal paper \cite{Nicolis:2008in}. For the transformation of the
couplings we get explicitly
\begin{equation}
d_{n}(\mathbf{\alpha }_{\mathrm{dS}}(\zeta ))=\sum_{m=n}^{d+1}\left(
\begin{array}{c}
m \\
n%
\end{array}%
\right) \left( 2\zeta \right) ^{m-n}d_{m}
\end{equation}%
Another example concerns the following matrix%
\begin{equation}
\mathbf{\alpha }_{L}=\left(
\begin{array}{cc}
0 & 1 \\
1 & 0%
\end{array}%
\right)
\end{equation}%
It results in the duality transformation%
\begin{equation*}
x=\partial ^{\prime }\phi ^{\prime },~~~\phi =-\phi ^{\prime }+x^{\prime
}\cdot \partial ^{\prime }\phi ^{\prime },~~~\partial \phi =x^{\prime }
\end{equation*}%
which can be rewritten in the more symmetric form as%
\begin{equation}
x\cdot x^{\prime }=\phi (x)+\phi ^{\prime }(x^{\prime })
\label{symmetric_form}
\end{equation}%
and which corresponds to the Legendre transformation. Duality properties of
the Galileon theory with respect to this transformation has been discussed
in detail in \cite{Curtright:2012gx}. Explicit form for the dual couplings
reads
\begin{equation}
d_{n}(\mathbf{\alpha }_{L})=-\frac{d-n+2}{n}d_{d-n+2}.
\end{equation}%
Let us now assume the diagonal matrix%
\begin{equation}
\mathbf{\alpha }_{S}(\lambda )=\left(
\begin{array}{cc}
\lambda & 0 \\
0 & \lambda ^{-\Delta -1}%
\end{array}%
\right)
\end{equation}%
corresponding to the scaling transformation ($\Delta $ is the Galileon
scaling dimension)%
\begin{equation}
x=\lambda x^{\prime },~~~\phi =\lambda ^{-\Delta }\phi ^{\prime
},~~~\partial \phi =\lambda ^{-\Delta -1}\partial ^{\prime }\phi ^{\prime }
\end{equation}%
for which the dual couplings simply scale according their dimension as
\begin{equation}
d_{n}(\mathbf{\alpha }_{S}(\lambda ))=\lambda ^{d+2-n(\Delta +2)}d_{n}.
\label{d_scaling}
\end{equation}%
More general scaling is also possible, namely%
\begin{equation}
\mathbf{\alpha }_{S}(\lambda ,\kappa )=\left(
\begin{array}{cc}
\lambda & 0 \\
0 & \kappa%
\end{array}%
\right)
\end{equation}%
for which
\begin{equation}
x=\lambda x^{\prime },~~~\phi =\lambda \kappa \phi ^{\prime },~~~\partial
\phi =\kappa \partial ^{\prime }\phi ^{\prime }  \label{general_scaling}
\end{equation}%
and in the dual theory
\begin{equation}
d_{n}(\mathbf{\alpha }_{S}(\lambda ,\kappa ))=\kappa ^{n}\lambda
^{d-n+2}d_{n}.
\end{equation}%
Let us assume now duality transformations induced by the matrices of the form%
\footnote{%
The rationale for the minus sign of the element $\alpha _{PB}$ is that with
this choice the infinitesimal form of this duality transformation is%
\begin{equation*}
\phi (x)=\phi ^{\prime }(x)+\theta \partial \phi ^{\prime }(x)\cdot \partial
\phi ^{\prime }(x).
\end{equation*}%
See Appendix \ref{bu} for bottom up construction of the finite duality
transformation from the infinitesimal one.}%
\begin{equation}
\mathbf{\alpha }_{D}(\theta )=\left(
\begin{array}{cc}
1 & -2\theta \\
0 & 1%
\end{array}%
\right)  \label{alpha_theta}
\end{equation}%
which represents a one-parameter subgroup%
\begin{equation}
\mathbf{\alpha }_{D}(\theta )\cdot \mathbf{\alpha }_{D}(\theta ^{\prime })=%
\mathbf{\alpha }_{D}(\theta ^{\prime })\cdot \mathbf{\alpha }_{D}(\theta )=%
\mathbf{\alpha }_{D}(\theta +\theta ^{\prime }).
\end{equation}%
The corresponding coordinate and field transformation reads%
\begin{equation}
x=x^{\prime }-2\theta \partial ^{\prime }\phi ^{\prime },~~~\phi =\phi
^{\prime }-\theta \partial ^{\prime }\phi ^{\prime }\cdot \partial ^{\prime
}\phi ^{\prime },~~~\partial \phi =\partial ^{\prime }\phi ^{\prime }.
\label{one_parametric_duality}
\end{equation}%
Such a type of duality (with special value of the parameter $\theta $) has
been discussed in the papers \cite{Fasiello:2013woa, deRham:2013hsa,
deRham:2014zqa} and its one-parametric group structure has been recognized
in a very recent paper \cite{deRham:2014lqa}. The couplings transform
according to%
\begin{equation}
d_{n}(\mathbf{\alpha }_{D}(\theta ))=\frac{1}{n}\sum_{m=1}^{n}m\left(
\begin{array}{c}
d-m+1 \\
n-m%
\end{array}%
\right) \left( -2\theta \right) ^{n-m}d_{m}.  \label{d_theta}
\end{equation}%
It is obvious, that any duality transformation can be obtained as a
combination of the above elementary types of transformations. Indeed, for
general matrix $\mathbf{\alpha }$ we can write the following decomposition%
\begin{equation}
\left(
\begin{array}{cc}
\alpha _{PP} & \alpha _{PB} \\
\alpha _{BP} & \alpha _{BB}%
\end{array}%
\right) =\left(
\begin{array}{cc}
\alpha _{PP} & 0 \\
0 & \alpha _{PP}^{-1}\det \left( \mathbf{\alpha }\right)%
\end{array}%
\right) \left(
\begin{array}{cc}
1 & 0 \\
\alpha _{PP}\alpha _{BP}\det^{-1}\left( \mathbf{\alpha }\right) & 1%
\end{array}%
\right) \left(
\begin{array}{cc}
1 & \alpha _{PP}^{-1}\alpha _{PB} \\
0 & 1%
\end{array}%
\right) .  \label{duality_decomposition}
\end{equation}%
Let us give another simple example of such a type of decomposition. For
instance, we have
\begin{equation}
\mathbf{\alpha }_{D}(\theta )=\left[ \mathbf{\alpha }_{S}(1,\left( 2\theta
\right) ^{-1})\cdot \mathbf{\alpha }_{\mathrm{dS}}(-2^{-1})\right] \cdot
\mathbf{\alpha }_{\mathrm{L}}\cdot \left[ \mathbf{\alpha }_{\mathrm{dS}%
}(2^{-1})\cdot \mathbf{\alpha }_{S}(1,-2\theta )\right] ,
\label{D_decomposition}
\end{equation}%
and therefore we can understand the one-parametric duality (\ref%
{one_parametric_duality}) as a Legendre transformation of the function $\psi
^{\prime }(x^{\prime })$ into $\psi (x)$ where\footnote{%
The function $\psi ^{\prime }(x^{\prime })$ can be obtained by means of
application of the dual transformation corresponding to the product of
matrices in the second square brackets in (\ref{D_decomposition}), similarly
for $\psi (x)$.}
\begin{equation}
\psi ^{\prime }(x^{\prime })=\frac{1}{2}x^{\prime 2}-2\theta \phi ^{\prime
}(x^{\prime }),~~~\psi (x)=~\frac{1}{2}x^{2}+2\theta \phi (x)
\end{equation}%
which can be written in the symmetric form (cf. (\ref{symmetric_form})) as%
\begin{equation}
x\cdot x^{\prime }=\psi (x)+\psi ^{\prime }(x^{\prime }).
\end{equation}

\subsection{Duality invariants}

As we will see in the following sections, the one parameter subgroup of
duality transformations (\ref{one_parametric_duality}) is the most
interesting one relevant from the point of view of physical applications.
Let us briefly comment on some properties of its representation on the
Galileon theory space $D_{d+1}$(see (\ref{theory_space})). Any classical
Galileon Lagrangian $\mathcal{L}=\sum_{n=1}^{d+1}d_{n}\mathcal{L}_{n}$
corresponds to the $d+1$-tuple $\mathbf{d}$ of the couplings $d_{n}$
\begin{equation}
\mathbf{d=}\left(
\begin{array}{c}
d_{1} \\
d_{2} \\
\vdots \\
d_{d+1}%
\end{array}%
\right) \in D_{d+1}
\end{equation}%
and the one parametric subgroup (\ref{one_parametric_duality}) of duality
transformation $\mathbf{\alpha }_{D}(\theta )$ acts on this tuple linearly
via the matrix $\mathbf{A}(\theta )\equiv $ $A_{nm}(\mathbf{\alpha }%
_{D}(\theta ))$ (see (\ref{d_theta}))%
\begin{equation}
\mathbf{A}(\theta )_{nm}=\frac{1}{n}\sum_{m=1}^{n}m\left(
\begin{array}{c}
d-m+1 \\
n-m%
\end{array}%
\right) \left( -2\theta \right) ^{n-m}.
\end{equation}%
As we will see in the next sections, some of the relevant physical
quantities (e.g. the $S$ matrix) are invariant with respect to the duality
transformations $\mathbf{\alpha }_{D}(\theta )$. Such quantities are
therefore functions of the invariant combinations of the couplings $d_{n}$.
Here we will give a classification of such invariants built from $d_{n}$.
The main idea behind this classification is to identify these invariants
with conserved integrals of motion of a system of first order differential
equations for $\mathbf{d}(\theta )=\mathbf{A}(\theta )\cdot \mathbf{d}$ with
$\theta $ taken as the evolution parameter.

First, because the matrices $\mathbf{A}(\theta )$ are lower triangular
matrices, any subspace $D_{d+1}^{(k)}\subset D_{d+1}$spanned by the $d+1$%
-tuples with first $k$ couplings equal to zero (i.e. $D_{d+1}^{(k)}=\{%
\mathbf{d}|d_{n}=0$ for $n\leq k\}$) is left invariant by $\mathbf{A}(\theta
)$\textbf{. }We can therefore restrict ourselves to some fixed $%
D_{d+1}^{(k)} $ in what follows\footnote{%
For the physical applications it is natural to set $d_{1}=0$ in order to
avoid tadpoles and assume therefore the subspace $D_{d+1}^{(1)}$.}.

Note also that $\mathbf{\alpha }_{D}(\theta )$ is a one-parametric subgroup
and thus the matrices $\mathbf{A}(\theta )$ satisfy a differential equation%
\begin{equation}
\frac{\mathrm{d}}{\mathrm{d}\theta }\mathbf{A}(\theta )=\boldsymbol{T}\cdot
\mathbf{A}(\theta )
\end{equation}%
where%
\begin{equation}
T_{mn}=\frac{\mathrm{d}}{\mathrm{d}\theta }A_{nm}(\mathbf{\alpha }%
_{D}(\theta ))|_{\theta =0}=-2\frac{n-1}{n}(d-n+2)\delta _{n,m+1}.
\end{equation}%
Consequently we get for $d+1-k$-tuple $\mathbf{d}(\theta )\equiv d_{n}(%
\mathbf{\alpha }_{D}(\theta ))\in $\ $D_{d+1}^{(k)}$%
\begin{equation}
\frac{\mathrm{d}}{\mathrm{d}\theta }\mathbf{d}(\theta )=\boldsymbol{T}\cdot
\mathbf{d}(\theta ).  \label{diff_d}
\end{equation}%
This is a system of $d-k$ nontrivial ordinary differential equations (note
that the first of the equations (\ref{diff_d}) is trivial
\begin{equation*}
\frac{\mathrm{d}}{\mathrm{d}\theta }d_{k+1}(\theta )=0,
\end{equation*}%
i.e. $d_{k+1}$ can be taken as fixed\footnote{%
For instance, for $k=1$ it is natural to set $d_{2}=1/12$ in order to
normalize the kinetic term of the Galileon as usual.} once for ever)
describing the \textquotedblleft running\textquotedblright\ of the couplings
with the change of the duality parameter $\theta $. Of course, the solutions
are just $d_{n}(\mathbf{\alpha }_{D}(\theta ))$ given by (\ref{d_theta})
with $d_{n}$, $n>k$ as the initial conditions at $\theta =0$. Such a system
have in general $d-k-1$ functionally independent integrals of motion (we
denote them $I_{k+3},I_{k+4},\ldots ,I_{d+1}$ for reason which will be clear
from the construction) which do not depend explicitly on $\theta $. Once
these are known, any other such an integral of motion can be then expressed
as
\begin{equation}
I=f(I_{k+3},I_{k+4},\ldots ,I_{d+1}),
\end{equation}%
where $f$ is some function. The set $I_{k+3},I_{k+4},\ldots ,I_{d+1}$
represents therefore a basis of the $\mathbf{\alpha }_{D}(\theta )$ duality
invariants on the subspace $D_{d+1}^{(k)}$ of the Galileon theory space.

The set of independent invariants $I_{k+3},I_{k+4},\ldots ,I_{d+1}$ can be
constructed by means of elimination of the initial conditions and $\theta $
the from the solution (\ref{d_theta}). This can be done as follows. Note
that (\ref{d_theta}) for $n=k+2$ and $d_{n}=0$ for $n\leq k$ reads
\begin{equation}
d_{k+2}(\theta )=d_{k+2}-2\theta \frac{(k+1)(d-k)}{k+2}d_{k+1},
\end{equation}%
and thus we have unique solution $\theta ^{\ast }$ for $\theta $ such that $%
d_{k+2}(\theta ^{\ast })=0$. According to the group property we can rewrite
the solution of (\ref{diff_d}) in the form%
\begin{equation}
\mathbf{d}(\theta )=\mathbf{A}(\theta -\theta ^{\ast })\cdot \mathbf{d}%
(\theta ^{\ast })  \label{new_d_sol}
\end{equation}%
with new initial conditions $\mathbf{d}(\theta ^{\ast })$. Inverting (\ref%
{new_d_sol}) we get
\begin{equation}
\mathbf{d}(\theta ^{\ast })=\mathbf{A}(\theta ^{\ast }-\theta )\cdot \mathbf{%
d}(\theta )  \label{d_inversion}
\end{equation}%
the right hand side of which is $\theta $ independent. For $d_{k+2}$ the
equation (\ref{new_d_sol}) reads%
\begin{equation}
d_{k+2}(\theta )=-2(\theta -\theta ^{\ast })\frac{(k+1)(d-k)}{k+2}d_{k+1}
\end{equation}%
and thus we can easily eliminate $\theta -\theta ^{\ast }$ solely in terms
of $d_{k+2}(\theta )$. \ Inserting now this for the explicit $\theta-\theta
^{\ast}$ dependence into (\ref{d_inversion}) for $n=k+3,\ldots ,d+1$ we get
the desired integrals of motion $I_{l}(d_{k+2}(\theta ),\ldots
,d_{d+1}(\theta ))$. Their interpretation is clear, according to our
construction $I_{l}$ represents a value of couplings $d_{l}$ in the theory
dual with the original one such that in the dual theory the coupling $%
d_{k+2} $ is zero. These integrals form the basis of the $\mathbf{\alpha }%
_{D}(\theta )$ duality subgroup invariants on the Galileon theory subspace $%
D_{d+1}^{(k)}$ we started with.

Let us illustrate this general construction of $\ \mathbf{\alpha }%
_{D}(\theta )$ duality invariants in the case of three and four dimensional
Galileon theory. We will restrict ourselves to the theory subspaces $%
D_{4}^{(1)}$ and $D_{5}^{(1)}$, i.e. we set in both cases $d_{1}=0$, and we
further fix $d_{2}$ for $d=3,4$ as $1/4$ and $1/12$ respectively. For $d=3$
we get from (\ref{d_theta})
\begin{eqnarray}
d_{3}(\theta ) &=&d_{3}-\frac{2}{3}\theta  \notag \\
d_{4}(\theta ) &=&d_{4}-\frac{3}{2}d_{3}\theta +\frac{1}{2}\theta ^{2}
\label{3d_duallity}
\end{eqnarray}%
and according the general recipe, the only $\mathbf{\alpha }_{D}(\theta )$
duality invariant is $I_{4}=d_{4}(\theta ^{\ast })$ with $\theta ^{\ast
}=3d_{3}/2$, explicitly%
\begin{equation}
I_{4}=d_{4}-\frac{9}{8}d_{3}^{2}.  \label{3d_invariant}
\end{equation}%
For $d=4$ the $\mathbf{\alpha }_{D}(\theta )$ duality transformation reads%
\begin{eqnarray}
d_{3}(\theta ) &=&d_{3}-\frac{1}{3}\theta  \notag \\
d_{4}(\theta ) &=&d_{4}-3\theta d_{3}+\frac{1}{2}\theta ^{2}  \notag \\
d_{5}(\theta ) &=&d_{5}-\frac{8}{5}\theta d_{4}+\frac{12}{5}\theta ^{2}d_{3}-%
\frac{4}{15}\theta ^{3},  \label{4d_duality}
\end{eqnarray}%
and we have two independent duality invariants $I_{4,5}=d_{4,5}(\theta
^{\ast })$ where $\theta ^{\ast }=3d_{3}$, explicitly
\begin{eqnarray}
I_{4} &=&d_{4}-\frac{9}{2}d_{3}^{2}  \notag \\
I_{5} &=&d_{5}-\frac{24}{5}d_{3}d_{4}+\frac{72}{5}d_{3}^{3}.  \label{I_4_5}
\end{eqnarray}

\section{Applications\label{sec6applications}}

Two Galileon theories connected by duality are different theories. Therefore
all the properties of such theories cannot be the same. However, it does not
mean that the dual theories cannot be used to describe the same physical
reality. We have only to identify carefully those physical observables that
are dual to each other in both theories. Omitting this aspect of the duality
might lead to apparent paradoxes. A closely related aspect of the duality is
\textquotedblleft calculational\textquotedblleft . Because the duality
relates different Galileon theories, its main benefit is based on the
possibility to solve a given problem in the simplest exemplar of the set of
theories connected by duality. Then the result can be translated back to the
apparently more complex original theory for which the problem has been
formulated. In order to realize this approach effectively it is necessary to
establish the correct interrelation of the observables in both theories. In
subsections \ref{section_classical_solutions}, \ref{phase_space_duality} and %
\ref{section_fluctuations} we will discuss this issue in more detail. We
will show that the classical covariant phase spaces of two theories
connected by duality are (at least formally) in one-to-one correspondence.
The same is also true for classical observables for which the duality
transformation can be established.

Also the (off-shell) symmetries of the Galileon theories are realized
differently within the dual theories. Some of them are not directly visible
from the form of the classical dual Lagrangian, in this sense they are
hidden but still present in the dual theory. In subsection \ref%
{section_hidden_symmetries}\ we give some elementary examples of such hidden
symmetries.

An exceptional role play those physical observables that are invariants of
the duality. Only such observables are independent on the choice of the
representative in the class of theories connected by duality. As we have
mentioned in the previous section, the most useful duality is the
one-parametric subgroup $\mathbf{\alpha }_{D}(\theta )$, which is (together
with $\mathbf{\alpha }_{\mathrm{L}}$) the only one for which the field and
coordinate transformation is nontrivial. Therefore the class of observables
invariant with respect to this subgroup are the most important ones. These
observables give the same result on the whole class of Galileon theories
connected by $\mathbf{\alpha }_{D}(\theta )$ duality. In subsection \ref%
{S_duality} we will show that the tree level $S$ matrix belongs to the class
of $\mathbf{\alpha }_{D}(\theta )$ invariant quantities. This will enable us
to understand the structure of the results (\ref{M3}), (\ref{M4}) and (\ref%
{M5}) for the lowest scattering amplitudes (see subsection \ref{tree_level_A}%
). We will also give a simple alternative derivation of these results using
their properties under duality. As a next step we will discuss the loop
corrections in the Galileon theory in the framework of perturbative low
energy effective expansion and the properties of these corrections under
duality (see subsections \ref{CT} and \ref{one_loop_duality}).

Because the tree level $S$ matrix is invariant of the duality subgroup $%
\mathbf{\alpha }_{D}(\theta )$, it is useful to classify the Galileon
theories with respect to this subgroup, i.e. to find all nontrivial classes
of the Galileon theories modulo $\mathbf{\alpha }_{D}(\theta )$. At the same
time we get also classification of all the nontrivial $S$ matrices. In
subsection \ref{section_classification} we provide such a classification
using the duality invariants $I_{k+2},\dots ,I_{d+1}$ introduced in the
previous section in three and four dimensions.

All the above aspects of the duality will be illustrated using several
explicit examples both on classical and quantum levels.

In what follows we almost exclusively work in four dimensions with Minkowski
metric and in the Galileon Lagrangian we set $d_{1}=0$ to avoid the tadpole
and $d_{2}=1/12$ to get a canonical normalization of the kinetic term.

\subsection{Classical solutions\label{section_classical_solutions}}

As the calculational aspect of the duality is concerned, in some cases the
duality can help us\ to find solutions of the classical equation of motion
very efficiently. Let $\phi _{\alpha }(x)$ be the dual transformation of the
field configuration $\phi (x)$ under the $\alpha \in GL(2,\mathbf{R})$ (cf. (%
\ref{generalized_duality_alpha})). The definition of the corresponding dual
action $S_{\alpha }$ (cf. (\ref{dual_action_definition}))
\begin{equation}
S_{\alpha }[\phi ]=S[\phi _{\alpha }]  \label{S_theta}
\end{equation}%
then guarantees that, provided $\phi _{\ast }(x)$ is a minimum (or
stationary point) of the dual action $S_{\alpha }$, the dual configuration $%
\left( \phi _{\ast }\right) _{\alpha }(x)$ realizes a minimum (or stationary
point) of the original action $S$. In many cases we can choose the matrix $%
\alpha $ in such a way that we can solve easily the equation of motion for
the action $S_{\alpha }$ and find explicitly the dual of this solution
simultaneously. This gives us immediately the solution of the apparently
much more complicated equation of motion for the original action $S$. Such a
method for finding solutions of Galileon equation of motion is usually
efficient when we seek after a solution with additional symmetry which
effectively reduces the dimensionality of the space-time. It is known that
in such a case only limited subset of couplings $d_{n}$ enter the equation
of motion \cite{Deser:2012gm} and the duality transformation with properly
chosen matrix $\alpha $ can further reduce this subset. In the ideal case
the dual equation of motion becomes that of the free theory but also in
other cases such an approach might be useful.

More formally and in more detail, the functional derivative of (\ref{S_theta}%
) with respect to $\phi (x)$ gives
\begin{equation}
\frac{\delta S_{\alpha }[\phi ]}{\delta \phi (x)}=\int \mathrm{d}^{d}z\frac{%
\delta S[\phi _{\alpha }]}{\delta \phi _{\alpha }(z)}\frac{\delta \phi
_{\alpha }(z)}{\delta \phi (x)}.
\end{equation}%
Therefore we get the following relation between the stationary points of
both actions%
\begin{equation}
\frac{\delta S[\phi _{\alpha }]}{\delta \phi _{\alpha }(z)}|_{\left( \phi
_{\ast }\right) _{\alpha }}=0\Longrightarrow \frac{\delta S_{\alpha }[\phi ]%
}{\delta \phi (x)}|_{\phi _{\ast }}=0  \label{solution_relation}
\end{equation}%
and provided the duality transformation induced by the configuration $\phi
_{\ast }$ is invertible, also the reversed implication holds.

The invertibility of the duality transformation is related to the operator $%
\delta \phi _{\alpha }(z)/\delta \phi (x)$. For further convenience, let us
calculate $\delta \phi _{\alpha }(z)/\delta \phi (x)$ explicitly for the
case of the subgroup $\alpha _{D}(\theta )$ given by (\ref{alpha_theta}) and
(\ref{one_parametric_duality}). Note that the duality transformation under
the one parametric subgroup $\alpha _{D}\left( \theta \right) $ (here we
denote $\phi _{\alpha _{D}(\theta )}$ simply as $\phi _{\theta }$ and
analogously for $x_{\theta }$)%
\begin{eqnarray}
x_{\theta } &=&x-2\theta \partial \phi (x)  \notag \\
\phi _{\theta }(x_{\theta }) &=&\phi (x)-\theta \partial \phi (x)\cdot
\partial \phi (x)  \label{original_duality}
\end{eqnarray}%
can be rewritten in the inverted form%
\begin{eqnarray}
y &=&X[\phi ](y)-2\theta \partial \phi (X[\phi ](y))  \notag \\
\phi _{\theta }(y) &=&\phi (X[\phi ](y))-\theta \partial \phi (X[\phi
](y))\cdot \partial \phi (X[\phi ](y)).  \label{rewritten_alpha_duality}
\end{eqnarray}%
Here $X[\phi ](y)$ is the inversion of the coordinate transformation defined
as
\begin{equation}
x=X[\phi ](x_{\theta }),  \label{X[phi](x)}
\end{equation}%
and we have explicitly shown the functional dependence of this inversion on $%
\phi $. Then taking this implicit dependence into account we get%
\begin{eqnarray}
\frac{\delta \phi _{\theta }(z)}{\delta \phi (x)} &=&\delta ^{(d)}(X[\phi
](z)-x)+\left( \partial \phi \right) (X[\phi ](z))\cdot \frac{\delta X[\phi
](z)}{\delta \phi (x)}  \notag \\
&&-2\theta \partial \phi (X[\phi ](z))\cdot \partial \delta ^{(d)}(X[\phi
](z)-x)  \notag \\
&&-2\theta \partial \phi (X[\phi ](z))\cdot \partial \partial \phi (X[\phi
](z))\cdot \frac{\delta X[\phi ](z)}{\delta \phi (x)}.
\label{d_phi_theta/d_phi}
\end{eqnarray}%
But taking a functional derivative of the first equation of (\ref%
{rewritten_alpha_duality}) respect to $\phi (x)$ we get%
\begin{equation}
0=\frac{\delta X[\phi ](y)}{\delta \phi (x)}-2\theta \partial \delta
^{(d)}(X[\phi ](y)-x)-2\theta \partial \partial \phi (X[\phi ](y))\cdot
\frac{\delta X[\phi ](y)}{\delta \phi (x)}
\end{equation}%
and inserting this to (\ref{d_phi_theta/d_phi}) we get finally\footnote{%
Note, that the substitution $z\rightarrow z_{\theta }$ and the functional
derivative with respect to $\phi (x)$ do not commute. Provided we make this
replacement in (\ref{d_phi_theta/d_phi}) after the functional derivative is
taken, we get%
\begin{equation*}
\left( \frac{\delta \phi _{\theta }(z)}{\delta \phi (x)}\right)
|_{z\rightarrow z_{\theta }}=\delta ^{(d)}(X[\phi ](z_{\theta })-x)=\delta
^{(d)}(z-x).
\end{equation*}%
On the other hand, making this inserting before the functional
differentiation, we change the functional dependence of the differentiated
functional and the result is different, namely%
\begin{equation*}
\frac{\delta }{\delta \phi (x)}\left( \phi _{\theta }(z)|_{z\rightarrow
z_{\theta }}\right) =\frac{\delta }{\delta \phi (x)}\left( \phi (z)-\theta
\partial \phi (z)\cdot \partial \phi (z)\right) =\delta ^{(d)}(z-x)-2\theta
\partial \phi (z)\cdot \partial \delta ^{(d)}(z-x).
\end{equation*}%
}%
\begin{equation}
\frac{\delta \phi _{\theta }(z)}{\delta \phi (x)}=\delta ^{(d)}(X[\phi
](z)-x).  \label{phi_duality_jacobi_matrix}
\end{equation}%
This result can be used to find a direct relation between equations of
motions in both theories. We have (denoting $S_{\alpha _{D}(\theta )}\equiv
S_{\theta }$ for simplicity)
\begin{eqnarray}
\frac{\delta S_{\theta }[\phi ]}{\delta \phi (x)} &=&\int \mathrm{d}^{d}z%
\frac{\delta S[\phi _{\theta }]}{\delta \phi _{\theta }(z)}\delta
^{(d)}(X[\phi ](z)-x)  \notag \\
&=&\int \mathrm{d}^{d}z\det \left( \frac{\partial z_{\theta }}{\partial z}%
\right) \frac{\delta S[\phi _{\theta }]}{\delta \phi _{\theta }(z_{\theta })}%
\delta ^{(d)}(z-x)
\end{eqnarray}%
where we have substituted $z\rightarrow $ $z_{\theta }=z-2\theta \partial
\phi (z)$ and used (\ref{X[phi](x)}) when passing to the second line.
Therefore%
\begin{equation}
\frac{\delta S_{\theta }[\phi ]}{\delta \phi (x)}=\det \left( \frac{\partial
x_{\theta }}{\partial x}\right) \frac{\delta S[\phi _{\theta }]}{\delta \phi
_{\theta }(x_{\theta })}  \label{equation_relation}
\end{equation}%
and explicitly (cf. (\ref{eom}))%
\begin{equation}
\sum_{n=1}^{d+1}nd_{n}(\theta )\mathcal{L}_{n-1}^{\mathrm{der}}(\partial
\partial \phi (x))=\det \left( \frac{\partial x_{\theta }}{\partial x}%
\right) \sum_{n=1}^{d+1}nd_{n}\mathcal{L}_{n-1}^{\mathrm{der}}(\partial
_{\theta }\partial _{\theta }\phi _{\theta }(x_{\theta })).
\label{eom_duality_explicitly}
\end{equation}

Let us note, that the above discussion are in fact not restricted to the one
parametric subgroup $\alpha _{D}\left( \theta \right) $ but holds also for
general duality transformation with general $\mathbf{\alpha }\in GL(2,%
\mathbf{R})$ with the obvious replacement $S_{\theta }\rightarrow S_{\alpha
} $, $\phi _{\theta }\rightarrow \phi _{\alpha }$, $x_{\theta }\rightarrow
x_{\alpha }$ (cf. the general formulae (\ref{generalized_duality_alpha}) and
(\ref{S_alpha})). The derivation of the functional derivative $\delta \phi
_{\alpha }(x)/\delta \phi (y)$ follows the same logic as for $\delta \phi
_{\theta }(x)/\delta \phi (y)$ with minor changes caused by the more
complicated formulae for $x_{\alpha }$ and $\phi _{\alpha }$. The result is%
\begin{equation}
\frac{\delta \phi _{\alpha }(z)}{\delta \phi (x)}=\det \mathbf{\alpha ~}%
\delta ^{(d)}(X[\phi ](z)-x),
\end{equation}%
where now $X[\phi ](z)$ is the inversion of the coordinate transformation $%
x_{\alpha }=\alpha _{PP}x+\alpha _{PB}\partial \phi (x)$. In the general
case the formula (\ref{equation_relation}) reads%
\begin{equation}
\frac{\delta S_{\alpha }[\phi ]}{\delta \phi (x)}=\det \mathbf{\alpha }\det
\left( \frac{\partial x_{\alpha }}{\partial x}\right) \frac{\delta S[\phi
_{\alpha }]}{\delta \phi _{\alpha }(x_{\alpha })}.
\end{equation}

In what follows we give two explicit examples of the applicability of the
duality with respect to the subgroup $\alpha _{D}\left( \theta \right) $ for
finding the solutions of the classical equation of motion, namely the static
cylindrically symmetric solution \ and \ a point-like source.

\subsubsection{Cylindrically symmetric static solution}

As a first example of the calculational efficiency of duality, we will
illustrate its application on a simple and analytically solvable case. We
will consider a static axial-symmetric solution of Galilean equations with
an external source coupled to the Galileon field as%
\begin{equation}
S_{\mathrm{int}}=\int \mathrm{d}^{4}x\phi (x)T(x).  \label{external_source}
\end{equation}%
The source $T$ will be represented by an infinite \textquotedblleft cosmic
string\textquotedblright\ along the $x_{1}$-axis with linear density $\sigma
>0$%
\begin{equation}
T(x)=-\sigma \delta (x^{2})\delta (x^{3}).
\end{equation}%
Due to the symmetry, the problem is effectively two-dimensional and
therefore the quartic and quintic Galileon couplings are irrelevant (their
contributions to the classical equation of motion vanish). Exactly this
feature is the key ingredient which makes the duality efficient in this
case. By appropriate choice of the dual theory we can effectively eliminate
also the cubic coupling and solve the dual problem in the framework of free
theory.

Note however, that for general external source, the part $S_{\mathrm{int}}$
of the complete action violates duality. Therefore we cannot in general case
simply argue that the duality transformation of the classical solution in
the original theory is also a solution of the dual theory with the same
external source. However, our source term is very special being local and
therefore it modifies the equations of motion only on the set of points of
zero measure. As we shall explicitly see, for such a source\footnote{%
This remains true also for the point-like source studied in the next
subsection.} the duality works, which illustrates the conclusions made in
the very recent papers\footnote{%
See also \cite{deRham:2013hsa} for discussions of point-like sources.} \cite%
{deRham:2014lqa,Creminelli:2014zxa}.

Let us first consider the general Galileon theory with all the couplings
present. Our axial-symmetric ansatz is
\begin{equation}
\phi (x)\equiv \phi (z\overline{z})\,,
\end{equation}%
where we have introduced the complex coordinates $z$ and $\overline{z}$:
\begin{equation}
z=x^{2}+\mathrm{i}x^{3}\,,\qquad \overline{z}=x^{2}-\mathrm{i}x^{3}
\end{equation}%
i.e.%
\begin{equation}
\partial _{2}=\partial +\overline{\partial }\,,\qquad \partial _{3}=\mathrm{i%
}\partial -\mathrm{i}\overline{\partial }\,,\qquad \mathrm{d}^{2}z\equiv -%
\mathrm{i~d}\overline{z}\mathrm{d}z=2\mathrm{d}x^{2}\mathrm{d}x^{3}
\end{equation}%
In order to obtain the explicit form for the classical equations of motion
we will start with the following useful formula \cite{Curtright:2012gx}
\begin{equation}
\mathcal{L}_{4}^{\mathrm{der}}[\eta +w\partial \partial \phi ]=4!\det [\eta
+w\partial \partial \phi ]=\sum_{k=0}^{4}w^{k}\left(
\begin{array}{c}
4 \\
k%
\end{array}%
\right) \mathcal{L}_{k}^{\mathrm{der}}[\partial \partial \phi ]\,,
\label{nice_formula}
\end{equation}%
where we can easily work out the left hand side because the matrix $\eta
+w\partial \partial \phi $ is block-diagonal,
\begin{equation}
\det [\eta +w\partial \partial \phi ]=-1+4w\partial \overline{\partial }\phi
+4w^{2}\left[ \partial ^{2}\phi \overline{\partial }^{2}\phi -\left(
\partial \overline{\partial }\phi \right) ^{2}\right] \,.
\end{equation}%
Comparing this with the right hand side of (\ref{nice_formula}) we get%
\begin{eqnarray}
\mathcal{L}_{1}^{\mathrm{der}}[\partial \partial \phi ] &=&24\partial
\overline{\partial }\phi  \notag \\
\mathcal{L}_{2}^{\mathrm{der}}[\partial \partial \phi ] &=&16\left[ \partial
^{2}\phi \overline{\partial }^{2}\phi -\left( \partial \overline{\partial }%
\phi \right) ^{2}\right]  \notag \\
\mathcal{L}_{3}^{\mathrm{der}}[\partial \partial \phi ] &=&\mathcal{L}_{4}^{%
\mathrm{der}}[\partial \partial \phi ]=0
\end{eqnarray}%
and therefore the equation of motion with an external source $T$ is
\begin{equation}
\frac{\delta S}{\delta \phi }=\sum_{n}nd_{n}\mathcal{L}_{n-1}^{\mathrm{der}%
}+T=2d_{2}\mathcal{L}_{1}^{\mathrm{der}}[\partial \partial \phi ]+3d_{3}%
\mathcal{L}_{2}^{\mathrm{der}}[\partial \partial \phi ]+T=0\,,
\end{equation}%
where we will set $d_{2}=1/12$ in the following. In our case $T=-2\sigma
\delta ^{(2)}(z,\overline{z})$ so the equation of motion becomes
\begin{equation}
4\partial \overline{\partial }\phi +48d_{3}\left[ \partial ^{2}\phi
\overline{\partial }^{2}\phi -\left( \partial \overline{\partial }\phi
\right) ^{2}\right] =2\sigma \delta ^{(2)}(z,\overline{z})  \label{EOM}
\end{equation}%
First we can easily solve the theory for $d_{3}=0$. The axial symmetric
solution is (up to a constant term)
\begin{equation}
\phi (z\overline{z})=\frac{\sigma }{4\pi }\ln z\overline{z}.
\label{free_theory}
\end{equation}%
In the case when $d_{3}\neq 0$ we can first rewrite the equation of motion
to
\begin{equation}
\frac{1}{z}\overline{\partial }\left[ z\overline{z}\phi ^{\prime }(z%
\overline{z})-12d_{3}z\overline{z}\phi ^{\prime }(z\overline{z})^{2}\right] =%
\frac{\sigma }{2}\sigma \delta ^{(2)}(z,\overline{z})\,,
\end{equation}%
where the prime means a derivative with respect to $z\overline{z}$. By
further integration over the disc with $z\overline{z}\leq R^{2}$ and using
the Gauss theorem in two dimension we will arrive to\footnote{%
This is in fact an expected result. As we have mentioned above, the problem
is effectively two-dimensional and therefore posses two-dimensional
spherical symmetry. In any dimension the spherically symmetric Galileon
equation reduces to algebraic equation for the first derivative of the field.%
}
\begin{equation}
\phi ^{\prime }(R^{2})-12d_{3}\phi ^{\prime }(R^{2})^{2}=\frac{\sigma }{4\pi
R^{2}}\,,  \label{first_order_equation}
\end{equation}%
which can be algebraically solved to
\begin{equation}
\phi _{\pm }^{\prime }(R^{2})=\frac{1\pm \sqrt{1-12\frac{d_{3}\sigma }{\pi
R^{2}}}}{24d_{3}}\,.
\end{equation}%
The final result can be obtained by elementary integration. We have two
solutions, which is for $d_{3}>0$ defined only for $R^{2}>12d_{3}\sigma /\pi
$ (for $R^{2}<12d_{3}\sigma /\pi $ this solution has an imaginary part)%
\footnote{%
For $d_{3}<0$ the solution $\phi _{-}$ exhibits the Vainshtein mechanism
\cite{Vainshtein:1972sx} with Vainshtein radius $R_{V}^{2}=-12d_{3}\sigma
/\pi $. Indeed, outside and inside the Vainshtein radius we have
\begin{equation*}
\frac{\mathrm{d}}{\mathrm{d}R}\phi _{-}=\left\{
\begin{array}{c}
\frac{\sigma }{2\pi R}+O(R^{-3}),~~\mathrm{for}~R>R_{V}, \\
-\left( -\frac{\sigma }{12d_{3}\pi }\right) ^{1/2}+O(R),~\mathrm{for}%
~R<R_{V}.%
\end{array}%
\right.
\end{equation*}%
}. We will show in the following how this can be obtained using duality in a
much simpler and pure algebraical way.

The transformation of duality under the subgroup $\mathbf{\alpha }%
_{D}(\theta )$ can be expressed in our coordinates as
\begin{eqnarray}
z_{\theta } &=&z+4\theta \overline{\partial }\phi (z,\overline{z})  \notag \\
\overline{z}_{\theta } &=&\overline{z}+4\theta \partial \phi (z,\overline{z})
\notag \\
\phi _{\theta }(z_{\theta },\overline{z}_{\theta }) &=&\phi (z,\overline{z}%
)+4\theta \overline{\partial }\phi (z,\overline{z})\partial \phi (z,%
\overline{z})
\end{eqnarray}%
while the remaining coordinates $x^{0}$ and $x^{1}$are left unchanged (cf. (%
\ref{one_parametric_duality})). Let us assume that $\phi (z,\overline{z})$
is the solution of the theory (\ref{free_theory}) with $d_{3}=0$. The
duality transformation of $\phi (z,\overline{z})$ is then given implicitly
as
\begin{eqnarray}
z_{\theta } &=&z+\frac{\sigma \theta }{\pi \overline{z}}  \notag \\
\overline{z}_{\theta } &=&\overline{z}+\frac{\sigma \theta }{\pi z}  \notag
\\
\phi _{\theta }(z_{\theta },\overline{z}_{\theta }) &=&\frac{\sigma }{4\pi }%
\left( \ln z\overline{z}+\frac{\sigma \theta }{\pi z\overline{z}}\right) .
\label{resenifi}
\end{eqnarray}%
We have therefore%
\begin{equation}
z_{\theta }\overline{z}_{\theta }=z\overline{z}+\left( \frac{\sigma \theta }{%
\pi }\right) ^{2}\frac{1}{z\overline{z}}+2\frac{\sigma \theta }{\pi }.
\label{z_theta(z)}
\end{equation}%
Let us note that for $\theta >0$ the transformation $z\rightarrow z_{\theta
} $ double covers the complement of a circle $z_{\theta }\overline{z}%
_{\theta }<4\frac{\sigma \theta }{\pi }$; inside of this circle $\phi
_{\theta }$ is not defined. For $\theta <0$ this transformation double
covers the whole complex plane, the circle $\ z\overline{z}=-\frac{\sigma
\theta }{\pi }$ is mapped to the point $z_{\theta }=0$. The inversion of (%
\ref{z_theta(z)}) which shall be inserted to the right hand side of $\phi
_{\theta }(z_{\theta },\overline{z}_{\theta })$ is then
\begin{equation}
z\overline{z}=\frac{1}{2}\left( z_{\theta }\overline{z}_{\theta }-2\frac{%
\sigma \theta }{\pi }\pm \sqrt{z_{\theta }\overline{z}_{\theta }\left(
z_{\theta }\overline{z}_{\theta }-4\frac{\sigma \theta }{\pi }\right) }%
\right)  \label{dual_zzbar}
\end{equation}%
Now the duality means that%
\begin{equation}
S[\phi _{\theta }]=S_{\theta }[\phi ]
\end{equation}%
where $S$ and $S_{\theta }$ are the actions (without the external source
term $S_{\mathrm{int}}$) of the general Galileon theory and its $\mathbf{%
\alpha }_{D}(\theta )$ dual respectively. In our case we take the former to
be the general interacting theory (with $d_{3}\neq 0$) and the latter we
identify with its dual chosen in such a way that $d_{3}\left( \theta \right)
=0$. As we know from (\ref{4d_duality}) such a theory can be obtained from
the general one by duality transformation with $\theta =3d_{3}$ and thus for
this value the eq.(\ref{resenifi}) is expected to represent the wanted
solution of (\ref{EOM}, \ref{first_order_equation}). Let us now verify that
it is indeed the case.

Using the duality transformation of the derivatives (c.f. the last equation
of (\ref{one_parametric_duality}))
\begin{equation}
\partial \phi (z\overline{z})=\overline{z}\phi ^{\prime }(z\overline{z}%
)=\partial _{\theta }\phi _{\theta }(z_{\theta }\overline{z}_{\theta })=%
\overline{z}_{\theta }\phi _{\theta }^{\prime }(z_{\theta }\overline{z}%
_{\theta })
\end{equation}%
we obtain
\begin{equation}
\phi _{\theta }^{\prime }(z_{\theta }\overline{z}_{\theta })=\frac{z}{%
z_{\theta }}\phi ^{\prime }(z\overline{z})=\frac{\overline{z}}{\overline{z}%
_{\theta }}\phi ^{\prime }(z\overline{z})=\left( 1+\frac{\sigma \theta }{\pi
z\overline{z}}\right) ^{-1}\phi ^{\prime }(z\overline{z}).
\end{equation}%
Inserting this in the left hand side of (\ref{first_order_equation}) we get%
\begin{eqnarray}
\phi _{\theta }^{\prime }(z_{\theta }\overline{z}_{\theta })-12d_{3}\phi
_{\theta }^{\prime }(z_{\theta }\overline{z}_{\theta })^{2} &=&\left( 1+%
\frac{\sigma \theta }{\pi z\overline{z}}\right) ^{-1}\phi ^{\prime }(z%
\overline{z})-12d_{3}\left[ \left( 1+\frac{\sigma \theta }{\pi z\overline{z}}%
\right) ^{-1}\phi ^{\prime }(z\overline{z})\right] ^{2}  \notag \\
&=&\frac{\sigma }{4\pi z\overline{z}}\left( 1+\frac{\sigma \theta }{\pi z%
\overline{z}}\right) ^{-2}\left[ 1+\frac{\sigma (\theta -3d_{3})}{\pi z%
\overline{z}}\right]
\end{eqnarray}%
where in the last line we used the explicit form of $\phi (z\overline{z})$.
\ Therefore for $\theta =3d_{3}$ and expressing back $z\overline{z}$ in
terms of $z_{\theta }\overline{z}_{\theta }$ we get
\begin{equation}
\phi _{\theta }^{\prime }(z_{\theta }\overline{z}_{\theta })-4\theta \phi
_{\theta }^{\prime }(z_{\theta }\overline{z}_{\theta })^{2}=\frac{\sigma }{%
4\pi z\overline{z}}\left( 1+\frac{\sigma \theta }{\pi z\overline{z}}\right)
^{-2}=\frac{\sigma }{4\pi z_{\theta }\overline{z}_{\theta }}
\end{equation}%
which means that $\phi _{3d_{3}}$ is a solution of the equation (\ref%
{first_order_equation}).

\subsubsection{Point-like source}

As a next example, let us repeat the above strategy for the spherically
symmetric solution of the galileon equation of motion with point-like source
$T(x)=-4\pi M\delta ^{(3)}(\mathbf{x})$. Though the duality does not help us
much in solving the most general equation of motion, as we will see, it
might be useful in some special cases.

Due to the spherical symmetry the situation is similar to the previous
subsection. The problem is effectively three-dimensional and the quintic
Galileon coupling thus disappear from the problem. The ansatz for the
solution is%
\begin{equation}
\phi (x)\equiv \phi (r)
\end{equation}%
where $r=|\mathbf{x}|=\sqrt{x^{i}x^{i}}$. After some algebra we end up with
the equation of motion in the form (cf. \cite{Nicolis:2008in} for more
details)%
\begin{equation}
\frac{12d_{2}}{r^{2}}\left( r^{2}\phi ^{\prime }(r)\right) ^{\prime }-\frac{%
12d_{3}}{r^{2}}\left( r\phi ^{\prime }(r)^{2}\right) ^{\prime }+\frac{8d_{4}%
}{r^{2}}\left( \phi ^{\prime }(r)^{3}\right) ^{\prime }-4\pi M\delta ^{(3)}(%
\mathbf{x})=0.
\end{equation}%
Integrating over $\mathrm{d}^{3}\mathbf{x}=4\pi r^{2}\mathrm{d}r$ and
assuming canonically normalized kinetic term ($d_{2}=1/12$) we get
\begin{equation}
r^{2}\phi ^{\prime }(r)-12d_{3}r\phi ^{\prime }(r)^{2}+8d_{4}\phi ^{\prime
}(r)^{3}=M  \label{central_equation}
\end{equation}%
which is an algebraic equation for $\phi ^{\prime }(r)/r$. The duality
transformation of the spherically symmetric static solution $\phi (r)$ reads%
\begin{eqnarray}
x_{\theta }^{0} &=&x^{0},~~x_{\theta }^{i}=\left( 1+2\theta \frac{\phi
^{\prime }(r)}{r}\right) x^{i} \\
\phi _{\theta }(x_{\theta }) &=&\phi (r)+\theta \phi ^{\prime }(r)^{2}
\end{eqnarray}%
Therefore, provided $1+2\theta \phi ^{\prime }(r)/r>0$%
\begin{equation}
r_{\theta }=\left( 1+2\theta \frac{\phi ^{\prime }(r)}{r}\right) r,
\label{r_theta}
\end{equation}%
and thus $\phi _{\theta }(x_{\theta })$ is function of $r_{\theta }$ only.
From the general formula $\left( \partial \phi \right) _{\theta }(x_{\theta
})=\partial \phi (x)$ we get further%
\begin{equation}
\phi _{\theta }^{\prime }(r_{\theta })=\phi ^{\prime }(r).
\label{phi_prime_r}
\end{equation}%
Now let $\phi $ be a solution of (\ref{central_equation}) with $%
d_{i}\rightarrow d_{i}(\theta )$, i.e. let%
\begin{equation}
r^{2}\phi ^{\prime }(r)-12d_{3}(\theta )r\phi ^{\prime
}(r)^{2}+8d_{4}(\theta )\phi ^{\prime }(r)^{3}=M  \label{integrated_equation}
\end{equation}%
It is then easy to show, that $\phi _{\theta }(r_{\theta })$ is a solution
of (\ref{central_equation}). Indeed%
\begin{eqnarray*}
&&r_{\theta }^{2}\phi _{\theta }^{\prime }(r_{\theta })-12d_{3}r_{\theta
}\phi _{\theta }^{\prime }(r_{\theta })+8d_{4}^{3}\phi _{\theta }^{\prime
}(r_{\theta }) \\
&=&r^{2}\left( 1+2\theta \frac{\phi ^{\prime }(r)}{r}\right) ^{2}\phi
^{\prime }(r)-12d_{3}r\left( 1+2\theta \frac{\phi ^{\prime }(r)}{r}\right)
\phi ^{\prime }(r)^{2}+8d_{4}\phi ^{\prime }(r)^{3} \\
&=&r^{2}\phi ^{\prime }(r)-12\left( d_{3}-\frac{1}{3}\theta \right) r\phi
^{\prime }(r)^{2}+8\left( d_{4}-3d_{3}\theta +\frac{1}{2}\theta ^{2}\right)
\phi ^{\prime }(r)^{3} \\
&=&r^{2}\phi ^{\prime }(r)-12d_{3}(\theta )r\phi ^{\prime
}(r)^{2}+8d_{4}(\theta )\phi ^{\prime }(r)^{3}=M.
\end{eqnarray*}%
For $\theta =3d_{3}$ we can eliminate the cubic Galileon coupling and the
equation (\ref{integrated_equation}) becomes
\begin{equation}
r^{2}\phi ^{\prime }(r)+8I_{4}\phi ^{\prime }(r)^{3}=M,
\label{reduced_central_solution}
\end{equation}%
where $I_{4}$ is the invariant (\ref{I_4_5})%
\begin{equation}
I_{4}=\left( d_{4}-\frac{9}{2}d_{3}^{2}\right) .
\end{equation}%
Moreover, for special case $I_{4}=0$ we can find the solution of (\ref%
{reduced_central_solution}) simply (up to an additive constant) as
\begin{equation}
\phi (r)=-\frac{M}{r}.
\end{equation}%
Its dual given by%
\begin{eqnarray}
r_{\theta } &=&r\left( 1+2\theta \frac{M}{r^{3}}\right) \\
\phi _{\theta }(r_{\theta }) &=&-\frac{M}{r}+\theta \frac{M^{2}}{r^{4}}=-%
\frac{M}{r}\left( 1-\theta \frac{M}{r^{3}}\right) ,
\end{eqnarray}%
is for $\theta =3d_{3}$ a solution of equation
\begin{equation}
r^{2}\phi ^{\prime }(r)-12d_{3}r\phi ^{\prime }(r)^{2}+36d_{3}^{2}\phi
^{\prime }(r)^{3}=M,
\end{equation}%
which corresponds to two-parametric set of Galileon theories with parameters
$d_{3}$, $d_{5}$ with special quartic coupling $d_{4}=9d_{3}^{2}/2$.

Let us assume now a complementary application of duality. For $I_{4}<0$ we
can choose
\begin{equation}
\theta _{\pm }=3d_{3}\pm \sqrt{-2I_{4}},
\end{equation}%
and eliminate the quartic couplings in the dual action $S_{\theta _{\pm
}}[\phi ]$. The dual equation is then%
\begin{equation}
r^{2}\phi ^{\prime }(r)-12d_{3}(\theta _{\pm })r\phi ^{\prime }(r)^{2}=M
\end{equation}%
where%
\begin{equation}
d_{3}(\theta _{\pm })=\mp \frac{1}{3}\sqrt{-2I_{4}}.
\end{equation}%
For further convenience let us choose $\theta _{+}$ to ensure $d_{3}(\theta
_{+})<0$. The solution for the derivative is then simply%
\begin{equation}
\phi _{\pm }^{\prime }(r)=r\frac{1\pm \sqrt{1-48d_{3}(\theta _{+})\frac{M}{%
r^{3}}}}{24d_{3}(\theta _{+})}
\end{equation}%
and integration gives (cf. also \cite{Curtright:2012gx})%
\begin{equation}
\phi _{\pm }(r)=\phi _{\pm }(0)+\frac{r^{2}}{48d_{3}(\theta _{+})}\pm \left(
-\frac{Mr}{3d_{3}(\theta _{+})}\right) ^{1/2}~_{2}F_{1}\left( -\frac{1}{2},%
\frac{1}{6},\frac{7}{6};\frac{r^{3}}{48d_{3}(\theta _{+})M}\right)
\end{equation}%
According to (\ref{r_theta}, \ref{phi_prime_r}) we get for the dual
transformation of this solution
\begin{equation}
r_{\theta _{+}}=r\left( 1+\theta _{+}\frac{1\pm \sqrt{1-48d_{3}(\theta _{+})%
\frac{M}{r^{3}}}}{12d_{3}(\theta _{+})}\right)
\end{equation}%
\begin{eqnarray}
\phi _{\theta _{+}}(r_{\theta _{+}})_{\pm } &=&\phi _{\pm }(0)+\frac{r^{2}}{%
48d_{3}(\theta _{+})}\pm \left( -\frac{Mr}{3d_{3}(\theta _{+})}\right)
^{1/2}~_{2}F_{1}\left( -\frac{1}{2},\frac{1}{6},\frac{7}{6};\frac{r^{3}}{%
48d_{3}(\theta _{+})M}\right)  \notag \\
&&+\theta _{+}r^{2}\left( \frac{1\pm \sqrt{1-48d_{3}(\theta _{+})\frac{M}{%
r^{3}}}}{24d_{3}(\theta _{+})}\right) ^{2}
\end{eqnarray}%
which solves (\ref{central_equation}).

\subsection{Duality of classical observables\label{phase_space_duality}}

As we have seen in the previous subsection, the most general duality
transformation which corresponds to the matrix $\mathbf{\alpha }\in GL(2,%
\mathbf{R})$ assigns to each field configuration $\phi (x)$ its dual
configuration $\phi _{\alpha }[\phi ](x)$, which is given by the implicit
formulae (\ref{generalized_duality_alpha}). With help of this transformation
we can define a dual action $S_{\alpha }[\phi ]$ to the original action $%
S[\phi ]$ according to the \ prescription%
\begin{equation}
S_{\alpha }[\phi ]=S[\phi _{\alpha }[\phi ]].
\end{equation}%
We have shown that provided $\phi _{\ast }$ is solution of the equation of
motion corresponding to the dual action $S_{\alpha }[\phi ]$ the dual
configuration $\phi _{\alpha }[\phi _{\ast }]$ is a solution of the equation
of motion for the original action $S[\phi ]$. This can be interpreted that
there is (at least formally) a correspondence between the (classical)
covariant phase spaces $\Sigma $ and $\Sigma _{\alpha }$ of both theories,
which is mediated by duality. By covariant phase space we mean the space of
all solution of the classical equation of motion i.e. without any
constraints on the initial or final data (see \cite{Barnich:1991tc} for more
details), that means the following sets\footnote{%
The covariant phase space can be equipped by symplectic structure e.g. by
Peierls brackets.}
\begin{equation}
\Sigma =\left\{ \phi _{\ast }(x)|\frac{\delta S[\phi _{\ast }]}{\delta \phi
(x)}=0\right\} ,~~~~~\Sigma _{\alpha }=\left\{ \phi _{\ast }(x)|\frac{\delta
S_{\alpha }[\phi _{\ast }]}{\delta \phi (x)}=0\right\} .
\end{equation}%
The correspondence of the covariant phase spaces is then a mapping $\Sigma
_{\alpha }\rightarrow \Sigma $ according to the prescription $\phi _{\ast
}\in \Sigma _{\alpha }\rightarrow \phi _{\alpha }[\phi _{\ast }]\in \Sigma $.

\begin{figure}[t]
\begin{center}
\includegraphics[scale=0.5]{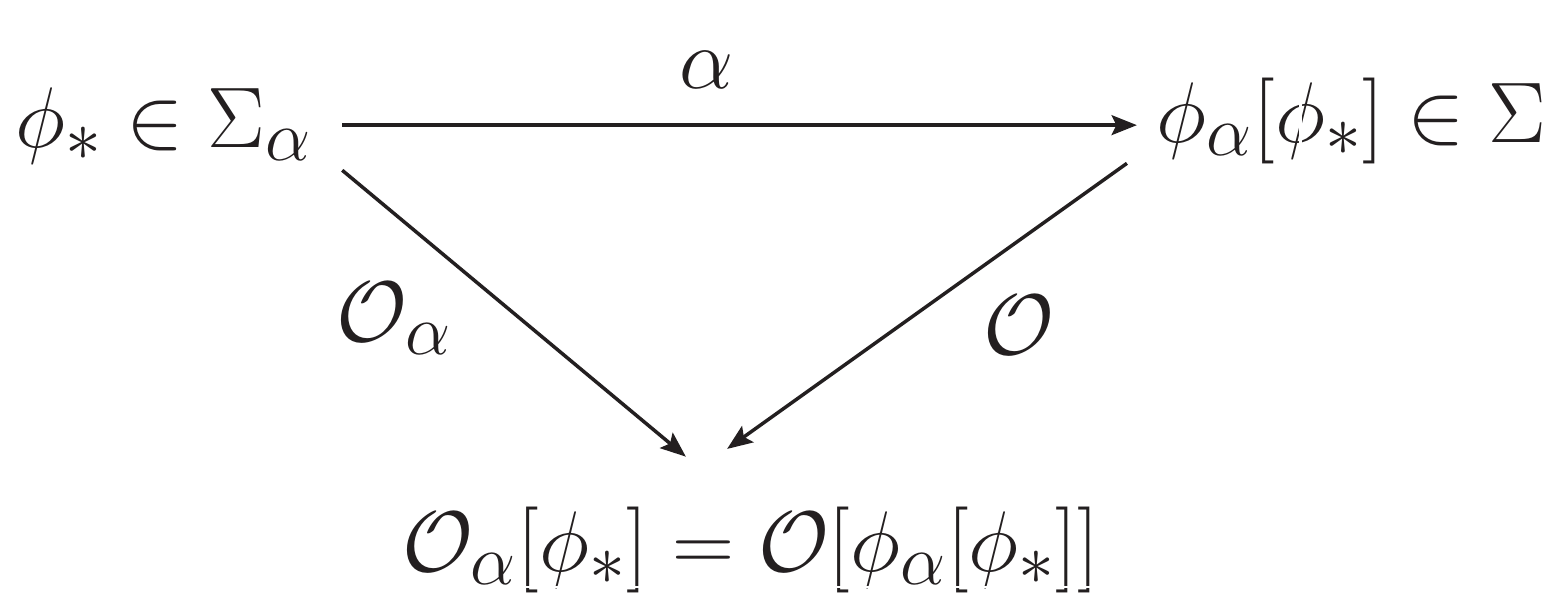} 
\end{center}
\caption{Graphical description of the definition of dual observable. To get
the value of the observable ${\mathcal{O}}$ we can either use the phase
space $\Sigma$ of the original theory or the dual phase space $\Sigma_{%
\protect\alpha}$ and dual observable ${\mathcal{O}}_{\protect\alpha}$}
\label{observables_figure}
\end{figure}

The physical observables are then real functionals on the covariant phase
space, i.e. the mappings $\Sigma \rightarrow \mathbf{R}$ (or $\Sigma
_{\alpha }\rightarrow \mathbf{R}$). They can be understood as a restriction
of the general functionals $\mathcal{O}[\phi ]$, defined on all admissible
field configurations $\phi $, to the space $\Sigma $ or $\Sigma _{\alpha }$.
Two such general functionals $\mathcal{O}$ and $\mathcal{O}^{\prime }$
define then the same observable on the space $\Sigma $ provided their
difference vanishes on $\Sigma $, i.e.%
\begin{equation}
\mathcal{O}[\phi _{\ast }]-\mathcal{O}^{\prime }[\phi _{\ast }]=0~~~~\mathrm{%
~for}~~~\phi _{\ast }\in \Sigma .  \label{observable_equivalence}
\end{equation}%
We can enlarge the duality transformation $\phi \rightarrow \phi _{\alpha
}[\phi ]$ to observables according to the prescription%
\begin{equation}
\mathcal{O}_{\alpha }[\phi ]=\mathcal{O}[\phi _{\alpha }[\phi ]]
\label{observables_duality}
\end{equation}%
(see Fig. \ref{observables_figure}) where the functional $\mathcal{O}[\phi ]$
defines observable on $\Sigma $ and $\mathcal{O}_{\alpha }[\phi ]$ defines%
\footnote{%
Note that this definition is consistent. Indeed, provided the functional $%
\mathcal{G}[\phi ]$ defines the same observable as $\mathcal{F}[\phi ]$
(i.e. (\ref{observable_equivalence}) is satisfied) then for $\phi _{\ast
}\in \Sigma _{\alpha }$ and $\phi _{\alpha }[\phi _{\ast }]=\left( \phi
_{\ast }\right) _{\alpha }\in \Sigma $ (cf. (\ref{solution_relation})).
Therefore
\begin{equation}
\mathcal{F}_{\alpha }[\phi _{\ast }]-\mathcal{G}_{\alpha }[\phi _{\ast }]=%
\mathcal{F}[\phi _{\alpha }[\phi _{\ast }]]-\mathcal{G}[\phi _{\alpha }[\phi
_{\ast }]]=0.
\end{equation}%
As a result, the functionals $\mathcal{F}_{\alpha }[\phi ]$ and $\mathcal{G}%
_{\alpha }[\phi ]$ define the same observable on $\Sigma _{\alpha }$.} the
corresponding dual observable on $\Sigma _{\alpha }$. \ We can therefore
freely calculate the value of given \textquotedblright
abstract\textquotedblright\ observable either within the dual theory using
the point $\phi _{\ast }\in \Sigma _{\alpha }$ or within the original theory
using the dual point $\phi _{\alpha }[\phi _{\ast }]\in \Sigma $ . However,
we have to take care to use corresponding observables $\mathcal{O}_{\alpha }$
or $\mathcal{O~}$\ within the dual and original theories respectively.

The above identification of the phase spaces and the corresponding algebras
of observables can be used in practical calculations. Note however, that in
some cases, a conservation of complexity might take place. Sometimes we end
up with dual theory the action $S_{\alpha }$ of which is much simpler than
the original one. However, to get a concrete value of some simple observable
$\mathcal{O}$ on $\Sigma $ in terms of the simpler dual theory, we have to
use much more complicated observable $\mathcal{O}_{\alpha }$ on $\Sigma
_{\alpha }$.

\subsection{Fluctuations of classical solutions\label{section_fluctuations}}

The duality can also be helpful when the small perturbations $\chi (x)$ of
solution $\phi _{\ast }(x)$ of the classical equation of motion are
investigated, i.e. when we set $\phi =\phi _{\ast }+\chi $ in the Galileon
action. In this section we will discuss how the duality transformation acts
on the field $\chi (x)$ in the linearized theory of fluctuations. We will
show that the solution of the linearized fluctuation equation of motion in
the dual theory is related by appropriate duality transformation to the
corresponding solution within the original theory. This means that the
covariant phase spaces and observables in these theories are related by
duality. We will also discuss the possible superluminal propagation of the
fluctuations in the theories connected by duality and argue that apparent
paradoxes (i.e. when healthy theory with (sub)luminal propagation of the
fluctuation is dual to superluminally propagating one) stem from the
inadequate identification of the dual observables. We will illustrate this
issue on explicitly solvable examples.

To start with, let us insert $\phi =\phi _{\ast }+\chi $ into the Galileon
action
\begin{equation}
S[\phi _{\ast }+\chi ]=S[\phi _{\ast }]+\frac{1}{2}\int \mathrm{d}^{d}x%
\mathrm{d}^{d}y\chi (x)\frac{\delta ^{2}S[\phi ]}{\delta \phi (x)\delta \phi
(y)}|_{\phi _{\ast }}\chi (y)+O\left( \chi ^{3}\right) .  \label{chi_action}
\end{equation}%
Here we used the equation of motion for $\phi _{\ast }$ to eliminate the
term linear in $\chi $. The second variation of the Galileon action (i.e.
the fluctuation operator) is a local second order differential operator of
the form
\begin{equation}
\frac{\delta ^{2}S[\phi ]}{\delta \phi (x)\delta \phi (y)}|_{\phi _{\ast
}}=-g[\phi _{\ast }]^{\mu \nu }\partial _{\mu }\partial _{\nu }\delta
^{(d)}(x-y)
\end{equation}%
where (cf. (\ref{eom}))%
\begin{equation}
g[\phi _{\ast }]^{\mu \nu }=-\sum_{n=2}^{d+1}n(n-1)d_{n}\varepsilon ^{\mu
\mu _{2}\ldots \mu _{d}}\varepsilon ^{\nu \nu _{2}\ldots \nu
_{d}}\prod\limits_{i=2}^{n-1}\partial _{\mu _{i}}\partial _{\nu _{i}}\phi
_{\ast }(x)\prod\limits_{j=n}^{d}\eta _{\mu _{j}\nu _{j}}.
\label{fluctuation_operator}
\end{equation}%
Note that the Minkowski tensor $g[\phi _{\ast }]^{\mu \nu }$ obeys the
following relation%
\begin{equation}
\partial _{\mu }g[\phi _{\ast }]^{\mu \nu }=\partial _{\nu }g[\phi _{\ast
}]^{\mu \nu }=0.
\end{equation}%
This enables us to rewrite the quadratic part of the action (\ref{chi_action}%
) equations of motion for $d>2$ in a form
\begin{equation}
S[\phi _{\ast }+\chi ]=S[\phi _{\ast }]+\frac{1}{2}\int \mathrm{d}^{d}x\sqrt{%
\left\vert G[\phi _{\ast }]\right\vert }G^{\mu \nu }[\phi _{\ast }]\partial
_{\mu }\chi \partial _{\nu }\chi +O\left( \chi ^{3}\right)
\label{diff_invariant}
\end{equation}%
where the effective metric $G^{\mu \nu }[\phi _{\ast }]$ is given in terms
of $g[\phi _{\ast }]^{\mu \nu }$ as%
\begin{equation}
G^{\mu \nu }[\phi _{\ast }]=\left\vert \det \left( g[\phi _{\ast }]^{\cdot
\cdot }\right) \right\vert ^{\frac{1}{2-d}}g^{\mu \nu }[\phi _{\ast }]
\label{effective_inverse_metric}
\end{equation}%
and $G[\phi _{\ast }]=\det \left( G_{\mu \nu }^{-1}[\phi _{\ast }]\right) $
with $G_{\mu \sigma }^{-1}[\phi _{\ast }]G^{\sigma \nu }[\phi _{\ast
}]=\delta _{\mu }^{\nu }$. The second term on the right hand side of (\ref%
{diff_invariant}) is manifestly invariant with respect to general coordinate
transformations $x\rightarrow x^{\prime }$ under which
\begin{equation}
\chi ^{\prime }(x^{\prime })=\chi (x),~~G^{\mu \nu }[\phi _{\ast }]~^{\prime
}(x^{\prime })=\partial _{\rho }x^{\prime \mu }\partial _{\sigma }x^{\prime
\nu }G^{\rho \sigma }[\phi _{\ast }](x)  \label{metric_transformation}
\end{equation}%
(i.e. we assume $\chi (x)$ to be a scalar with respect to the
diffeomorphisms). The linearized equation of motion for $\chi (x)$ then reads%
\begin{equation}
\int \mathrm{d}^{d}y\frac{\delta ^{2}S[\phi ]}{\delta \phi (x)\delta \phi (y)%
}|_{\phi _{\ast }}\chi (y)=-g[\phi _{\ast }]^{\mu \nu }(x)\partial _{\mu
}\partial _{\nu }\chi (x)=0.  \label{linearized_eom}
\end{equation}%
or in manifestly invariant form\footnote{%
Here \textquotedblleft ;\textquotedblright\ means the covariant derivative
with respect to the effective metric $G_{\mu \nu }^{-1}[\phi _{*}]$.}%
\begin{equation}
\chi _{;\mu }^{;\mu }=\frac{1}{\sqrt{\left\vert G[\phi _{\ast }]\right\vert }%
}\partial _{\mu }\left( \sqrt{\left\vert G[\phi _{\ast }]\right\vert }G^{\mu
\nu }[\phi _{\ast }]\partial _{\nu }\chi \right) =0
\label{eom_diff_invariant}
\end{equation}%
The fluctuations of the classical background $\phi _{\ast }$ propagate
therefore according to the massless Klein-Gordon equation in an effective
spacetime with a metric%
\begin{equation}
\mathrm{d}s^{2}=G_{\mu \nu }^{-1}[\phi _{\ast }]\mathrm{d}x^{\mu }\mathrm{d}%
x^{\nu }.  \label{effective_metric}
\end{equation}%
The situation is completely analogous to the case of the small perturbation
of the k-essence in a given classical background which has been discussed in
detail in \cite{Babichev:2007dw}. As explained there, the effective metric $%
G_{\mu \nu }^{-1}[\phi _{\ast }]$ defines the cone of influence of the
fluctuations $\chi (x)$ by the equation\footnote{%
Strictly speaking, the influence cone is given by the equation $g_{\mu \nu
}^{-1}[\phi _{\ast }]N^{\mu }N^{\nu }=0$ where $g_{\mu \nu }^{-1}[\phi
_{\ast }]$ is inverse to $g^{\mu \nu }[\phi _{\ast }]$. However according to
(\ref{effective_inverse_metric}) the metric $G_{\mu \nu }^{-1}[\phi _{\ast
}] $ is conformally equivalent to $g_{\mu \nu }^{-1}[\phi _{\ast }]$.}%
\begin{equation}
G_{\mu \nu }^{-1}[\phi _{\ast }]N^{\mu }N^{\nu }=0.  \label{influence_cone}
\end{equation}%
Provided this influence cone is larger than the Minkowski one, i.e. when $%
N^{2}<0$, the small fluctuations can propagate superluminally.

As we will show in what follows, we can relate the perturbation in original
and dual theory through simple duality transformation induced by the
background solution $\phi _{\ast }(x)$. This will enable us to translate the
solutions for the linearized equations of motion for perturbations from
original to dual theory and vice versa, and to study the effects of the
propagation of the perturbations in both theories. For simplicity we will
restrict our discussion to the one parameter subgroup $\mathbf{\alpha }%
_{D}(\theta )$, however, it can be easily modified for the general case.

The transformation formula can be formally obtained as follows. Taking the
second functional derivative of (\ref{S_theta}) we get%
\begin{eqnarray}
\frac{\delta ^{2}S_{\theta }[\phi ]}{\delta \phi (x)\delta \phi (y)} &=&\int
\mathrm{d}^{d}z\mathrm{d}^{d}w\frac{\delta ^{2}S[\phi _{\theta }]}{\delta
\phi _{\theta }(z)\delta \phi _{\theta }(w)}\frac{\delta \phi _{\theta }(z)}{%
\delta \phi (x)}\frac{\delta \phi _{\theta }(w)}{\delta \phi (y)}  \notag \\
&&+\int \mathrm{d}^{d}z\frac{\delta S[\phi _{\theta }]}{\delta \phi _{\theta
}(z)}\frac{\delta ^{2}\phi _{\theta }(z)}{\delta \phi (x)\delta \phi (y)}
\end{eqnarray}%
Inserting now $\phi \rightarrow \phi _{\ast }$ the solution of the equation
of motion for the action $S_{\theta }$, the second term on the right hand
side drops out\footnote{%
Here we tacitly assume the invertibility of the dual transformation.} \ (cf.
(\ref{equation_relation})) and we have
\begin{equation}
\frac{\delta ^{2}S_{\theta }[\phi ]}{\delta \phi (x)\delta \phi (y)}|_{\phi
_{\ast }}=\int \mathrm{d}^{d}z\mathrm{d}^{d}w\frac{\delta \phi _{\theta }(z)%
}{\delta \phi (x)}\frac{\delta ^{2}S[\phi _{\theta }]}{\delta \phi _{\theta
}(z)\delta \phi _{\theta }(w)}\frac{\delta \phi _{\theta }(w)}{\delta \phi
(y)}|_{\phi _{\ast }}.  \label{f_eom_relation}
\end{equation}%
As a consequence, provided the duality transformation induced by the
background $\phi _{\ast }$ is invertible, the linearized equation of motion (%
\ref{linearized_eom}) for the perturbation $\chi (x)$ around the background
of $\phi _{\ast }$ in the dual theory with action $S_{\theta }$ is
equivalent to%
\begin{equation}
\int \mathrm{d}^{d}y\frac{\delta ^{2}S[\phi _{\theta }]}{\delta \phi
_{\theta }(x)\delta \phi _{\theta }(y)}|_{\left( \phi _{\ast }\right)
_{\theta }}\chi _{\theta }(y)=0.
\end{equation}%
Here we have defined (cf. (\ref{phi_duality_jacobi_matrix})\ )
\begin{equation}
\chi _{\theta }(x)=\int \mathrm{d}^{d}z\frac{\delta \phi _{\theta }(x)}{%
\delta \phi (z)}|_{\phi _{\ast }}\chi (z)=\int \mathrm{d}^{d}z\delta
^{(d)}(X[\phi _{\ast }](x)-z)\chi (z)=\chi \left( X[\phi _{\ast }](x)\right)
.  \label{chi_duality}
\end{equation}%
$\chi _{\theta }(x)$ is therefore a solution of the linearized fluctuation
equation in the original theory around the classical configuration $\left(
\phi _{\ast }\right) _{\theta }(x)$. Formula (\ref{chi_duality}) is thus the
desired duality transformation for the perturbations $\chi (x)$.

Up to now we have interpreted all the duality transformations \emph{actively}%
, i.e. we assumed that both the original and dual theories live on the same
Minkowski spacetime and the fields $\phi _{\ast }(x)$, $\chi (x)$ and $%
\left( \phi _{\ast }\right) _{\theta }(x)$, $\chi _{\theta }(x)$ represent
\emph{different} field configurations within two \emph{different} theories
expressed in terms of the \emph{same} Minkowski coordinates $x$. However,
due to the geometrical nature of the fluctuation action and the
corresponding equation of motion (\ref{diff_invariant}), (\ref%
{eom_diff_invariant}) we can also change the point of view and interpret the
duality transformation of $\chi (x)$ \emph{passively}. Note, that we can
rewrite it equivalently as a transformation of both field and coordinates
\begin{equation}
x_{\theta }=x-2\theta \partial \phi _{\ast }(x),~~~~~~~\chi _{\theta
}(x_{\theta })=\chi (x).  \label{chi_duality1}
\end{equation}%
Unlike the original duality (\ref{original_duality}), the coordinate
transformation here does not depend on the transformed function $\chi (x)$
and is fixed by the classical solution $\phi _{\ast }(x)$. Therefore, in the
the linearized theory of perturbation the duality can be interpreted as a
special coordinate transformation (\ref{metric_transformation}) under which
the effective (inverse) metric $G_{\theta }^{\mu \nu }[\phi _{\ast }]$
corresponding to the dual action\footnote{%
I.e. it is given by (\ref{effective_inverse_metric}) and (\ref%
{fluctuation_operator}) with the substitution $d_{n}\rightarrow d_{n}(\theta
)$.} $S_{\theta }$ transforms according to%
\begin{equation}
G_{\theta }^{\mu \nu }[\phi _{\ast }]~^{\prime }(x_{\theta })=G_{\theta
}^{\rho \sigma }[\phi _{\ast }](x)\partial _{\rho }x_{\theta }^{\mu
}\partial _{\sigma }x_{\theta }^{\nu }=G^{\rho \sigma }[\left( \phi _{\ast
}\right) _{\theta }](x_{\theta })  \label{effective_metric_duality}
\end{equation}%
where $G^{\rho \sigma }[\left( \phi _{\ast }\right) _{\theta }]$ refers to
the original action $S$. Under this interpretation, the fields $\chi (x)$
and $\chi _{\theta }(x_{\theta })$ represent the \emph{same} physical
(geometrical) object within the \emph{same} theory expressed in terms of two
\emph{different} systems of coordinates on a (generally curved) effective
spacetime with the (inverse) metric $G_{\theta }^{\mu \nu }[\phi _{\ast }]$.
Especially, the influence cone (\ref{influence_cone}) does not change within
the passive interpretation.

On the other hand, within the active interpretation, when both $x$ and $%
x_{\theta }$ are Minkowski coordinates and the original and dual theories
are taken to be different, we can relate the influence cones in both
theories. According to (\ref{effective_metric_duality}) we get
\begin{equation}
G_{\mu \nu }^{-1}[\phi _{\ast }]_{\theta }(x)N^{\mu }N^{\nu }=G_{\rho \sigma
}^{-1}[\left( \phi _{\ast }\right) _{\theta }](x_{\theta })\partial _{\mu
}x_{\theta }^{\rho }\partial _{\nu }x_{\theta }^{\sigma }N^{\mu }N^{\nu }.
\label{cone_duality}
\end{equation}%
Therefore, provided $N$ is a propagation vector forming the influence cone
at point $x$ in the dual theory, we get propagation vector $N_{\theta }$ at
point $x_{\theta }$ in the original one as
\begin{equation}
N_{\theta }^{\mu }=N^{\sigma }\partial _{\sigma }x_{\theta }^{\mu
}(x)=N^{\sigma }-2\theta N^{\sigma }\partial _{\sigma }\partial ^{\mu }\phi
_{\ast }(x).
\end{equation}%
The actively interpreted duality transformation can thus deform the
influence cone and e.g. connect healthy (sub)luminally propagating theory
with pathological superluminally propagating one.

This might seem to be a paradox, however, in fact there is nothing unnatural
in it. Let us remind the discussion of the duality of observables in the
subsection \ref{phase_space_duality}. Provided we would like to describe the
small fluctuations in the original theory in terms of the dual theory, we
have to use the corresponding dual observable to the fluctuation operator $%
\delta ^{2}S[\phi ]/\delta \phi \delta \phi $ in the sense of the definition
(\ref{observables_duality}). \ The point is that, while the actions $S[\phi
] $ and $S_{\theta }[\phi ]$ are dual observables in the sense of (\ref%
{observables_duality}), the corresponding fluctuation operators $\delta
^{2}S[\phi ]/\delta \phi \delta \phi $ and $\delta ^{2}S_{\theta }[\phi
]/\delta \phi \delta \phi $ are not. Note that the dual observable $\left(
\delta ^{2}S[\phi ]/\delta \phi \delta \phi \right) _{\theta }$ to $\delta
^{2}S[\phi ]/\delta \phi \delta \phi $ is according to (\ref%
{observables_duality})
\begin{equation}
\left( \frac{\delta ^{2}S}{\delta \phi (x)\delta \phi (y)}\right) _{\theta
}[\phi ]=\left( \frac{\delta ^{2}S}{\delta \phi (x)\delta \phi (y)}\right)
[\phi _{\theta }[\phi ]]=-g[\phi _{\theta }[\phi ]]^{\mu \nu }(x)\partial
_{\mu }\partial _{\nu }\delta ^{(d)}(x-y)
\end{equation}%
and for the background $\phi _{\ast }\in \Sigma _{\theta }$ within the dual
theory generates \emph{the same} effective metric $G_{\rho \sigma
}^{-1}[\left( \phi _{\ast }\right) _{\theta }](x)$ as $\left( \phi _{\ast
}\right) _{\theta }\in \Sigma $ in the original theory. On the other hand
the fluctuation operator in the dual theory has the form%
\begin{equation}
\frac{\delta ^{2}S_{\theta }[\phi ]}{\delta \phi (x)\delta \phi (y)}\equiv
-g_{\theta }^{\alpha \beta }[\phi ](x)\partial _{\mu }\partial _{\nu }\delta
^{(d)}(x-y)
\end{equation}%
where $g_{\theta }^{\alpha \beta }[\phi ]$ is given by formula (\ref%
{fluctuation_operator}) with $d_{n}\rightarrow d_{n}(\theta )$. The
corresponding effective metric is (cf. (\ref{effective_metric_duality}))%
\footnote{%
Using the inversion of the formula (\ref{f_eom_relation}) (cf. also (\ref%
{phi_duality_jacobi_matrix}) and definition of $X[\phi ](x)$) we can also
relate directly the fluctuation operators for $\phi _{\ast }\in \Sigma
_{\theta }$
\begin{equation}
\left( \frac{\delta ^{2}S}{\delta \phi (x)\delta \phi (y)}\right) _{\theta
}[\phi _{\ast }]=-\mathrm{det}^{-1}\left( \frac{\partial x_{\theta }}{%
\partial x}\right) \partial _{\alpha }x_{\theta }^{\mu }\partial _{\beta
}x_{\theta }^{\nu }g_{\theta }^{\alpha \beta }[\phi _{\ast }](X[\phi _{\ast
}](x))\partial _{\mu }\partial _{\nu }\delta ^{(d)}(x-y).
\end{equation}%
Note that, while $\delta ^{2}S_{\theta }[\phi ]/\delta \phi \delta \phi $ is
a local functional of $\phi _{\ast }$ depending only on the second
derivatives of $\phi _{\ast }$ at point $x$, the dual $\left( \delta
^{2}S[\phi ]/\delta \phi \delta \phi \right) _{\theta }$ is nonlocal and can
be formally expanded into an infinite series which includes all the
derivatives of $\phi _{\ast }$ at $x$. It is however a local functional of $%
\phi _{\theta }[\phi _{\ast }]$.}%
\begin{eqnarray}
G_{\mu \nu }^{-1}[\phi _{\ast }]_{\theta }(x) &=&G_{\rho \sigma
}^{-1}[\left( \phi _{\ast }\right) _{\theta }](x_{\theta }(x))\partial _{\mu
}x_{\theta }^{\rho }(x)\partial _{\nu }x_{\theta }^{\sigma }(x)  \notag \\
&=&\left( \eta _{\mu }^{\rho }-2\theta \partial _{\mu }\partial ^{\rho }\phi
_{\ast }(x)\right) \left( \eta _{\nu }^{\sigma }-2\theta \partial _{\nu
}\partial ^{\sigma }\phi _{\ast }(x)\right)  \notag \\
&&\times \sum_{n=0}^{\infty }\frac{(-2\theta )^{n}}{n!}\partial ^{\mu
_{1}}\phi _{\ast }(x)\ldots \partial ^{\mu _{n}}\phi _{\ast }(x)\partial
_{\mu _{1}}\ldots \partial _{\mu _{n}}G_{\rho \sigma }^{-1}[\left( \phi
_{\ast }\right) _{\theta }](x),
\end{eqnarray}%
which differs from $G_{\rho \sigma }^{-1}[\left( \phi _{\ast }\right)
_{\theta }](x)$ at the same point. As a result, the right dual observable to
$\delta ^{2}S[\phi ]/\delta \phi \delta \phi $ differs form $\delta
^{2}S_{\theta }[\phi ]/\delta \phi \delta \phi $. The fluctuation operators $%
\delta ^{2}S[\phi ]/\delta \phi \delta \phi $ and $\delta ^{2}S_{\theta
}[\phi ]/\delta \phi \delta \phi $ represent therefore two different
observables in two different theories. Because in the linearized case the
small fluctuations correspond to the zero modes of these fluctuation
operators, we cannot in general expect that they must necessarily propagate
with the same (front, group) velocity.

On the other hand, we can construct a solution of the linearized fluctuation
equation in the theory with action $S[\phi ]$ around the classical solution $%
\left( \phi _{\ast }\right) _{\theta }$ from the corresponding solution of
the dual theory with action $S_{\theta }[\phi ]$ by means of duality
transformation (\ref{chi_duality}, \ref{chi_duality1}). This means, that the
covariant phase spaces of both linearized theories of fluctuations are dual
to each other, and using the dual observables $\mathcal{O}_{\theta }[\chi ]=%
\mathcal{O}[\chi _{\theta }[\chi ]]$ in the framework of the dual theory
enables us to get results for the observables $O[\chi ]$ in the original
theory. See also Appendix \ref{appendix_fluctuation_duality}\ for further
details.

In the following subsections we show two examples of the duality
transformation of the perturbation of the classical solutions. We will
demonstrate explicitly the usefulness of the formula (\ref{chi_duality}) for
solving the linearized fluctuation equation of motion.

\subsubsection{Fluctuations of the plane wave background}

The first example is the fluctuation of the plane wave background (see e.g.
\cite{Evslin:2011vh}, \cite{deRham:2010eu}). In this case the background $%
\phi _{\ast }$ has the form
\begin{equation}
\phi _{\ast }(x)=F(n\cdot x)  \label{plane_wave_background}
\end{equation}%
where $F(\cdot )$ is some twice differentiable function. For light-like $%
n=(1,\mathbf{n})$ with $\mathbf{n}^{2}=1$, such $\phi _{\ast }(x)$ is a
classical solution of the general Galileon equation of motion\footnote{%
Here we take $d_{1}=0$ but the other $d_{i}$ we assume to be arbitrary.} (%
\ref{eom}). The reason is that, for the ansatz (\ref{plane_wave_background}%
), the problem is effectively one-dimensional and therefore the interaction
terms in the equation of motion vanish automatically\footnote{%
Let us remind that the equation of motion reads%
\begin{equation*}
\sum_{n=1}^{d+1}nd_{n}\mathcal{L}_{n-1}^{\mathrm{der}}(\partial \partial
\phi _{\ast })=0.
\end{equation*}%
We have for $n>2$%
\begin{equation*}
\mathcal{L}_{n-1}^{\mathrm{der}}(\partial \partial \phi _{\ast })=\mathcal{L}%
_{n-1}^{\mathrm{der}}(nnF^{\prime \prime }(n\cdot x))=0.
\end{equation*}%
For $n=2$%
\begin{equation*}
\mathcal{L}_{1}^{\mathrm{der}}(nnF^{\prime \prime }(n\cdot
x))=(-1)^{d-1}(d-1)!\left( n\cdot n\right) F^{\prime \prime }(n\cdot x)
\end{equation*}%
which requires $n$ to be light-like $n^{2}=0$.}.

The linearized equation of motion for the fluctuation of this background has
the form (\ref{linearized_eom}), explicitly
\begin{equation}
\left[ d_{2}\square -d_{3}F^{\prime \prime }(n\cdot x)\left( n\cdot \partial
\right) ^{2}\right] \chi (x)=0  \label{plane_wave_fluctuation}
\end{equation}%
and only the cubic coupling matters. The effective metric is in this case%
\begin{equation}
G_{\mu \nu }^{-1}[\phi _{\ast }]=\frac{1}{d_{2}}\left( \eta _{\mu \nu }+%
\frac{d_{3}}{d_{2}}F^{\prime \prime }(n\cdot x)n_{\mu }n_{\nu }\right)
,~~~~G[\phi _{\ast }]=-\frac{1}{d_{2}^{4}}
\label{plane_wave_effective_metric}
\end{equation}

The linearized theory of fluctuations on the plane wave background is known
to be a prominent example of a possible classical superluminal propagation,
which is sometimes interpreted as a pathology of the Galileon theory. In the
short wave-length (eikonal) approximation $\chi = A\exp (\mathrm{i}S)$
(where $S$ is assumed to be large) we get%
\begin{equation}
d_{2}\partial S\cdot \partial S-d_{3}F^{\prime \prime }(n\cdot x)\left(
n\cdot \partial S\right) ^{2}=0.
\end{equation}%
Denoting $\partial S=\left( \omega (\mathbf{k}),\mathbf{k}\right) $ we get
for the wave front velocity%
\begin{equation}
v_{front}^{2}=\frac{\omega (\mathbf{k})^{2}}{\mathbf{k}^{2}}=1+\frac{d_{3}}{%
d_{2}}F^{\prime \prime }(n\cdot x)\left( n\cdot \partial S\right) ^{2},
\end{equation}%
and thus the equation (\ref{plane_wave_fluctuation}) leads locally to
superluminal propagation of the fluctuations provided (see also \cite%
{deRham:2010eu} for more details)
\begin{equation}
\frac{d_{3}}{d_{2}}F^{\prime \prime }(n\cdot x)>0.
\end{equation}%
In this case the lightcone corresponding to the effective metric (\ref%
{plane_wave_effective_metric}) is wider then the Minkowski one.

Here we will show that duality can help us to find easily the explicit
solution of (\ref{plane_wave_fluctuation}). The duality transformation
induced by $\phi _{\ast }(x)=F(n\cdot x)$ reads%
\begin{eqnarray}
x_{\theta } &=&x-2\theta nF^{\prime }(n\cdot x)  \label{x_F_duality} \\
\left( \phi _{\ast }\right) _{\theta }(x_{\theta }) &=&F(n\cdot x).
\label{F_duality}
\end{eqnarray}%
Therefore
\begin{equation}
n\cdot x=n\cdot x_{\theta }  \label{nx_nx_theta}
\end{equation}%
and $\left( \phi _{\ast }\right) _{\theta }(x)=F(n\cdot x)$, while the
fluctuations $\chi (x)$ in the dual theory are governed by the equation (\ref%
{plane_wave_fluctuation}) with the exchange $d_{2,3}\rightarrow
d_{2,3}(\theta )$. With the choice
\begin{equation}
\theta _{\ast }=\frac{d_{3}}{4d_{2}}  \label{chi_star}
\end{equation}%
for the duality transformation (\ref{one_parametric_duality}) we get $%
d_{3}(\theta _{\ast })=0$ and the fluctuation equation is simply $\square
\chi (x)=0$ with general solution%
\begin{equation}
\chi (x)=\int \frac{\mathrm{d}^{3}\mathbf{k}}{(2\pi )^{3}2|\mathbf{k}|}\left[
a(\mathbf{k})\exp (-\mathrm{i}k\cdot x)+h.c\right]
\end{equation}%
where $k=(|\mathbf{k}|,\mathbf{k})$. The dual $\chi _{\theta _{\ast }}(x)$
to $\chi (x)$ given by (\ref{chi_duality}) yields then the solution of the
general fluctuation equation (\ref{plane_wave_fluctuation}). With help of (%
\ref{nx_nx_theta}) we can easily find the inversion $X[\phi _{\ast }](x)$ of
(\ref{x_F_duality}), namely%
\begin{equation}
X[\phi _{\ast }](x)=x+2\theta nF^{\prime }(n\cdot x),
\end{equation}%
and using the general prescription \ (\ref{chi_duality}) we get finally the
general solution of (\ref{plane_wave_fluctuation}) in the form
\begin{equation}
\chi _{\theta _{\ast }}(x)=\int \frac{\mathrm{d}^{3}\mathbf{k}}{(2\pi )^{3}2|%
\mathbf{k}|}\left\{ a(\mathbf{k})\chi _{\mathbf{k}}(x)+h.c.\right\} ,
\label{general_solution_chi}
\end{equation}%
where the basis of the solutions is%
\begin{equation}
\chi _{\mathbf{k}}(x)=\exp \left[ -\mathrm{i}k\cdot X[\phi _{\ast }](x)%
\right] =\exp \left[ -\mathrm{i}k\cdot x-2\mathrm{i}\theta _{\ast }\left(
n\cdot k\right) F^{\prime }(n\cdot x)\right].  \label{distorted_plane_wave}
\end{equation}

Let us now discuss the physical properties of this solution, namely the
conditions under which we get superluminal propagation. Assume that the
coefficient function $a(\mathbf{k})$ has a sharp peak at $\mathbf{k}=%
\overline{\mathbf{k}}$ and is nonzero only in as small vicinity of this
point. Then we can write approximately%
\begin{equation}
\chi _{\theta _{\ast }}(x)\approx \mathrm{e}^{-\mathrm{i}\overline{k}\cdot
X[\phi _{\ast }]}\mathcal{A}\left( \mathbf{X}[\phi _{\ast }]-\widehat{%
\mathbf{k}}X^{0}[\phi _{\ast }]\right) +h.c.
\end{equation}%
where $\overline{k}=(|\overline{\mathbf{k}}|,\overline{\mathbf{k}})$, $%
\widehat{\mathbf{k}}=\overline{\mathbf{k}}/|\overline{\mathbf{k}}|$ and the
shape of the wave packet is given by
\begin{equation}
\mathcal{A}\left( \mathbf{y}\right) =\int \frac{\mathrm{d}^{3}\mathbf{k}}{%
(2\pi )^{3}2|\mathbf{k}|}a(\mathbf{k+}\overline{\mathbf{k}})\mathrm{e}^{%
\mathrm{i}\mathbf{k}\cdot \mathbf{y}}.
\end{equation}%
The group velocity can be now obtained in a standard way by differentiating
the equation%
\begin{equation}
\mathbf{X}[\phi _{\ast }]-\widehat{\mathbf{k}}X^{0}[\phi _{\ast }]=\mathbf{%
const}.  \label{wave_packet_center}
\end{equation}%
with respect to $t\equiv x^{0}$. After some algebra (see Appendix \ref%
{v_group_appendix}) we find%
\begin{equation}
\mathbf{v}_{group}(x)=\frac{\widehat{\mathbf{k}}-2\theta _{\ast }(1-\mathbf{n%
}\cdot \widehat{\mathbf{k}})F^{\prime \prime }(n\cdot x)\mathbf{n}}{%
1-2\theta _{\ast }F^{\prime \prime }(n\cdot x)(1-\mathbf{n}\cdot \widehat{%
\mathbf{k}})}
\end{equation}%
and thus%
\begin{equation}
v_{group}(x)=\left( 1+\frac{4\theta _{\ast }F^{\prime \prime }(n\cdot x)(1-%
\mathbf{n}\cdot \widehat{\mathbf{k}})^{2}}{\left[ 1-2\theta _{\ast
}F^{\prime \prime }(n\cdot x)(1-\mathbf{n}\cdot \widehat{\mathbf{k}})\right]
^{2}}\right) ^{1/2}.
\end{equation}%
Therefore the superluminal propagation is possible in the space-time regions
with $\theta _{\ast }F^{\prime \prime }(n\cdot x)=3d_{3}F^{\prime \prime
}(n\cdot x)>0$. The phase velocity is given by%
\begin{equation}
v_{phase}(x)=\frac{\frac{\mathrm{d}}{\mathrm{d}t}\overline{k}\cdot X[\phi
_{\ast }]}{|\nabla \overline{k}\cdot X[\phi _{\ast }]|}=\left( 1-\frac{%
4\theta _{\ast }F^{\prime \prime }(n\cdot x)(1-\mathbf{n}\cdot \widehat{%
\mathbf{k}})^{2}}{\left[ 1+2\theta _{\ast }F^{\prime \prime }(n\cdot x)(1-%
\mathbf{n}\cdot \widehat{\mathbf{k}})\right] ^{2}}\right) ^{-1/2}
\end{equation}%
and is superluminal under the same condition as $v_{group}$.

The explicit knowledge of the basis $\chi _{\mathbf{k}}(x)$ allows us to
discuss also the quantum aspects of the fluctuations on the operator level.
Because of a special role of the variable $n\cdot x$, which naturally plays
a role of the evolution parameter in the problem, the most convenient
prescription for the quantization of the field $\chi $ is the Dirac
front-form one. Let us introduce the light cone coordinates as%
\begin{equation}
x^{+}=n\cdot x,~~~~x^{-}=\widetilde{n}\cdot x,~~~\mathbf{x}_{\bot }=\mathbf{x%
}-\mathbf{n}\left( \mathbf{n}\cdot \mathbf{x}\right) ~
\end{equation}%
where $\widetilde{n}=(1,-\mathbf{n)}$ and analogously for any other vector.
Then e.g. the solutions $\chi _{\mathbf{k}}(x)$ can be rewritten in the form
\begin{equation*}
\chi _{\mathbf{k}}(x)=\exp \left[ -\frac{\mathrm{i}}{2}k^{-}x^{+}-\frac{%
\mathrm{i}}{2}k^{+}x^{-}+\mathrm{i}\mathbf{k}_{_{\bot }}\cdot \mathbf{x}%
_{\bot }-2\mathrm{i}\theta _{\ast }k^{+}F^{\prime }(x^{+})\right]
\end{equation*}%
and the condition on the vector $k$ to be on-shell and positive-energy is
then expressed as
\begin{equation}
k^{-}=\frac{\mathbf{k}_{\bot }^{2}}{k^{+}},~~k^{+}\geq 0.
\end{equation}%
It is easy to prove that the elements of the basis $\chi _{\mathbf{k}}(x)$
are orthogonal with respect to the indefinite scalar product defined on the
solution of (\ref{plane_wave_fluctuation}) as\footnote{%
For the solutions $\chi _{1,2}$ of the fluctuation equation of motion the
scalar product is independent on the choice of $x^{+}$.}
\begin{equation}
\langle \chi _{1},\chi _{2}\rangle =\mathrm{i}\int_{x^{+}=const.}\mathrm{d}%
x^{-}\mathrm{d}^{2}\mathbf{x}_{\bot }\chi _{1}^{\ast }\overleftrightarrow{%
\partial _{-}}\chi _{2},
\end{equation}%
namely%
\begin{eqnarray}
\langle \chi _{\mathbf{k}},\chi _{\mathbf{q}}\rangle &=&-\langle \chi _{%
\mathbf{k}}^{\ast },\chi _{\mathbf{q}}^{\ast }\rangle =\left( 2\pi \right)
^{3}2k^{+}\delta \left( k^{+}-q^{+}\right) \delta ^{(2)}(\mathbf{k}_{\bot }-%
\mathbf{q}_{\bot })  \notag \\
\langle \chi _{\mathbf{k}}^{\ast },\chi _{\mathbf{q}}\rangle &=&\langle \chi
_{\mathbf{k}},\chi _{\mathbf{q}}^{\ast }\rangle =0.
\label{chi_orthogonality}
\end{eqnarray}%
Let us write the operator solution $\widehat{\chi }(x)$ of the equation (\ref%
{plane_wave_fluctuation}) in the form (see (\ref{general_solution_chi}) and (%
\ref{distorted_plane_wave}))
\begin{equation}
\widehat{\chi }(x)=\int_{k^{+}>0}\frac{\mathrm{d}k^{+}\mathrm{d}^{2}\mathbf{k%
}_{\bot }}{\left( 2\pi \right) ^{3}2k^{+}}\left( a(k^{+},\mathbf{k}_{\bot
})\chi _{\mathbf{k}}(x)+a^{+}(k^{+},\mathbf{k}_{\bot })\chi _{\mathbf{k}%
}^{\ast }(x)\right) .  \label{chi_operator}
\end{equation}%
According to the general quantization prescription, the operators $\widehat{%
\chi }(x)$ and their canonically conjugated momenta in the front-form
formalism%
\begin{equation}
\widehat{\pi }(x)=2\partial _{-}\widehat{\chi }(x)
\end{equation}%
have to satisfy the canonical commutation relations%
\begin{eqnarray}
\left[ \widehat{\chi }(x),\widehat{\chi }(y)\right] _{x^{+}=y^{+}} &=&-\frac{%
\mathrm{i}}{4}\varepsilon \left( x^{-}-y^{-}\right) \delta ^{(2)}(\mathbf{x}%
_{\bot }-\mathbf{y}_{\bot })  \notag \\
\left[ \widehat{\chi }(x),\widehat{\pi }(y)\right] _{x^{+}=y^{+}} &=&\mathrm{%
i}\delta \left( x^{-}-y^{-}\right) \delta ^{(2)}(\mathbf{x}_{\bot }-\mathbf{y%
}_{\bot }).
\end{eqnarray}%
This implies with help of (\ref{chi_orthogonality}) the canonical
commutation relations
\begin{eqnarray}
\left[ a(k^{+},\mathbf{k}_{\bot }),a^{+}(q^{+},\mathbf{q}_{\bot })\right]
&=&\left( 2\pi \right) ^{3}2k^{+}\delta \left( k^{+}-q^{+}\right) \delta
^{(2)}(\mathbf{k}_{\bot }-\mathbf{q}_{\bot })  \notag \\
\left[ a(k^{+},\mathbf{k}_{\bot }),a(q^{+},\mathbf{q}_{\bot })\right] &=&%
\left[ a^{+}(k^{+},\mathbf{k}_{\bot }),a^{+}(q^{+},\mathbf{q}_{\bot })\right]
=0.  \label{CCR}
\end{eqnarray}%
Now it is easy to calculate the commutator of the fields with the result
\begin{equation}
\left[ \widehat{\chi }(x),\widehat{\chi }(y)\right] =-\frac{i}{(2\pi )}%
\varepsilon (x^{+}-y^{+})\delta \left( \lambda \right)
\end{equation}%
where $\varepsilon (z)=\theta (z)-\theta (-z)$ is the sign function and%
\begin{equation*}
\lambda =(x-y)^{2}+4\theta _{\ast }(x^{+}-y^{+})\left( F^{\prime
}(x^{+})-F^{\prime }(y^{+})\right) .
\end{equation*}%
The commutator is nonzero only for
\begin{equation}
(x-y)^{2}=-4\theta _{\ast }(x^{+}-y^{+})^{2}\frac{F^{\prime
}(x^{+})-F^{\prime }(y^{+})}{x^{+}-y^{+}}=-4\theta _{\ast
}(x^{+}-y^{+})^{2}F^{\prime \prime }(\xi ^{+})
\end{equation}%
where $\xi ^{+}$ is located between $x^{+}$ and $y^{+}$. \ Provided $\theta
_{\ast }F^{\prime \prime }(\xi ^{+})>0$, the commutator is nonzero also
outside the light cone and causality is violated.

The commutation relations (\ref{CCR}) can be represented on the Fock space
built on the vacuum state $|0\rangle $ which is annihilated by the operators
$a(k^{+},\mathbf{k}_{\bot })$. Let us suppose that there exist finite limits%
\begin{equation}
\lim_{x^{+}\rightarrow \pm \infty }F^{\prime }(x^{+})\equiv \psi _{\pm }.
\end{equation}%
Then in the limit $t\rightarrow \pm \infty $, ($\mathbf{x}$ fixed) we get%
\begin{eqnarray}
\widehat{\chi }(x) &\rightarrow &\widehat{\chi }_{\mathrm{out,in}%
}(x)=\int_{k^{+}>0}\frac{\mathrm{d}k^{+}\mathrm{d}^{2}\mathbf{k}_{\bot }}{%
\left( 2\pi \right) ^{3}2k^{+}}\left( a_{\mathrm{out,in}}(k^{+},\mathbf{k}%
_{\bot })\mathrm{e}^{-\mathrm{i}k\cdot x}+a_{\mathrm{out,in}}^{+}(k^{+},%
\mathbf{k}_{\bot })\mathrm{e}^{\mathrm{i}k\cdot x}\right) .  \notag \\
&&
\end{eqnarray}%
Here we have identified the out and in creation and annihilation operators%
\begin{eqnarray}
a_{\mathrm{out,in}}(k^{+},\mathbf{k}_{\bot }) &=&a(k^{+},\mathbf{k}_{\bot
})\exp \left[ -2\mathrm{i}\theta _{\ast }k^{+}\psi _{\pm }\right]  \notag \\
a_{\mathrm{out,in}}^{+}(k^{+},\mathbf{k}_{\bot }) &=&a^{+}(k^{+},\mathbf{k}%
_{\bot })\exp \left[ 2\mathrm{i}\theta _{\ast }k^{+}\psi _{\pm }\right] .
\end{eqnarray}%
Note that these operators satisfy again the commutation relations (\ref{CCR}%
). The fields $\widehat{\chi }_{\mathrm{out,in}}(x)$ are therefore free
fields which create asymptotic in and out states from the Fock vacuum%
\begin{equation}
|\mathbf{k}_{\left( 1\right) },\ldots ,\mathbf{k}_{\left( m\right) };\mathrm{%
out,in}\rangle =a_{\mathrm{out,in}}^{+}(k_{\left( 1\right) }^{+},\mathbf{k}%
_{(1)\bot })\ldots a_{\mathrm{out,in}}^{+}(k_{\left( m\right) }^{+},\mathbf{k%
}_{(m)\bot })|0\rangle
\end{equation}%
These are eigenstates of the momentum operators%
\begin{eqnarray}
P^{+} &=&\int_{k^{+}>0}\frac{\mathrm{d}k^{+}\mathrm{d}^{2}\mathbf{k}_{\bot }%
}{\left( 2\pi \right) ^{3}2k^{+}}k^{+}a^{+}(k^{+},\mathbf{k}_{\bot })a(k^{+},%
\mathbf{k}_{\bot })  \notag \\
P^{-} &=&\int_{k^{+}>0}\frac{\mathrm{d}k^{+}\mathrm{d}^{2}\mathbf{k}_{\bot }%
}{\left( 2\pi \right) ^{3}2k^{+}}\frac{\mathbf{k}_{\bot }^{2}}{k^{+}}%
a^{+}(k^{+},\mathbf{k}_{\bot })a(k^{+},\mathbf{k}_{\bot })  \notag \\
\mathbf{P}_{\bot } &=&\int_{k^{+}>0}\frac{\mathrm{d}k^{+}\mathrm{d}^{2}%
\mathbf{k}_{\bot }}{\left( 2\pi \right) ^{3}2k^{+}}\mathbf{k}_{\bot
}a^{+}(k^{+},\mathbf{k}_{\bot })a(k^{+},\mathbf{k}_{\bot })
\end{eqnarray}%
with eigenvalues $\sum_{i=1}^{m}k_{(i)}$ where $k_{(i)}=\left( k_{\left(
i\right) }^{+},~k_{\left( i\right) }^{-},\mathbf{k}_{(i)\bot }\right) $
satisfying $k_{(i)}^{2}=0$ and correspond therefore to the $m$
non-interacting massless excitations. \ The $S$-matrix defined as%
\begin{equation}
a_{\mathrm{out}}(k^{+},\mathbf{k}_{\bot })=a_{\mathrm{in}}(k^{+},\mathbf{k}%
_{\bot })\exp \left[ 2\mathrm{i}\theta _{\ast }k^{+}\left( \psi _{-}-\psi
_{+}\right) \right] \equiv S^{+}a_{\mathrm{in}}(k^{+},\mathbf{k}_{\bot })S
\end{equation}%
is then expressed simply as a translation in the $x^{-}$ direction%
\begin{equation}
S=\exp \left[ 2\mathrm{i}\theta _{\ast }P^{+}\left( \psi _{-}-\psi
_{+}\right) \right] .
\end{equation}%
The only nontrivial connected scattering amplitude is the two-point one%
\footnote{%
Here we use the normalization
\begin{equation*}
S_{fi}=\langle \mathbf{k}^{\prime }|\mathbf{k}\rangle +2\pi \mathrm{i}\delta
(E^{\prime }-E)\mathcal{M}(\mathbf{k}^{\prime },\mathbf{k})
\end{equation*}%
}%
\begin{equation}
\mathcal{M}(\mathbf{k},\mathbf{k}^{\prime })=\left( 4\pi \right) ^{2}\frac{%
\mathrm{e}^{2\mathrm{i}\delta _{\mathbf{k}}}-1}{2\mathrm{i}|\mathbf{k}|}%
\delta ^{(2)}(\widehat{\mathbf{k}}-\widehat{\mathbf{k}}^{\prime })
\label{2pt_amplitude}
\end{equation}%
which describes scattering of the individual excitations on the background
resulting in a phase shift $\delta _{\mathbf{k}}\mathbf{=}\theta _{\ast
}k^{+}\left( \psi _{-}-\psi _{+}\right) $. This amplitude can be also
obtained by means of summation of the perturbative series generated by the
Feynman rules depicted in Fig. \ref{fig:F_rules} (see Appendix \ref{appendix_amplitude}).

Note that the $S$ matrix is trivial for $\left( \psi _{-}-\psi _{+}\right)
=0 $, i.e. especially when $F(x^{+})$ has compact support. This is in accord
with the discussion in \cite{Creminelli:2014zxa}, where it has been
demonstrated that for localized classical background the displacement of the
null geodetics with respect to the effective metric vanish asymptotically.

\begin{figure}[t]
\includegraphics[scale=0.8]{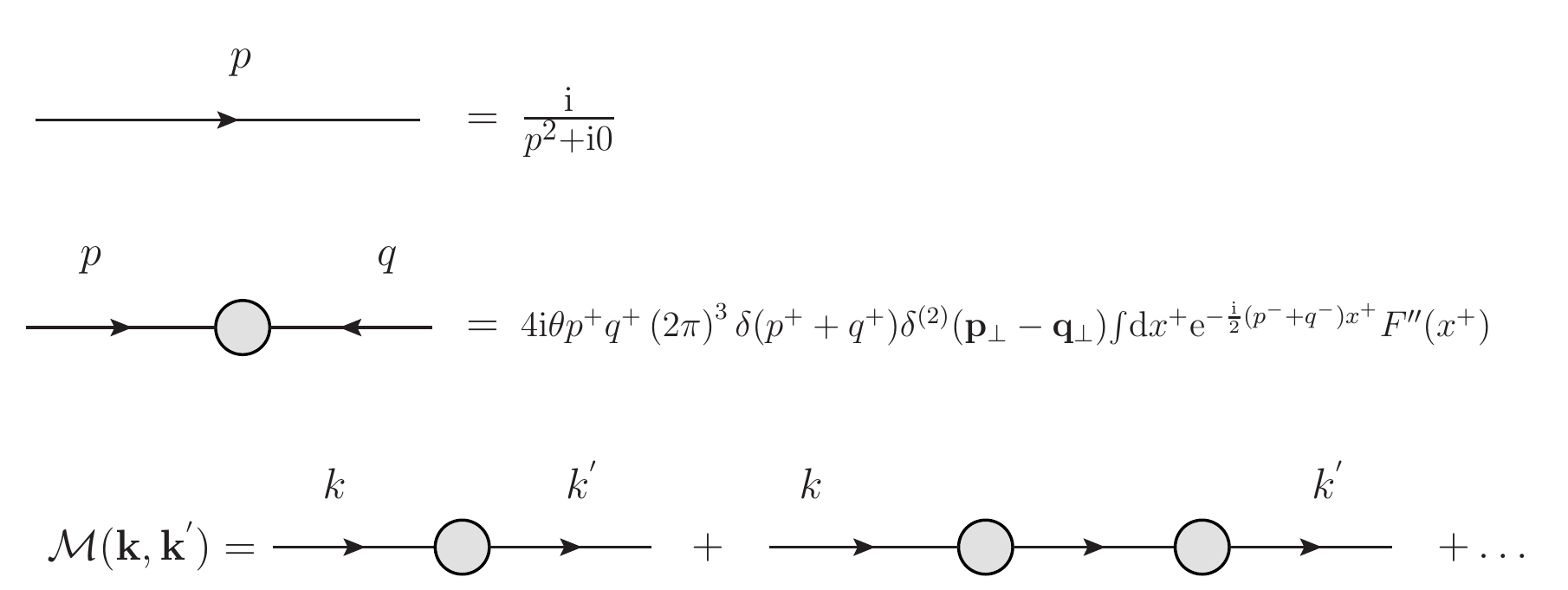} 
\caption{The Feynman rules for the perturbative calculation of the
one-particle amplitude ${\mathcal{M}}(\mathbf{k},\mathbf{k}^{^{\prime }})$
are given in the first two rows of this figure. This amplitude is given by
the sum of Feynman graphs depicted in the last row.}
\label{fig:F_rules}
\end{figure}

To conclude, we have shown that within the original theory with $d_{3}\neq 0$
the fluctuations in general interact with the plane wave background and
might be scattered by it provided its profile $F(x^{+})$ does not fall
rapidly enough for $x^{+}\rightarrow \infty $. \ Also the commutator of the
operators $\widehat{\chi }(x)$ might locally violate causality when $\theta
_{\ast }F^{\prime \prime }(x^{+})>0$ in some region. This might be felt as a
paradox since for the dual theory for which $d_{3}=0$ the linearized
fluctuation theory is free, causal, and its $S$ matrix is trivial. However,
we are comparing here two different theories the relation of which must be
taken with care. As we have argued, only comparison of dual observables
makes sense. E.g. provided we would like to describe the correlators of the
fluctuations of the original theory using the dual (free) one, we have not
to use the free operators
\begin{equation}
\widehat{\chi }_{0}(x)=\int_{k^{+}>0}\frac{\mathrm{d}k^{+}\mathrm{d}^{2}%
\mathbf{k}_{\bot }}{\left( 2\pi \right) ^{3}2k^{+}}\left( a(k^{+},\mathbf{k}%
_{\bot })\mathrm{e}^{-\mathrm{i}k\cdot x}+a^{+}(k^{+},\mathbf{k}_{\bot })%
\mathrm{e}^{\mathrm{i}k\cdot x}\right) ,
\end{equation}%
but their duals $\widehat{\chi }(x)$ given by (\ref{chi_operator}), see also
Appendix \ref{appendix_fluctuation_duality}. These, though live on the same
Hilbert space, are not unitarily equivalent\footnote{%
I.e. there does not exist any unitary operator $U$ on the Fock space such
that $\widehat{\chi }(x)$ $=U\widehat{\chi }_{0}(x)U^{+}$.} to $\widehat{%
\chi }_{0}(x)$. Instead we have
\begin{equation*}
\widehat{\chi }(x)=\exp \left[ -2\mathrm{i}\theta _{\ast }P^{+}F^{\prime
}(x^{+})\right] \widehat{\chi }_{0}(x)\exp \left[ 2\mathrm{i}\theta _{\ast
}P^{+}F^{\prime }(x^{+})\right]
\end{equation*}%
and thus the dual asymptotic in and out states defined by the dual operators
$\widehat{\chi }(x)$ differ from that defined by $\widehat{\chi }_{0}(x)$
when $\left( \psi _{-}-\psi _{+}\right) \neq 0$.

\subsubsection{Fluctuations of the cylindrically symmetric solution}

As the second example let us consider briefly the fluctuations of the
cylindrically symmetric static solution $\phi (z,\overline{z})$ discussed in
the previous subsection. The matrix $g[\phi ]^{\mu \nu }$ is in this case
given in the form (see Appendix \ref{cylindric_g_mu_nu} for details)%
\begin{eqnarray}
g[\phi ] &=&12d_{2}(\theta )\eta  \notag \\
&&-12d_{3}(\theta )\left(
\begin{array}{cccc}
4\partial \overline{\partial }\phi & 0 & 0 & 0 \\
0 & -4\partial \overline{\partial }\phi & 0 & 0 \\
0 & 0 & \left( \partial -\overline{\partial }\right) ^{2}\phi & \mathrm{i}%
\left( \partial ^{2}-\overline{\partial }^{2}\right) \phi \\
0 & 0 & \mathrm{i}\left( \partial ^{2}-\overline{\partial }^{2}\right) \phi
& -\left( \partial +\overline{\partial }\right) ^{2}\phi%
\end{array}%
\right)  \notag \\
&&-24d_{4}(\theta )\left(
\begin{array}{cccc}
4\left[ \partial ^{2}\phi \overline{\partial }^{2}\phi -\left( \partial
\overline{\partial }\phi \right) ^{2}\right] & 0 & 0 & 0 \\
0 & -4\left[ \partial ^{2}\phi \overline{\partial }^{2}\phi -\left( \partial
\overline{\partial }\phi \right) ^{2}\right] & 0 & 0 \\
0 & 0 & 0 & 0 \\
0 & 0 & 0 & 0%
\end{array}%
\right) .  \label{general_g[phi]}
\end{eqnarray}%
Due to the cylindrical symmetry $g[\phi ]$ does not depend on the quintic
coupling $d_{5}$. Let us now show the application of the duality to solving
the general equation (\ref{linearized_eom}). \ As in the above examples,
within the dual theory this equation simplifies. Setting $d_{2}=1/12$ as
usual and passing to the dual theory with $d_{3}(\theta )=0$ we get in terms
of the invariant $I_{4}$ (cf. (\ref{I_4_5}))
\begin{equation}
g_{\theta }[\phi ]=\left(
\begin{array}{cc}
\left\{ 1-96I_{4}\left[ \partial ^{2}\phi \overline{\partial }^{2}\phi
-\left( \partial \overline{\partial }\phi \right) ^{2}\right] \right\}
\sigma ^{3} & 0 \\
0 & -\mathbf{1}%
\end{array}%
\right)
\end{equation}%
where $\sigma ^{3}=\mathrm{diag}\left( 1,-1\right) $ is the third Pauli
matrix. Thus the equation (\ref{linearized_eom}) becomes%
\begin{equation}
\left\{ 1-96I_{4}\left[ \partial ^{2}\phi \overline{\partial }^{2}\phi
-\left( \partial \overline{\partial }\phi \right) ^{2}\right] \right\}
\left( \overset{..}{\chi }-\chi ^{\prime \prime }\right) -4\partial
\overline{\partial }\chi =0
\end{equation}%
where $\chi \equiv \chi (t,x,z,\overline{z})$, dots and dashes means
derivative with respect to $t$ and $x^{1}$ respectively. In the dual theory
the static solution $\phi _{\ast }$ is given by (\ref{free_theory}) and for $%
z\neq 0$ we get explicitly the following dual fluctuation equation%
\begin{equation}
\left[ 1-6I_{4}\left( \frac{\sigma }{\pi }\right) ^{2}\frac{1}{\left( z%
\overline{z}\right) ^{2}}\right] \left( \overset{..}{\chi }-\chi ^{\prime
\prime }\right) -4\partial \overline{\partial }\chi =0.
\end{equation}%
For%
\begin{equation*}
1-6I_{4}\left( \frac{\sigma }{\pi }\right) ^{2}\frac{1}{\left( z\overline{z}%
\right) ^{2}}>0
\end{equation*}%
we can rewrite it equivalently as%
\begin{equation}
\chi _{;\mu }^{;\mu }=0
\end{equation}%
where $\chi _{;\mu }^{;\mu }$ is the Laplace-Beltrami operator acting on $%
\chi $ and corresponding to the effective metric metric $G_{\mu \nu
}^{-1}[\phi _{\ast }]_{\theta }$ given explicitly as%
\begin{equation}
\mathrm{d}s^{2}=(\mathrm{d}t^{2}-\mathrm{d}x^{2})-\left[ 1-6I_{4}\left(
\frac{\sigma }{\pi }\right) ^{2}\frac{1}{\left( z\overline{z}\right) ^{2}}%
\right] \mathrm{d}z\mathrm{d}\overline{z}.
\end{equation}%
Similarly to the case of the point-like source, \ the duality arguments help
us to easily solve the equation (\ref{linearized_eom}) for the class of
theories for which $I_{4}=0$. In such a case we have to solve a free
equation $\square \chi =0$ with general solution (with luminal propagation)%
\begin{equation}
\chi (t,x,z,\overline{z})=\int \frac{\mathrm{d}^{3}\mathbf{k}}{(2\pi )^{3}2|%
\mathbf{k}|}\left\{ a(\mathbf{k})\exp \left[ -\mathrm{i}\left( |\mathbf{k}%
|t-k^{1}x-z\overline{w}-\overline{z}w\right) \right] +h.c\right\}
\end{equation}%
with $k=(|\mathbf{k}|,\mathbf{k})$ and
\begin{eqnarray}
w &=&\frac{1}{2}(k^{2}+\mathrm{i}k^{3})  \notag \\
\overline{w} &=&\frac{1}{2}(k^{2}-\mathrm{i}k^{3}).
\end{eqnarray}%
Its dual $\chi _{\theta }$ given by \ (\ref{chi_duality}) is then
\begin{equation}
\chi _{\theta }(t,x^{1},z,\overline{z})=\int \frac{\mathrm{d}^{3}\mathbf{k}}{%
(2\pi )^{3}2|\mathbf{k}|}\left\{ a(\mathbf{k})\exp \left[ -\mathrm{i}\left( |%
\mathbf{k}|t-k^{1}x-\zeta (z,\overline{z})\overline{w}-\overline{\zeta (z,%
\overline{z})}w\right) \right] +h.c\right\}
\end{equation}%
where (see (\ref{resenifi}) and (\ref{dual_zzbar}) )%
\begin{eqnarray}
\zeta (z,\overline{z}) &=&z\left( 1+\frac{\sigma \theta }{\pi \rho (z,%
\overline{z})}\right) ^{-1}  \notag \\
\rho (z,\overline{z}) &=&\frac{1}{2}\left( z\overline{z}-2\frac{\sigma
\theta }{\pi }\pm \sqrt{z\overline{z}\left( z\overline{z}-4\frac{\sigma
\theta }{\pi }\right) }\right) .
\end{eqnarray}%
According to the discussion in the introduction to this subsection, $\chi
_{\theta }(t,x,z,\overline{z})$ is for $\theta =3d_{3}$ the solution of the
general fluctuation equation (\ref{linearized_eom}) with $\ g[\phi ]$ given
by (\ref{general_g[phi]}) where $d_{4}=9d_{3}^{2}/2$.

\subsection{Hidden symmetries\label{section_hidden_symmetries}}

The Galileon duality often interrelates apparently very different theories.
For instance, let us assume a Galileon theory with additional $Z_{2}$
symmetry which corresponds to the intrinsic parity, namely%
\begin{equation}
\phi \rightarrow \phi ^{P}=-\phi .  \label{Z2}
\end{equation}%
On the Lagrangian level this symmetry requires $d_{n}=0$ for all $n$ odd.
Under the general duality transformation such a $Z_{2}$ invariant theory
might be mapped onto a dual with some $d_{2k-1}\neq 0$ and therefore the $%
Z_{2}$ symmetry ceases to be manifest in the dual theory. In this section we
will study the way how the symmetries of the original Lagrangian are
realized on the dual one.

Let us remind the definition of the dual action corresponding to the matrix $%
\mathbf{\alpha }$ $\ $
\begin{equation}
S_{\mathbf{\alpha }}[\phi ]=S[\phi _{\alpha }]  \label{action_duality}
\end{equation}%
where $\phi _{\alpha }$ is the duality transformation of the field $\phi $
given by (cf. (\ref{generalized_duality_alpha}))
\begin{eqnarray}
x_{\alpha } &=&\alpha _{PP}x+\alpha _{PB}\partial \phi (x)  \notag \\
\phi _{\alpha }(x_{\alpha }) &=&\det \left( \mathbf{\alpha }\right) \phi (x)
\notag \\
&&+\frac{1}{2}\left( \alpha _{PB}\alpha _{BB}\partial \phi (x)\cdot \partial
\phi (x)+2\alpha _{PB}\alpha _{BP}x\cdot \partial \phi (x)+\alpha
_{PP}\alpha _{BP}x^{2}\right)  \notag \\
\left( \partial \phi \right) _{\alpha }(x_{\alpha }) &=&\alpha _{BB}\partial
\phi (x)+\alpha _{BP}x.  \label{direct_duality}
\end{eqnarray}%
Here we have denoted\footnote{%
In the previous formulae and in what follows we suppress the Lorentz
indices, the index $\alpha $ in $x_{\alpha }$ refers to the matrix $\mathbf{%
\alpha }.$} $\left( \partial \phi \right) _{\alpha }\equiv \partial \phi
_{\alpha }/\partial x_{\alpha }$. Within this notation the group property of
the duality transformations can be formally expressed as
\begin{equation}
\left( Y_{\alpha }\right) _{\beta }=Y_{\beta \cdot \alpha },~~~~~Y=\left(
x,~\phi ,~\partial \phi \right) .
\end{equation}%
The inverse of the transformation (\ref{direct_duality}) has the same form
with the exchange
\begin{equation*}
\mathbf{\alpha }\rightarrow \mathbf{\alpha }^{-1}=\det \left( \mathbf{\alpha
}\right) ^{-1}\left(
\begin{array}{cc}
\alpha _{BB} & -\alpha _{PB} \\
-\alpha _{BP} & \alpha _{PP}%
\end{array}%
\right)
\end{equation*}%
\ and can be written symbolically as
\begin{equation}
Y=\left( Y_{\alpha }\right) _{\alpha ^{-1}}~~~~~Y=\left( x,~\phi ,~\partial
\phi \right) .
\end{equation}%
Using this notation the formula (\ref{action_duality}) can be rewritten in
the form%
\begin{equation}
S[\phi ]=S_{\alpha }[\phi _{\alpha ^{-1}}]  \label{action_duality_1}
\end{equation}%
Now any transformation of the general form
\begin{equation}
Y_{\alpha }\rightarrow \left( Y_{\alpha }\right) ^{\prime }=\left( \mathcal{F%
}^{x}(Y_{\alpha }),\mathcal{F}^{\phi }(Y_{\alpha }),\mathcal{F}^{\partial
\phi }(Y_{\alpha })\right) ,~~\ ~Y_{\alpha }=\left( x_{\alpha },~\phi
_{\alpha },~\left( \partial \phi \right) _{\alpha }\right) ,
\label{general_symmetry}
\end{equation}%
where $\mathcal{F}^{Y}$, $\ $($Y=x,~\phi ,~\partial \phi $), are local
functions\footnote{%
Here ${\mathcal{F}}^{\partial \phi }$ is in fact not independent because it
has to be compatible with the remaining two functions in such a way that ${%
\mathcal{F}}^{\partial \phi }(Y)=\partial ^{\prime }\phi ^{\prime }$} of $%
Y_{\alpha }$, is realized in terms of the variables $Y$ as
\begin{equation}
Y\rightarrow Y^{\prime }=\left( \left( Y_{\alpha }\right) ^{\prime }\right)
_{\alpha ^{-1}}=\left( \mathcal{F}^{x}(Y_{\alpha }),\mathcal{F}^{\phi
}(Y_{\alpha }),\mathcal{F}^{\partial \phi }(Y_{\alpha })\right) _{\alpha
^{-1}}.  \label{general_symmetry_1}
\end{equation}%
Provided the original action is symmetric with respect to the transformation
(\ref{general_symmetry}) we have using (\ref{action_duality_1})%
\begin{equation}
S_{\alpha }[\phi ^{\prime }]=S_{\alpha }[\left( \left( \phi _{\alpha
}\right) ^{\prime }\right) _{\alpha ^{-1}}]=S[\left( \phi _{\alpha }\right)
^{\prime }]=S[\phi _{\alpha }]=S_{\alpha }[\phi ]
\end{equation}%
and the dual action is invariant with respect to (\ref{general_symmetry_1}).

Let us now give some explicit examples of these general formulae. The first
example is the intrinsic parity transformation mentioned in the introduction
to this section. In this case, the formula (\ref{general_symmetry_1})
simplifies considerably. Let us note that the intrinsic parity
transformation (\ref{Z2}) can be treated as a special case of the duality
transformations (\ref{general_scaling}) with a matrix
\begin{equation}
\mathbf{\alpha }_{P}\equiv \mathbf{\alpha }_{S}(1,-1)=\left(
\begin{array}{cc}
1 & 0 \\
0 & -1%
\end{array}%
\right) .  \label{alpha_P}
\end{equation}%
Therefore $\phi ^{P}$ $=\phi _{\alpha _{P}}$ and (\ref{general_symmetry_1})
has the form
\begin{equation}
Y^{P}=\left( \left( Y_{\alpha }\right) ^{P}\right) _{\alpha ^{-1}}=\left(
Y_{\alpha _{P}\cdot \alpha }\right) _{\alpha ^{-1}}=Y_{\alpha ^{-1}\cdot
\alpha _{P}\cdot \alpha }
\end{equation}%
and the $Z_{2}$ symmetry is realized in the dual theory with action $%
S_{\alpha }[\phi ]$ as a duality transformation associated with the matrix%
\begin{equation}
\mathbf{\beta }_{P}(\mathbf{\alpha })=\mathbf{\alpha }^{-1}\cdot \mathbf{%
\alpha }_{P}\cdot \mathbf{\alpha }=\det \left( \mathbf{\alpha }\right)
^{-1}\left(
\begin{array}{cc}
\alpha _{PP}\alpha _{BB}+\alpha _{PB}\alpha _{BP} & 2\alpha _{PB}\alpha _{BB}
\\
-2\alpha _{BP}\alpha _{PP} & -\alpha _{PP}\alpha _{BB}-\alpha _{PB}\alpha
_{BP}%
\end{array}%
\right) ,  \label{beta_P}
\end{equation}%
or explicitly%
\begin{eqnarray}
x^{P} &=&\det \left( \mathbf{\alpha }\right) ^{-1}\left[ \left( \alpha
_{PP}\alpha _{BB}+\alpha _{PB}\alpha _{BP}\right) x+2\alpha _{PB}\alpha
_{BB}\partial \phi (x)\right]  \label{Z_2_nonlinear} \\
\phi ^{P}(x^{P}) &=&-\phi (x)-\det \left( \mathbf{\alpha }\right) ^{-2}\left[
\alpha _{PB}\alpha _{BB}\left( \alpha _{PP}\alpha _{BB}+\alpha _{PB}\alpha
_{BP}\right) \partial \phi (x)\cdot \partial \phi (x)\right.  \notag \\
&&\left. +4\alpha _{PB}\alpha _{BP}\alpha _{PP}\alpha _{BB}x\cdot \partial
\phi (x)+\alpha _{PP}\alpha _{BP}\left( \alpha _{PP}\alpha _{BB}+\alpha
_{PB}\alpha _{BP}\right) x^{2}\right]  \notag \\
\left( \partial \phi \right) ^{P}(x^{P}) &=&-\det \left( \mathbf{\alpha }%
\right) ^{-1}\left[ 2\alpha _{BP}\alpha _{PP}x+\left( \alpha _{PP}\alpha
_{BB}+\alpha _{PB}\alpha _{BP}\right) \partial \phi (x)\right] .
\end{eqnarray}%
The transformation corresponding to the intrinsic parity is therefore
realized in the dual theory non-linearly and non-locally as a simultaneous
transformation of both space-time coordinates and fields.

In the same way we can find the dual realization of the Galileon symmetry (%
\ref{galileon_symmetry}). The general formula (\ref{general_symmetry_1})
reads in this case%
\begin{equation}
\left( x,\phi ,\partial \phi \right) ^{\prime }=\left( x_{\alpha },\phi
_{\alpha }+a+b\cdot x_{\alpha },\left( \partial \phi \right) _{\alpha
}+b\right) _{\alpha ^{-1}},
\end{equation}%
or explicitly%
\begin{eqnarray}
x^{\prime } &=&x-\det \left( \mathbf{\alpha }\right) ^{-1}\alpha _{PB}b
\notag \\
\phi ^{\prime } &=&\phi +a\det \left( \mathbf{\alpha }\right) ^{-1}-\frac{1}{%
2}\det \left( \mathbf{\alpha }\right) ^{-2}\alpha _{PB}\alpha
_{PP}b^{2}+\det \left( \mathbf{\alpha }\right) ^{-1}\alpha _{PP}b\cdot x
\notag \\
\left( \partial \phi \right) ^{\prime } &=&\partial \phi +\det \left(
\mathbf{\alpha }\right) ^{-1}\alpha _{PP}b.
\end{eqnarray}%
A dual Galileon transformation is therefore superposition of space-time
translation and Galileon transformation with special values of parameters,
as was recognized for the duality of the type (\ref{one_parametric_duality})
in \cite{deRham:2013hsa}.

Let us now discuss the space-time symmetries. The duality transformation
respects the Lorentz symmetry, therefore its realization in the dual theory
is the same as in the original one. Indeed, restoring the Lorentz structure
in the matrix notation (\ref{matrix_notation1}), (\ref{matrix_notation2})
and (\ref{matrix_notation3}) we can write%
\begin{equation}
\mathbf{\alpha }=\left(
\begin{array}{cc}
\alpha _{PP} & \alpha _{PB} \\
\alpha _{BP} & \alpha _{BB}%
\end{array}%
\right) \otimes \mathbf{1,~~}~\widehat{\mathbf{\alpha }}=\left(
\begin{array}{cc}
\alpha _{BP} & 0 \\
0 & \alpha _{PB}%
\end{array}%
\right) \otimes \mathbf{1}  \label{alpha_lorentz}
\end{equation}%
where the second factor $\mathbf{1}\equiv \delta _{\nu }^{\mu }$ in the
tensor product acts to the vector indices of \ $X$. Using the same notation,
the Lorentz transformation $\Lambda \equiv $ $\Lambda _{~\nu }^{\mu }$can be
described by the formula (\ref{matrix_notation1}) with the matrix\footnote{%
Strictly speaking for scalar $\phi $ this is true only for proper Lorentz
transformation with $\det \Lambda =1$. Other alternatives (pseudoscalar $%
\phi $ or improper Lorentz transformation) can be discussed analogously with
minor changes.}
\begin{equation}
\mathbf{\alpha }_{L}=\mathbf{1}\otimes \Lambda
\end{equation}%
which commutes with (\ref{alpha_lorentz}) and, according to the general
prescription (\ref{general_symmetry_1}), the dual realization corresponds to
the matrix (cf. also (\ref{beta_P}))
\begin{equation*}
\mathbf{\beta }_{L}=\mathbf{\alpha }^{-1}\cdot \mathbf{\alpha }_{L}\cdot
\mathbf{\alpha =\alpha }_{L}.
\end{equation*}%
The last example is the space-time shift%
\begin{eqnarray}
x &\rightarrow &x+b  \notag \\
\phi (x) &\rightarrow &\phi (x-b),
\end{eqnarray}%
for which, according to (\ref{general_symmetry_1}), we get%
\begin{equation}
(x^{\prime },\phi ^{\prime }(x^{\prime }),\partial ^{\prime }\phi ^{\prime
}(x^{\prime }))=(x_{\alpha }+b,\phi _{\alpha }(x_{\alpha }),\left( \partial
\phi \right) _{\alpha }(x_{\alpha }))_{\alpha ^{-1}}
\end{equation}%
or explicitly%
\begin{eqnarray}
x^{\prime } &=&x+\alpha _{BB}\det (\mathbf{\alpha })^{-1}b  \notag \\
\phi ^{\prime } &=&\phi -\frac{1}{2}\alpha _{BB}\alpha _{BP}\det (\mathbf{%
\alpha })^{-2}b^{2}+\alpha _{BP}\det (\mathbf{\alpha })^{-1}b\cdot x  \notag
\\
\left( \partial \phi \right) ^{\prime } &=&\partial \phi +\alpha _{BP}\det (%
\mathbf{\alpha })^{-1}b\text{. }
\end{eqnarray}%
The dual realization of the space-time shift is therefore a composition of
shift and general Galileon transformation with $\mathbf{\alpha }$ and $b-$%
dependent parameters.

\subsection{Duality of the $S$ matrix\label{S_duality}}

On the quantum level the most important object is the $S$ matrix. In this
section we will discuss its properties with respect to the Galileon duality
and show that it is invariant with respect to the subgroups of duality
transformations $\mathbf{\alpha }_{D}\left( \theta \right) $ and $\mathbf{%
\alpha }_{S}\left( 1,\kappa \right) $.

Let us first briefly remind the well known equivalence theorem which makes a
statement about the invariance of the $S$ matrix with respect to the field
redefinitions (see e.g. \cite{Coleman:1969sm}). The $S$ matrix can be
obtained by means of LSZ formulae from the generating functional $Z[J]$ of
the Green functions which can be expressed in terms of the functional
integral.
\begin{equation}
Z[J]=\int \mathcal{D}\phi \exp \left( \frac{\mathrm{i}}{\hbar }S[\phi ]+%
\frac{\mathrm{i}}{\hbar }\left\langle J\phi \right\rangle \right) \,.
\label{Z_J}
\end{equation}%
In this formula we tacitly assume appropriate regularization which preserves
the properties of the action with respect to the Galileon symmetry and
duality transformations. The action can be expanded in powers of $\hbar $%
\begin{equation}
S[\phi ]=\sum_{n=0}^{\infty }\hbar ^{n}S_{n}[\phi ]\equiv S_{0}[\phi
]+S_{CT}[\phi ],  \label{action_h_expansion}
\end{equation}%
where $S_{0}[\phi ]$ is the Galileon Lagrangian (\ref{galileon_Lagrangian}).
The higher order terms $S_{n}[\phi ]$ in the expansion (\ref%
{action_h_expansion}) summed up in $S_{CT}[\phi ]$ represent the
counterterms which are necessary in order to renormalize the UV divergences
stemming form the $n$-loop graphs. The discussion of these counterterms we
postpone to the Section \ref{CT}, here we only stress that, because of the
derivative structure of the Galileon interaction vertices, the counterterms $%
S_{n}[\phi ]$ have more derivatives per field than the basic action $%
S_{0}[\phi ]$, and that under our assumptions on the regularization the
counterterms should respect the Galileon symmetry.

In the functional integral the field $\phi $ is a dummy variable and can be
freely changed by means of the field redefinition $\phi \rightarrow $ $%
\mathcal{F}[\phi ]$ according to%
\begin{equation}
Z[J]=\int \mathcal{D}\phi \det \left( \frac{\delta \mathcal{F}[\phi ]}{%
\delta \phi }\right) \exp \left( \frac{\mathrm{i}}{\hbar }S[\mathcal{F}[\phi
]]+\frac{\mathrm{i}}{\hbar }\left\langle J\mathcal{F}[\phi ]\right\rangle
\right) ,  \label{Z_transformed}
\end{equation}%
where we have abbreviated\footnote{%
In what follows we will often use this notation without further comments
unless it shall lead to misinterpretation.} $\left\langle \cdot
\right\rangle \equiv \int \mathrm{d}^{d}x\left( \cdot \right) $. This should
be compared with the generating functional $Z_{\mathcal{F}}[J]$ in the
theory with the action $S_{\mathcal{F}}[\phi ]\equiv S[\mathcal{F}[\phi ]]$%
\begin{equation}
Z_{\mathcal{F}}[J]=\int \mathcal{D}\phi \exp \left( \frac{\mathrm{i}}{\hbar }%
S_{\mathcal{F}}[\phi ]+\frac{\mathrm{i}}{\hbar }\left\langle J\mathcal{\phi }%
\right\rangle \right)
\end{equation}%
Ignoring the Jacobian on the right hand side of (\ref{Z_transformed}) for a
moment, the sufficient condition for the perturbative equivalence of the $S$
matrices in the theories with actions $S[\phi ]$ and $S_{\mathcal{F}}[\phi ]$
is that the Fourier transforms of the Green functions of the operators $\phi
(x)$ and $\mathcal{F}[\phi ](x)$ have the same one-particle poles at $%
p_{i}^{2}=0$ up to a simple re-scaling of the residues\footnote{%
Implicitly this means that the operator $\mathcal{F}[\phi ](x)$ is
translation invariant, i.e. $\mathcal{F}[\phi ](x)=e^{-iP\cdot x}\mathcal{F}%
[\phi ](0)e^{iP\cdot x}$. Violation of this condition might prevent the
applicability of the LSZ reduction formulae when passing from correlators to $%
S$ matrix elements.}. This is achieved provided $\mathcal{F}[0]=0$ and
\begin{equation}
\left\langle 0|\mathcal{F}[\phi ](0)|p\right\rangle =\mathcal{Z}_{\mathcal{F}%
}\left\langle 0|\mathcal{\phi }(0)|p\right\rangle
\end{equation}%
with $\mathcal{Z}_{\mathcal{F}}\neq 0$. This requirement is respected by the
Galileon duality transformation which are represented by the upper
triangular matrices with $\alpha _{PP}=1$. To prove this, it is sufficient
to investigate the dualities corresponding to $\mathbf{\alpha }_{D}\left(
\theta \right) $ and $\mathbf{\alpha }_{S}\left( 1,\kappa \right) $
separately because of the decomposition (\ref{duality_decomposition}). In
the former case we have%
\begin{equation}
\mathcal{F}[\phi ](x)\equiv \phi _{\theta }(x)=\phi (x)+\theta \partial \phi
(x)\cdot \partial \phi (x)+O(\theta ^{2},\phi ^{4})
\label{theta_phi_expansion}
\end{equation}%
and therefore\footnote{%
The $O(\hbar )$ part stems from the contributions of the terms bilinear and
higher in the derivatives of the field $\phi $ on the right hand side of (%
\ref{theta_phi_expansion}). These terms start to contribute to $\mathcal{Z}_{%
\mathcal{F}}$ only at the loop level.}%
\begin{equation}
\mathcal{Z}_{\mathcal{F}}=1+O(\hbar )
\end{equation}%
while in the latter case we trivially\footnote{%
The case of $\mathbf{\alpha }_{S}\left( \lambda ,\kappa \right) $ is more
complicated, because
\begin{equation}
\mathcal{F}[\phi ](x)=\lambda \kappa \phi (\lambda ^{-1}x),~~~\int \mathrm{d}%
^{d}x\mathrm{e}^{\mathrm{i}p\cdot x}\lambda \kappa \phi (\lambda
^{-1}x)=\lambda ^{d+1}\kappa \int \mathrm{d}^{d}x\mathrm{e}^{\mathrm{i}%
\lambda p\cdot x}\phi (x)
\end{equation}%
and the behaviour of the Green functions under the re-scaling of the momenta
is governed by the renormalization group. This is the rationale for the
constraint $\alpha _{PP}=1$. However, at the tree level this simplifies to
scaling respecting the canonical dimensions.} get $\mathcal{Z}_{\mathcal{F}%
}=\kappa $. Thus the only obstruction which prevents us to make a statement
on the equivalence of the on-shell $S$ matrices in both theories also at the
loop level is the possible nontrivial functional determinant on the right
hand side of (\ref{Z_transformed}). Its actual value depends on the
regularization. In what follows we will show that using dimensional
regularization the functional determinant equals to one for the duality
transformations $\mathbf{\alpha }_{D}\left( \theta \right) $ (the case $%
\mathbf{\alpha }_{S}\left( 1,\kappa \right) $ is of course trivial).

For the infinitesimal $\theta $ we can expand the functional determinant
according to (\ref{theta_phi_expansion}) as
\begin{equation}
\det \left( \frac{\delta \phi _{\theta }(x)}{\delta \phi (y)}\right)
=1+2\theta \mathrm{Tr}\left( \partial \phi (x)\partial \right) \,.
\end{equation}%
The trace can be further expressed in a standard way (introducing the
operators $\widehat{X}\equiv x$, $\widehat{K}\equiv -\mathrm{i}\partial $
and their eigenvectors $|x)$ and $|k)$ respectively) as
\begin{eqnarray}
\mathrm{Tr}\left( \partial \phi (x)\partial \right) &=&\mathrm{i}\int
\mathrm{d}^{d-2\varepsilon }x(x|\partial \phi (\widehat{X})\cdot \widehat{K}%
|x)=\mathrm{i}\int \mathrm{d}^{{d-2\varepsilon }}x\mathrm{d}^{{%
d-2\varepsilon }}k(x|\partial \phi (\widehat{X})\cdot \widehat{K}|k)(k|x)
\notag \\
&=&\mathrm{i}\int \mathrm{d}^{{d-2\varepsilon }}x\partial \phi (x)\cdot \int
\frac{\mathrm{d}^{{d-2\varepsilon }}k}{(2\pi )^{^{d-2\varepsilon }}}k
\end{eqnarray}%
The first factor equals to zero for well behaved $\phi $ while the second
one vanishes within the the dimensional regularization. We have thus
\begin{equation}
\det \left( \frac{\delta \phi _{\theta }(x)}{\delta \phi (y)}\right)
=1+O(\theta ^{2})
\end{equation}%
For the finite transformation we can use the fact that the transformation
forms a one-parametric group and thus
\begin{equation}
\phi _{\theta +\xi }[\phi ]=\phi _{\theta }[\phi _{\xi }]=\phi _{\xi }[\phi
_{\theta }]\,,
\end{equation}%
which implies
\begin{equation}
\det \left( \frac{\delta \phi _{\theta +\Delta \theta }}{\delta \phi
_{\theta }}\right) =1+O(\left( \Delta \theta \right) ^{2})\,.
\end{equation}%
Using the formula
\begin{equation}
\frac{\delta \phi _{\theta +\Delta \theta }(x)}{\delta \phi (y)}=\int
\mathrm{d}^{d}z\frac{\delta \phi _{\theta +\Delta \theta }(x)}{\delta \phi
_{\theta }(z)}\frac{\delta \phi _{\theta }(z)}{\delta \phi (y)}
\end{equation}%
we get formally%
\begin{equation}
\det \left( \frac{\delta \phi _{\theta +\Delta \theta }}{\delta \phi }%
\right) =\det \left( \frac{\delta \phi _{\theta +\Delta \theta }}{\delta
\phi _{\theta }}\right) \det \left( \frac{\delta \phi _{\theta }}{\delta
\phi }\right)
\end{equation}%
and therefore
\begin{equation}
\frac{\partial }{\partial \theta }\ln \det \left( \frac{\delta \phi _{\theta
}}{\delta \phi }\right) =0\,,
\end{equation}%
By means of integration from $0$ to $\theta $ this leads to the desired
statement\footnote{%
Let us note that without the knowledge that the Jacobian of this
transformation is equal one, in a standard way, we can introduced the ghost
fields which would reproduce the studied determinant. At the end, however,
one would find that propagators of such ghosts are proportional to 1, and
thus every integration over ghost loop with momentum $l$ is of the type:
\begin{equation*}
\int \mathrm{d}^{d-2\varepsilon }l\times \mathrm{Polynomial}(l)=0
\end{equation*}%
which is true for the dimensional regularization.}
\begin{equation}
\det \left( \frac{\delta \phi _{\theta }}{\delta \phi }\right) =1\,.
\end{equation}%
The on-shell $S$ matrices in theories with actions $S_{\mathcal{F}}[\phi ]$
and $S[\phi ]$ are therefore formally equivalent for the above duality
transformations. This statement, however, must be taken with care. The first
reason is that the counterterm part $S_{CT}[\phi ]$ of the action has not
the form of the Galileon Lagrangian and transforms therefore highly
nontrivially with respect to the duality (note that the duality
transformation is in general non-local and involves infinite number of
terms, cf. Appendix \ref{bu}). The second reason is that though we have
formally established equivalence of the on-shell $S$ matrices, the off-shell
Green functions stay to be different in both theories. Indeed, we have in
fact only proved that Green functions of operators $\phi (x)$ in original
theory and those of operators $\mathcal{F}[\phi ](x)$ in the dual theory
coincide\footnote{%
Analogous statement can be made also for any other (composite) operator $%
\mathcal{O}[\phi ]$. The correlators of a string of such operators within
the original theory coincide with correlators of corresponding dual
operators $\mathcal{O}[\mathcal{F}[\phi ]]$ calculated within the dual
theory. This is in accord with our discussion of the dual observables on the
classical level.}. Therefore, the recursive construction of the counterterms
in the dual theory starting with the dual basic action $S_{0}[\mathcal{F}%
[\phi ]]$ will lead to counterterm action $S_{CT}^{\mathcal{F}}[\phi ]$
different from $S_{CT}[\mathcal{F}[\phi ]]$. On the other hand, the
counterterms from $S_{CT}[\mathcal{F}[\phi ]]$ will be sufficient to cancel
the divergences of the on-shell amplitudes in the dual theory. Before we
proceed to the more detailed discussion of quantum correction and
counterterms (we postpone it to the Sections \ref{CT} and \ref%
{one_loop_duality}) we will illustrate the consequences of duality in the
case of tree level scattering amplitudes.

\subsection{Tree level amplitudes\label{tree_level_A}}

As we have mentioned in Section~\ref{motivations}, the tree level amplitudes
up to the $5$pt one have surprisingly simple structure though they are sums
of a large number of nontrivial contributions stemming from individual
Feynman graphs with different topologies. Therefore large cancellations
between different contribution have to occur the reason of which is not
transparent on the Lagrangian level. In this subsection we will show on an
elementary level how these results can be understood better with the help of
the duality.

As we will demonstrate, the mechanism of the above mentioned cancellation is
a consequence of the invariance of the $S$ matrix with respect to the
duality subgroup $\mathbf{\alpha }_{D}\left( \theta \right) $. \ The key
observation is that the tree amplitudes in the dual theory are polynomials
in the parameter $\theta $, however, due to the invariance, they are in fact
$\theta $ independent. Therefore the coefficients of the above mentioned
polynomial at the positive powers of $\theta $ have to vanish which gives
nontrivial relations between different contributions to the amplitude.

We will also show that though the $S$ matrix is not invariant under the
duality subgroup $\mathbf{\alpha }_{S}(\lambda )$, the tree level amplitudes
have simple transformation properties which can be used to relate also the $%
S $ matrices in theories dual with respect to $\mathbf{\alpha }_{S}(\lambda
) $.

Let us first remind some of the general properties of the tree level
amplitudes. For general tree amplitudes we have
\begin{equation}
I=V-1\,,\quad \sum_{n}n\alpha _{n}=2I+E\,,
\end{equation}%
where $I$ and $E$ represents number of internal and external lines
respectively, $V$ is number of vertices and $\alpha _{n}$ is number of
vertices with $n$ legs;\ putting this together we get
\begin{equation}
\sum_{n}(n\alpha _{n}-2)=E-2\,.  \label{alpha_n}
\end{equation}%
It is clear that any amplitude must be represented by a linear combination
of the monomials $\prod_{n}d_{n}^{\alpha _{n}}$ with $d_{n}$- independent
kinematical coefficients, which carry the information on the momentum
dependence of the amplitudes, explicitly\footnote{%
Her we abbreviate $M(p_{1},p_{2},\ldots ,p_{E};d_{n})$ by $M(1,\ldots
,E;d_{n})$ and similarly for the $d_{n}$ independent kinematical
coefficients.}
\begin{equation}
M(1,\ldots ,E;d_{n})=\sum\limits_{\{\alpha _{n}\}}M_{\{\alpha
_{n}\}}(1,\ldots ,E)\prod\limits_{n}d_{n}^{\alpha _{n}}.  \label{d3d4d5_form}
\end{equation}%
Here the sum is over the sequences $\left\{ \alpha _{n}\right\} _{n=3}^{d+1}$
which satisfy the condition (\ref{alpha_n}) and the coefficients $%
M_{\{\alpha _{n}\}}(1,\ldots ,E)$ represent the sum of Feynman diagrams with
$\alpha _{n}$ vertices with $n$ legs. In what follows we restrict ourselves
to case $d=4$, i.e. the sum in (\ref{d3d4d5_form}) is over the ordered
triplets $\{\alpha _{3},\alpha _{4},\alpha _{5}\}$.

As we have seen in Subsection~\ref{S_duality}, the tree-level $S$ matrix is
invariant with respect to the duality $\mathbf{\alpha }_{D}(\theta )$,
therefore the amplitudes have to satisfy the following condition
\begin{equation}
\frac{\partial }{\partial \theta }M(1,\ldots ,E;d_{n}(\theta ))=0,
\label{invariance_S}
\end{equation}%
where $d_{n}(\theta )$ are given by (\ref{4d_duality}). This gives us
non-trivial constraints on the form of the coefficients $M_{\{\alpha
_{3},\alpha _{4},\alpha _{5}\}}(1,\ldots ,E)$. Let us now study the impact
of these constraint on individual amplitudes. For $E=3$ the only allowed
sequence in (\ref{d3d4d5_form}) is $\ \{1,0,0\}$. Inserting (\ref{4d_duality}%
) into (\ref{d3d4d5_form}) we get
\begin{equation}
M(1,2,3;d_{n}(\theta ))=\left( d_{3}-\frac{1}{3}\theta \right)
M_{\{1,0,0\}}(1,2,3),
\end{equation}%
and thus from (\ref{invariance_S}) we get without any calculation (cf. (\ref%
{M3}))
\begin{equation}
M_{\{1,0,0\}}(1,2,3)=0.
\end{equation}%
For $E=4$ we have in the same way
\begin{equation}
M(1,2,3,4;d_{n}(\theta ))=\left( d_{3}-\frac{1}{3}\theta \right)
^{2}M_{\{2,0,0\}}(1,2,3,4)+\left( d_{4}-3\theta d_{3}+\frac{1}{2}\theta
^{2}\right) M_{\{0,1,0\}}(1,2,3,4)
\end{equation}%
and nullifying the coefficient at different powers of $\theta $ we get the
constraint
\begin{equation}
M_{\{2,0,0\}}(1,2,3,4)=-\frac{9}{2}M_{\{0,1,0\}}(1,2,3,4)
\end{equation}%
and thus
\begin{equation}
M(1,2,3,4;d_{n})=\left( d_{4}-\frac{9}{2}d_{3}^{2}\right)
M_{\{0,1,0\}}(1,2,3,4)\,.
\end{equation}%
As $M_{\{0,1,0\}}(1,2,3,4)$ is just the Feynman rule for the four-point
vertex (cf. (\ref{Feynman_rule})) we may immediately write
\begin{equation}
M_{\{0,1,0\}}(1,2,3,4)=4!G(1,2,3).
\end{equation}%
Together this yields
\begin{equation*}
M(1,2,3,4;d_{n})=12\left( 2d_{4}-9d_{3}^{2}\right) G(1,2,3),
\end{equation*}%
in agreement with (\ref{M4}). We can continue with further amplitudes and
find out that the duality simplifies significantly the calculation. For
instance for $E=5$
\begin{eqnarray}
M(1,2,3,4,5;d_{n}(\theta ))
&=&d_{3}^{3}M_{\{3,0,0\}}(1,2,3,4,5)+d_{3}d_{4}M_{\{1,1,0\}}(1,2,3,4,5)
\notag \\
&+&d_{5}M_{\{0,0,1\}}(1,2,3,4,5)
\end{eqnarray}%
and the duality constraints are now
\begin{eqnarray}
M_{\{1,1,0\}}(1,2,3,4,5) &=&-\frac{24}{5}M_{\{0,0,1\}}(1,2,3,4,5)  \notag \\
M_{\{3,0,0\}}(1,2,3,4,5) &=&\frac{72}{5}M_{\{0,0,1\}}(1,2,3,4,5).
\end{eqnarray}%
As a consequence%
\begin{equation}
M(1,2,3,4,5;d_{n})=\left( \frac{72}{5}d_{3}^{3}-\frac{24}{5}%
d_{3}d_{4}+d_{5}\right) M_{\{0,0,1\}}(1,2,3,4,5).
\end{equation}%
Again $M_{\{0,0,1\}}(1,2,3,4,5)$ is just the Feynman rule%
\begin{equation}
M_{\{0,0,1\}}(1,2,3,4,5)=-5!G(1,2,3,4)
\end{equation}%
and we conclude without calculations (cf. (\ref{M5}))%
\begin{equation}
M(1,2,3,4,5;d_{n})=-24\left( 72d_{3}^{3}-24d_{3}d_{4}+5d_{5}\right)
G(1,2,3,4).
\end{equation}%
As a last example we take $E=6$, the computer calculation of which though
possible gives rather lengthy and untransparent final output so it is
difficult to reveal any regular structure hidden in it. As we will see also
here the duality helps considerably.

There are four kinematical factors in this case corresponding to the
sequences $\{4,0,0\}$, $\{0,2,0\}$, $\{2,1,0\}$ and $\{1,0,1\}$. The duality
constraints are
\begin{eqnarray}
M_{\{4,0,0\}}(1,\ldots ,6) &=&\frac{81}{4}M_{\{0,2,0\}}(1,\ldots ,6)  \notag
\\
M_{\{2,1,0\}}(1,\ldots ,6) &=&-9M_{\{0,2,0\}}(1,\ldots ,6)  \notag \\
M_{\{1,0,1\}}(1,\ldots ,6) &=&0,
\end{eqnarray}%
and thus when inserted to the formula (\ref{d3d4d5_form}) we get finally
\begin{eqnarray}
M(1,\ldots ,6;d_{n}) &=&\left( \frac{81}{4}%
d_{3}^{4}+d_{4}^{2}-9d_{3}^{2}d_{4}\right) M_{\{0,2,0\}}(1,\ldots ,6)  \notag
\\
&=&\left( d_{4}-\frac{9}{2}d_{3}^{2}\right) ^{2}M_{\{0,2,0\}}(1,\ldots ,6).
\end{eqnarray}%
Here $M_{\{0,2,0\}}$ is the sum of graphs with two four-point vertices
connected by one propagator and can be therefore written in the form%
\begin{equation}
M_{\{0,2,0\}}(1,\ldots ,6)=-16\sum_{\sigma \in S_{6}}\frac{G(\sigma
(1),\sigma (2),\sigma (3))G(\sigma (4),\sigma (5),\sigma (6))}{(p_{\sigma
(1)}+p_{\sigma (2)}+p_{\sigma (3)})^{2}},
\end{equation}%
where we sum over the permutations of the external momenta.

Of course these results are not surprising. The tree-level $S$ matrix being
an invariant of the duality subgroup $\mathbf{\alpha }_{D}(\theta )$ can
depend on $d_{n}$ only as a function of the the basic $\mathbf{\alpha }%
_{D}(\theta )$ duality invariants $I_{4}$, $I_{5}$ given by (\ref{I_4_5}).
Because these invariants can be interpreted as $d_{4}$ and $d_{5}$ in a dual
theory with $d_{3}=0$, the tree-level amplitudes must have the form%
\begin{equation}
M(1,\ldots ,E;d_{k})=\sum_{\{m,n\}\geq 0}M_{\{0,m,n\}}(1,\ldots
,E)I_{4}^{m}I_{5}^{n}  \label{I4I5_form}
\end{equation}%
where the summation over $m$ and $n$ must fulfil (\ref{alpha_n}), i.e.
\begin{equation}
2m+3n=E-2.
\end{equation}%
This general structure can be easily recognized in all the above examples.
Let us note that in general case
\begin{equation}
I_{4}^{m}I_{5}^{n}=\sum\limits_{\{\alpha _{k}\}}c_{\{\alpha
_{k}\}}^{mn}\prod\limits_{k}d_{k}^{\alpha _{k}}
\end{equation}%
where $c_{\{\alpha _{k}\}}^{mn}$ are rational numbers. Then comparing the
coefficients (\ref{d3d4d5_form}) and (\ref{I4I5_form}) we get the above
discussed constraints on the individual contributions to the amplitude in a
general form
\begin{equation}
M_{\{\alpha _{k}\}}(1,\ldots ,E)=\sum_{m,n}c_{\{\alpha
_{k}\}}^{mn}M_{\{0,m,n\}}(1,\ldots ,E).
\end{equation}

As we have illustrated above, the tree-level amplitudes are invariants of
the subgroup $\mathbf{\alpha }_{D}(\theta )$ but also their transformation
properties with respect to the scalings $\mathbf{\alpha }_{S}(\lambda )$ and
more generally $\mathbf{\alpha }_{S}(\lambda ,\kappa )$ are transparent. Let
us remind that, under the $\mathbf{\alpha }_{S}(\lambda )$ the couplings $%
d_{n}$ scale according its dimension (cf. (\ref{d_scaling}) with $\Delta
=(d-2)/2$, which is the canonical dimension of the field $\phi $)
\begin{equation}
d_{n}(\mathbf{\alpha }_{S}(\lambda ))\equiv d_{n}(\lambda )=\lambda ^{-\frac{%
1}{2}(d+2)(n-2)}d_{n},
\end{equation}%
which just corresponds to the re-scaling of the units. Note that, for $d$
even we can generalize the above scaling also to $\lambda <0$. From the
homogeneity of the tree\footnote{%
Note that, at the loop level, we have additional dependence of the
amplitudes on additional dimensionfull parameters, namely on the counterterm
couplings as well as on the renormalization scale.} amplitudes
\begin{eqnarray}
M(\lambda p_{1},\ldots ,\lambda p_{n};\lambda ^{\dim d_{k}}d_{k}) &=&\lambda
^{\dim M(p_{1},\ldots ,p_{n};d_{k})}M(p_{1},\ldots ,p_{n};d_{k})  \notag \\
&=&\lambda ^{d-n\frac{d-2}{2}}M(p_{1},\ldots ,p_{n};d_{k}),
\end{eqnarray}%
it follows that two amplitudes with $d_{n}$ and $d_{n}(\lambda )$ are
connected by
\begin{equation}
M(p_{1},\ldots ,p_{n};d_{k}(\lambda ))=\lambda ^{d-n\frac{d-2}{2}}M(\lambda
^{-1}p_{1},\ldots ,\lambda ^{-1}p_{n};d_{k}).
\end{equation}%
At tree\footnote{%
As we will see in the subsequent sections, the loop amplitudes have higher
degree of homogeneity with respect to re-scaling the momenta.} level on the
other hand
\begin{equation}
M(\lambda ^{-1}p_{1},\ldots ,\lambda ^{-1}p_{n};d_{k})=\lambda
^{-2(n-1)}M(p_{1},\ldots ,p_{n};d_{k}),
\end{equation}%
as we will show in the Section \ref{CT} and therefore we get finally
\begin{equation}
M(p_{1},\ldots ,p_{n};d_{k}(\lambda ))=\lambda ^{\frac{1}{2}%
(d+2)(2-n)}M(p_{1},\ldots ,p_{n};d_{k})\,.  \label{M_scaling}
\end{equation}%
Therefore not only that it is sufficient to know the amplitudes for some
representant of the group orbit of $\mathbf{\alpha }_{D}(\theta )$ in the
theory subspace $D_{d+1}^{(1)}$but we can also travel between different (but
qualitatively similar) orbits using the formula (\ref{M_scaling}).

\subsection{Classification of the Galileon theories\label%
{section_classification}}

As we have shown, some physical consequences of the Galileon theories are
not directly visible from the Galileon Lagrangian. This concerns e.g. the
cancellations of the various contributions to the tree-level amplitudes as
well as the hidden $Z_{2}$ symmetry of the Galileon action discussed in the
previous sections. However, as was seen in the latter case, such properties
are usually shared by the theories which are connected by the group of
duality transformations (or by some of its subgroup). It is therefore
important to describe the equivalence classes of the Galileon theories with
respect to the duality.

In what follows we will classify in this sense the Galileon theories in $d=3$
and $4$. We will restrict ourselves to the theory subspace $D_{d+1}^{(1)}$
with $d_{2}=1/4$ and $1/12$ respectively and we will consider only the
dualities corresponding the upper triangular matrices $\mathbf{\alpha }$
which make sense also in the quantum case. According to the results of the
previous sections, such a classification is at the same time also a
classification of the nontrivial tree level $S$ matrices because these are
either invariants of the duality with respect to the subgroup $\mathbf{%
\alpha }_{D}\left( \theta \right) $ or trivially scale with respect to $%
\mathbf{\alpha }_{S}(\lambda )$.

\subsubsection{Galileons in $d=4$}

The properties of the theory which belongs to the theory subspace $%
D_{5}^{(1)}$with constants $d_{3}$, $d_{4}$ and $d_{5}$ are governed by the
invariants of the duality transformation $I_{4}$ and $I_{5}$ given by (\ref%
{I_4_5}). Let us remind that $I_{n}$ represents the value of the constant $%
d_{n}$ in the theory which is dual to the original one but satisfies the
condition $d_{3}=0 $. We have the following cases (see Fig. \ref%
{parametricspace}):

\begin{itemize}
\item $I_{4}=I_{5}=0$, in this case the theory is dual to free theory

\item $I_{5}=0$, in this case the theory is $Z_{2}$ invariant (it is dual to
the theory with $d_{3}=d_{5}=0$). The $Z_{2}$ invariance is realized by (\ref%
{Z_2_nonlinear}) with $\mathbf{\alpha }=\mathbf{\alpha }_{D}\left(
-3d_{3}\right) $. The only non-zero amplitudes are those with even number of
legs

\item $I_{4}=0$, the theory is a dual to quintic Galileon (where $%
d_{3}=d_{4}=0$)

\item both $I_{4},I_{5}\neq 0$, the theory is dual to $d_{3}=0$, $%
d_{4,5}\neq 0$
\end{itemize}

Let us now summarize the cases for which a concrete coupling $d_{n}$ can be
removed by duality transformation. The following conditions are easily
derived as the conditions for the existence of the solutions of the
equations $d_{n}(\theta )=0$ with respect to $\theta $ (cf. (\ref{4d_duality}%
))

\begin{itemize}
\item Every theory is dual to theory with $d_{3}=0$

\item $I_{4}<0$, then theory is dual to just two theories with odd
interactions where $d_{4}=0$ (this can be achieved by duality transformation
corresponding to $\mathbf{\alpha }_{D}\left( \theta _{\pm }\right) $ for $%
\theta _{\pm }=3d_{3}\pm \sqrt{-2I_{4}}$)

\item $I_{4}>0$, then there is no dual with $d_{4}=0$

\item For $(8I_{4})^{3}+(15I_{5})^{2}>0$ theory is dual to exactly one
theory with $d_{5}=0$

\item For $(8I_{4})^{3}+(15I_{5})^{2}<0$ theory is dual to exactly three
theories with $d_{5}=0$
\end{itemize}

The invariants $I_{4}$ and $I_{5}$ scale as $d_{4}$ and $d_{5}$, namely%
\begin{equation}
I_{4}(\lambda )=\lambda ^{-6}I_{4},~~I_{5}(\lambda )=\lambda ^{-9}~I_{5}.
\end{equation}%
Therefore by means of the scaling $\mathbf{\alpha }_{S}(\lambda )$ we can
always arrange either $I_{4}=\pm 1$ or $I_{5}=1$ (with $\lambda <0$ when
necessary). To summarize, non-trivial theories (i.e. those which are not
connected by dualities $\mathbf{\alpha }_{D}\left( \theta \right) $ or $%
\mathbf{\alpha }_{S}(\lambda )$) are

\begin{itemize}
\item $I_{4}=I_{5}=0$ --- free theory

\item $I_{4}=\pm 1$, $I_{5}=0$ --- $Z_{2}$ invariant theory (only even
amplitudes are non-zero)

\item $I_{5}=1$, $I_{4}=0$ --- quintic Galileon

\item $I_{5}=1$, $I_{4}\neq 0$ --- general case
\end{itemize}

\begin{figure}[t]
\begin{center}
\includegraphics[scale=0.8]{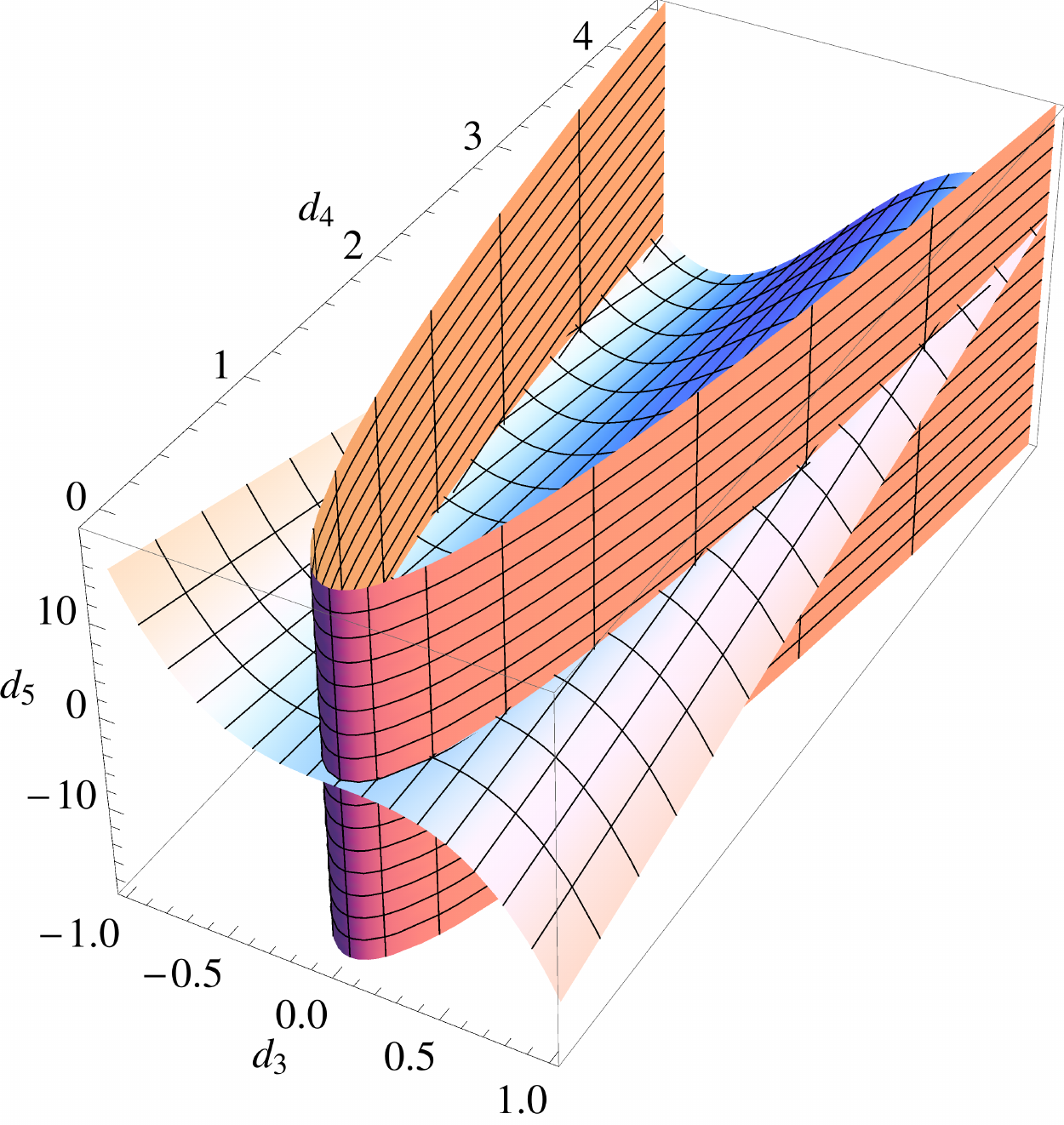}
\end{center}
\caption{The surfaces $I_{4}=0$ (cylindrical one corresponding to duals of
the quintic Galileon) and $I_{5}=0$ ($Z_{2}$ symmetric Galileons) in the
theory space $D_{5}^{(1)}$ with $d_{2}=1/12$ fixed. The intersection of
these surfaces corresponds to the duals of a free theory. Both surfaces are
invariant with respect to the scaling.}
\label{parametricspace}
\end{figure}

\subsubsection{Galileons in $d=3$}

The situation in three dimension is even simpler. There is only one
invariant of the $\mathbf{\alpha }_{D}\left( \theta \right) $ duality%
\begin{equation}
I_{4}=d_{4}-\frac{9}{8}d_{3}^{2}
\end{equation}%
From the previous it follows readily
\begin{equation}
d_{3}(\lambda )=\lambda ^{-\frac{5}{2}}d_{3},~~~~d_{4}(\lambda )=\lambda
^{-5}d_{4},~~~I_{4}(\lambda )=\lambda ^{-5}I_{4}
\end{equation}%
For $I_{4}<0$ we can remove $d_{4}$ by a duality $\mathbf{\alpha }_{D}\left(
\theta _{\pm }\right) $ with
\begin{equation}
\theta _{\pm }=\frac{3}{4}d_{3}\pm \frac{1}{2}\sqrt{-2I_{4}}
\end{equation}%
Provided $I_{4}>0$ there is no $\mathbf{\alpha }_{D}\left( \theta \right) $
dual with $d_{4}(\theta )=0$ and naively there is no possibility to change
the sign of $I_{4}$ by simple scaling because $\lambda <0$ is not allowed in
odd dimension for theory with $d_{2k-1}\neq 0$. However, we can first remove
$d_{3}$ by $\mathbf{\alpha }\left( 3d_{3}/2\right) $ and only then scale
with $\lambda <0$ to arrange $I_{4}=1$, because there is no odd vertex in
the dual theory. This leads to the following classification

\begin{itemize}
\item $I_{4}=0$ --- free theory

\item $I_{4}=1$ --- $Z_{2}$ invariant theory
\end{itemize}

To summarize, up to the above described $\mathbf{\alpha }_{D}\left( \theta
\right) $ and $\mathbf{\alpha }_{S}(\lambda )$ dualities there is only one
non-trivial Galileon theory in three dimension the only nonzero amplitudes
of which are the even ones.

\subsection{Counterterms\label{CT}}

From the results of the previous sections it seems possible to use the
duality relations also at the quantum level. However, this is true only
provided the quantum level makes sense. Starting with the basic (i.e. the
tree-level) Galileon Lagrangian and choosing an appropriate regularization
prescription which preserves the Galileon symmetry (in what follows we use
exclusively dimensional regularization), we can construct one-loop diagrams.
Such diagrams will be divergent and will thus need to be renormalized by the
counterterms. From a simple dimensional consideration it is clear (and it
will be explicitly shown below) that it is not possible to create such
counterterms using the basic tree-level Lagrangian. We will thus have to add
qualitatively new terms in the action constrained in their form only by the
Galileon symmetry. In fact at any order of the loop expansion an infinite
tower of new counterterms is necessary. This is of course nothing new, such
a mechanism is well studied in many different effective theories e.g. in the
Chiral perturbation theory (ChPT) \cite{Gasser:1983yg, Gasser:1984gg}. The
problem of construction of higher order Lagrangians (e.g.
next-to-leading-order and next-to-next-to-leading order as is the nowadays
status in ChPT) is the problem by itself. Here we will merely classify the
order (i.e. the degree of homogeneity in the external momenta or the number
of derivatives) of the graphs and the corresponding counterterms at the
given loop level.

Let us start with the Weinberg formula \cite{Weinberg:1978kz} in $d$%
-dimension for the number of derivatives in the counterterm for a given
graph with $L$ loops and vertices $V$ with $d_{V}$ derivatives\footnote{%
This formula holds provided the dimensional regularization or any other
regularization without dimensionfull cutoff parameter is used to regulate
the UV divergences. Note also that in our case of massless theory $D$ is
also the superficial degree of divergence of the given graph.}
\begin{equation}
D=2+(d-2)L+\sum_{V}\left( d_{V}-2\right) .  \label{weinberg_formula}
\end{equation}%
The number of external legs $E$ and internal lines $I$ is connected via
\begin{equation}
\sum_{V}n_{V}=2I+E\,,
\end{equation}%
where $n_{V}$ is the number of legs for the given vertex $V$. We can also
simply extract number of loops
\begin{equation}
L=I-V+1\,.
\end{equation}%
Together with the previous relation this leads to
\begin{equation}
E=2+\sum_{V}\left( n_{V}-2\right) -2L,  \label{E_in_L_n_V}
\end{equation}%
and thus
\begin{equation}
D-2\left( E-1\right) =(d+2)L+\sum_{V}\left( d_{V}-2\left( n_{V}-1\right)
\right) .  \label{index}
\end{equation}%
Let us now define an index of general vertex $\delta _{V}$ as a surplus of
the derivatives for the general vertex in comparison with the basic
Lagrangian, namely
\begin{equation}
\delta _{V}=d_{V}-2\left( n_{V}-1\right)
\end{equation}%
(i.e. for all the vertices of the basic Lagrangian $\delta _{V}=0$). In
terms of such a defined index we can rewrite the formula (\ref{index}) in
the form
\begin{equation}
\delta _{CT}=(d+2)L+\sum_{V}\delta _{V}\equiv \delta _{\Gamma }
\label{CT_index}
\end{equation}%
the right hand side of which defines the index $\delta _{\Gamma }$ of a $L$%
-loop graph $\Gamma $ built from the vertices with indices $\delta _{V}$.
This formula is in fact an Galileon analogue of the Weinberg formula for ChPT
and represents thus the connection between the loop expansion and expansion
in the diagram index $\delta _{\Gamma }$, which is the order of the diagram
homogeneity in momenta (modulo logs) relative to tree-level diagrams
constructed from the basic Lagrangian.

Note that according to the formula (\ref{CT_index}) each loop contributes
with an additional $d+2$ term in the counterterm index $\delta _{CT}$. This
means that the counterterms induced by the loops have $\delta _{CT}>0$ and
therefore (because for the vertices of the basic Lagrangian $\delta _{V}=0$)
they must be different form the terms of the basic Lagrangian. In other
words the basic Galileon Lagrangian is not renormalized by loops. This
proves what is often meant in the literature as the non-renormalization
theorem \cite{Luty:2003vm, Hinterbichler:2010xn, deRham:2012ew}.

Let us note that similarly to the Weinberg formula for the ChPT, the formula
(\ref{CT_index}) itself cannot be used for the proof of the generalized
renormalizability. Note that the restriction $\delta _{CT}=N=\mathrm{const}.$
constrains only the number of derivatives $d$ according to%
\begin{equation}
d=2\left( n-1\right) +N,  \label{CT_constraint}
\end{equation}%
but it does not constrain the number $n$ of fields. In the case of ChPT the
additional principle is a chiral symmetry which ensures that the infinite
number of counterterms differing by the number of fields at each order
combine into a finite number of chiral invariant operators. In our case we
have only the Galileon symmetry at our disposal. As we have discussed above,
it tells us only that the most general Galileon invariant Lagrangian is
built from the building blocks $\partial _{\mu _{1}}\partial _{\mu
_{2}}\ldots \partial _{\mu _{k}}\phi $ where $k\geq 2$, therefore the
general counterterm with $n$ legs satisfying (\ref{CT_constraint}) has the
general form%
\begin{equation}
\mathcal{L}_{CT}^{(n)}=\sum_{l,~k_{i}\geq 2,~\sum_{i}k_{i}=2\left(
n-1\right) +N}c_{k_{1}k_{2}\ldots k_{n}}^{(l)}T_{(l)}^{\mu _{1}^{1}\mu
_{2}^{1}\ldots \mu _{k_{1}}^{1}\ldots \mu _{1}^{n}\ldots \mu
_{k_{n}}^{n}}\prod\limits_{j=1}^{n}\partial _{\mu _{1}^{j}}\partial _{\mu
_{2}^{j}}\ldots \partial _{\mu _{k_{j}}^{j}}\phi .  \label{general_CT}
\end{equation}%
with couplings $c_{k_{1}k_{2}\ldots k_{n}}^{(l)}$ and Lorentz invariant
tensors $T_{(l)}^{\mu _{1}^{1}\mu _{2}^{1}\ldots \mu _{k_{1}}^{1}\ldots \mu
_{1}^{n}\ldots \mu _{k_{n}}^{n}}$. Though for $n$ fixed we have finite
number of terms, $\ $the number $n$ increases to infinity and already at the
one loop level (where $N=d+2$) we get infinite number of independent terms.

Let us note that quantum Galileon is in principal a two scale theory. At the
classical level there is a scale (let us denote it $F$), which is
responsible for the hierarchy of the nonlinearities in the basic classical
Lagrangian (\ref{galileon_Lagrangian})\cite{Nicolis:2008in}. Restoring the
correct dimensions of the tree-level Galileon couplings we can write for the
general term of the basic Lagrangian schematically (up to $O(1)$
dimensionless constant, cf. (\ref{alternative_basic_L}))
\begin{equation}
\mathcal{L}_{b}^{(n)}\sim \left( \partial \phi \right) ^{2}\left( \frac{%
\partial \partial \phi }{F^{d_{\phi }+2}}\right) ^{n-2}\sim F^{-\frac{\left(
d+2\right) (n-2)}{2}}
\end{equation}%
where $d_{\phi }=(d-2)/2$ is the canonical dimension of the Galileon field.
On the other hand, according to the formula (\ref{CT_index}), the quantum
corrections can be organized as a expansion in powers of the characteristic
quantum scale, let us denote it $\Lambda $. Each counterterm with index $%
\delta _{CT}$ is suppressed by a power $\Lambda ^{-\delta _{CT}}$.
Schematically (up to $O(1)$ dimensionless coupling constant, cf. (\ref%
{general_CT}))%
\begin{equation}
\mathcal{L}_{CT}^{(n,\delta _{CT})}\sim F^{d}\left( \frac{F}{\Lambda }%
\right) ^{2}\prod\limits_{j=1}^{n}\left( \frac{\partial }{\Lambda }\right)
^{k_{i}}\left( \frac{\partial \partial \phi }{F^{d_{\phi }+2}}\right) \sim
F^{-\frac{\left( d+2\right) (n-2)}{2}}\Lambda ^{-\delta _{CT}},
\end{equation}%
where we have assumed the constraint $\sum_{i}\left( k_{i}+2\right) =2\left(
n-1\right) +\delta _{CT}$ . Note that we have to restore the correct
dimension of $\mathcal{L}_{CT}^{(n,\delta _{CT})}$ using the scale $F$ in
order not to disturb the hierarchy of the counterterms.

The requirement of consistency of the Galileon as an effective theory at the
quantum level puts however constraint on the above two scales. The point is
that the renormalized loop graphs and the corresponding counterterms
contributions have to be numerically of the same order. To get this
constraint let us assume a general graph $\Gamma $ with $L$ loops, $E$
external legs and $V$ vertices (each vertex has index $\delta _{V}$ and $%
n_{V}$ legs). Its contribution is schematically\footnote{%
Here the factor $(4\pi )^{-2}$ is generic for each loop momentum integration.%
}
\begin{equation}
\Gamma \sim \left( \frac{1}{4\pi }\right) ^{2L}\frac{1}{\Lambda ^{\Delta
_{\Gamma }^{\Lambda }}}\frac{1}{F^{\Delta _{\Gamma }^{F}}}
\end{equation}%
where%
\begin{eqnarray}
\Delta _{\Gamma }^{\Lambda } =\sum_{V}\delta _{V}, \,\,\,\,\, \Delta
_{\Gamma }^{F} =\sum_{V}\frac{\left( d+2\right) }{2}(n_{V}-2)=\frac{\left(
d+2\right) }{2}\left( 2L+E-2\right)
\end{eqnarray}%
and we have used (\ref{E_in_L_n_V}). For the counterterm contribution we get%
\begin{equation}
CT\sim \frac{1}{\Lambda ^{\delta _{CT}}}\frac{1}{F^{\frac{\left( d+2\right)
(E-2)}{2}}},
\end{equation}%
where $\delta _{CT}$ is given by (\ref{CT_index}). Requiring $\Gamma \sim CT$
gives then the desired relation between the classical and quantum scales%
\begin{equation}
\Lambda \sim \left( 4\pi \right) ^{\frac{2}{d+2}}F,  \label{Lambda_F}
\end{equation}%
or $\Lambda \sim 2.3F$ for $d=4$; both scales are therefore forced to be
roughly of the same order of magnitude in this case. This is analogue of the
formula known in ChPT which relates the pion decay constant with the scale
of the chiral symmetry breaking%
\begin{equation*}
\Lambda _{\chi PT}\sim \left( 4\pi \right) ^{\frac{2}{d-2}}F_{\pi }
\end{equation*}%
which however for $d=4$ requires $\Lambda _{\chi PT}$ to be one order of
magnitude larger than $F_{\pi }$.

\subsection{Examples of one-loop order duality\label{one_loop_duality}}

In the previous sections we have explicitly calculated the tree-level
scattering amplitudes of the Galileon fields up to six particles. The
non-trivial results start with the four-point scattering. In this section we
will focus on this process and will study it at one-loop order. Of course,
as mentioned above, such a full calculation would necessary need inclusion
of so-far undefined Lagrangian $\mathcal{L}_{CT}^{(4)}$, which would play a
role of counterterms in this process. However, our main motivation is to
explicitly show that the duality is not spoiled at the quantum level (i.e.
by loop contributions) at least for the graphs with the vertices from the
basic Galileon Lagrangian. We will thus first calculate dimensionally
regularized individual contributions to 4-pt scattering at one-loop order in
one Galileon theory and show that the final result is related to other
Galileon theory connected by duality.

In the Table \ref{amps} we summarize the one-loop diagrams to be calculated
and their corresponding divergent parts in $d=4$ dimension (the full results
in $d$ dimension for $A_{1-6}$ are summarized in Appendix \ref{A_Sigma}).
\begin{table}[t]
\begin{tabular}{lm{4cm}m{10cm}}
$\mathrm{i}A_{1}\equiv $ & \includegraphics[scale=0.8]{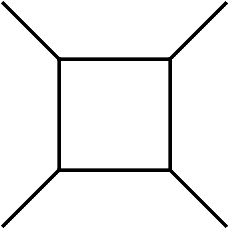} & $=-\frac{243%
\mathrm{i}}{40}d_{3}^{4}\Lambda (s^{2}+t^{2}+u^{2})^{3}$ \\
$\mathrm{i}A_{2}\equiv $ & \includegraphics[scale=0.8]{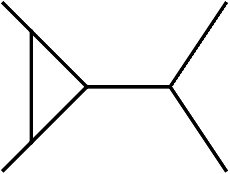} & $=81\mathrm{%
i}d_{3}^{4}\Lambda (s^{6}+t^{6}+u^{6})$ \\
$\mathrm{i}A_{3}\equiv $ & \includegraphics[scale=0.8]{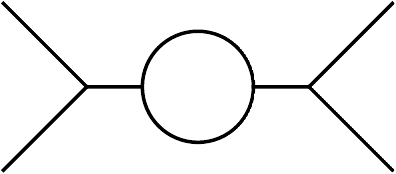} & $=-81%
\mathrm{i}d_{3}^{4}\Lambda (s^{6}+t^{6}+u^{6})$ \\
$\mathrm{i}A_{4}\equiv $ & \includegraphics[scale=0.8]{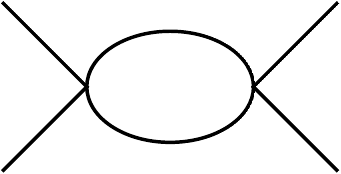} & $=-\frac{3%
\mathrm{i}}{10}d_{4}^{2}\Lambda (s^{2}+t^{2}+u^{2})^{3}$ \\
$\mathrm{i}A_{5}\equiv $ & \includegraphics[scale=0.8]{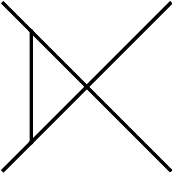} & $=-\frac{9%
\mathrm{i}d_{4}d_{3}^{2}}{10}\Lambda \lbrack
20(s^{6}+t^{6}+u^{6})-3(s^{2}+t^{2}+u^{2})^{3}]$ \\
$\mathrm{i}A_{6}\equiv $ & \includegraphics[scale=0.8]{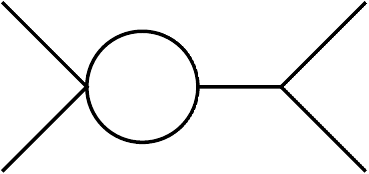} & $=18\mathrm{%
i}d_{4}d_{3}^{2}\Lambda (s^{6}+t^{6}+u^{6})$ \\
$\mathrm{i}A_{7}\equiv $ & \includegraphics[scale=0.8]{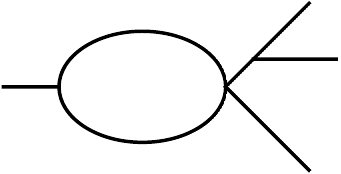} & $=0$ \\
$\mathrm{i}A_{8}\equiv $ & \includegraphics[scale=0.8]{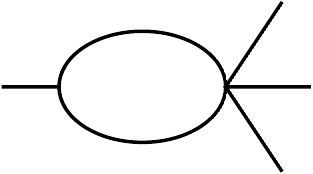} & $=0$%
\end{tabular}%
\caption{One loop graphs contributing to the 4pt amplitude and their
divergent parts.}
\label{amps}
\end{table}
Here we have used the standard Mandelstam variables for four-point
scattering:
\begin{equation}
s=(p_{1}+p_{2})^{2},\qquad t=(p_{1}+p_{3})^{2},\qquad u=(p_{1}+p_{4})^{2}
\end{equation}%
where all momenta are ingoing and on-shell so that $s+t+u=0$. The
singularity in $d=4$ dimension is given by
\begin{equation}
\Lambda =\frac{1}{(4\pi )^{2}}\frac{1}{d-4}\,.
\end{equation}
Due to specific form of the of 3-pt vertex in the Galileon theory which can
be rewritten in the form%
\begin{equation}
\mathcal{V}_{3}(q_{1},q_{2},q_{3})=6d_{3}\left[ (q_{1}\cdot
q_{2})q_{3}^{2}+(q_{1}\cdot q_{3})p_{2}^{2}+(q_{2}\cdot q_{3})p_{1}^{2}%
\right]  \label{V_3}
\end{equation}%
the contributions $A_{7}$ and $A_{8}$ (corresponding to graphs for which $%
\mathcal{V}_{3}$ is one of the two vertices of a bubble) are zero also for
general $d$. Indeed, with external momenta on shell, the only term of (\ref%
{V_3}) which could contribute is schematically $(p_{\mathrm{ext}}\cdot l)(p_{%
\mathrm{ext}}-l)^{2}$ where $p_{\mathrm{ext}}$ is one of the external
momenta and $l$ is the loop momentum. Therefore the $(p_{\mathrm{ext}%
}-l)^{2} $ factor cancels one of the bubble propagators which thus
degenerate in a massless tadpole and the latter is zero in dimensional
regularization.

Summing the diagrams together, we get that the divergent part of the
amplitude for the 4-pt galileon-scattering at the one-loop order is
\begin{equation}
A^{\mathrm{div}}=\sum_{i}A_{i}^{\mathrm{div}}=-\frac{3}{40}\Lambda
(9d_{3}^{2}-2d_{4})^{2}(s^{2}+t^{2}+u^{2})^{3}\,=-\frac{3}{10}\Lambda
I_{4}^{2}(s^{2}+t^{2}+u^{2})^{3}.
\end{equation}%
Note that the degree of homogeneity in external momenta is in accord with
the formula (\ref{CT_index}). As we have expected, the singular part (and in
fact also the complete result (\ref{fullA}), cf. Appendix \ref{A_Sigma})
depend on the $\mathbf{\alpha }_{D}\left( \theta \right) $ duality invariant
$I_{4}$ which illustrates the conclusions of Section \ref{S_duality} that $%
\mathbf{\alpha }_{D}\left( \theta \right) $ dual theories produce the same $%
S $-matrices. This offers also another possibility how to use the duality
relations similar to that we have discussed for the tree amplitudes in
Section \ref{tree_level_A}. Because $I_{4}$ is the coupling $d_{4}$ in the
dual Galileon theory with new constants $d_{i}(\theta ^{\ast })$ such as $%
d_{3}(\theta ^{\ast })=0$ we can effectively eliminate 3pt vertices by
passing to this dual theory. The only diagram which is left to calculate\ in
such a dual theory is $A_{4}$ which simplifies the calculations considerably.

Let us present another simple example of the one-loop calculation concerning
the self-energy correction for the Galileon field. The relevant graph is
depicted in Fig.\ref{SE} and the explicit result for the divergent part
reads (see Appendix \ref{A_Sigma} for the complete result)%
\begin{equation}
\Sigma ^{\mathrm{div}}(p)=9\Lambda d_{3}^{4}\left( p^{2}\right) ^{4},
\end{equation}
cf. also \cite{dePaulaNetto:2012hm}.
\begin{figure}[t]
\begin{center}
\includegraphics[scale=0.8]{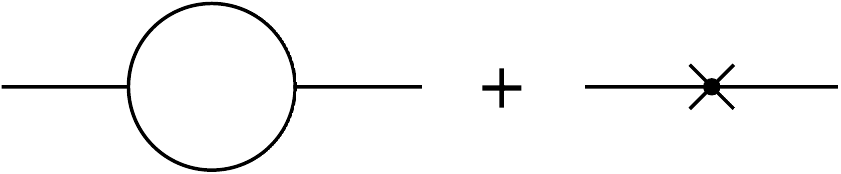}
\end{center}
\caption{The Feynman graph for the Galileon self-energy and its counterterm.}
\label{SE}
\end{figure}
Therefore, on one hand, in Galileon theories with $d_{3}\neq 0$ we need a
counterterm $\mathcal{L}_{CT}^{(2)}$ of the general form (\ref{general_CT})
to renormalize this divergence. The corresponding Feynman rule reads
\begin{equation}
\mathcal{V}_{2}(p_{1})=\mu ^{d-4}\left( -9\Lambda d_{3}^{4}+C^{r}(\mu
)\right) \left( p^{2}\right) ^{4},  \label{V3}
\end{equation}%
where $\mu $ is the dimensional regularization scale which is necessary to
restore the canonical dimension of the loop integration and where $C^{r}(\mu
)$ is a linear combination of the finite parts of the counterterm couplings $%
c_{k_{1}k_{2}}^{(l)}$ in (\ref{general_CT}) renormalized at scale $\mu $. On
the other hand, in the above mentioned dual theory with $d_{3}(\theta ^{\ast
})=0$ such a divergence does not occur. This is consequence of the fact that
off-shell Green functions are not invariants with respect to the duality as
discussed in Section \ref{S_duality}.

Extreme example of such non-invariance of the counterterms is the case of
the free theory and some of its $\mathbf{\alpha }_{D}\left( \theta \right) $
duals. While the free theory does not need any counterterm of the above
type, its dual always\footnote{%
Note that any such dual has $d_{3}=-\theta /3\neq 0$.} does. However, as far
as the $S$ matrix is concerned, no counterterms are needed for the graphs
with the vertices from the basic dual Lagrangian because these graphs have
to combine into the trivial $S$ matrix of the original free theory which is
trivially divergence-free. Therefore, the contribution of the divergent part
of (\ref{V3}) and analogical counterterms (which are needed to renormalize
the divergent subgraphs in the Table \ref{amps}) has to cancel in the final
result. This is, however, not true for the finite part of the counterterms
the couplings of which are in principle independent. E.g. the
renormalization of the bubble subgraph in the graphs $A_{3}$ in Table \ref%
{amps} brings about the contribution (for $d\rightarrow 4$)%
\begin{equation}
A_{3}^{CT}=-9\mathrm{i}C^{r}(\mu )\left( s^{6}+t^{6}+u^{6}\right) .
\end{equation}%
Therefore, the only possibility how to recover the free theory $S$ matrix
also in the dual theory with counterterms is to set at some scale all the
renormalized counterterm coupling constants equal to zero. Because the
couplings run with the renormalization scale, this might seem to be
insufficient, because at another scale the finite parts of the counterterms
are in general nonzero. However, because in the amplitude all the
contributions of the divergent parts of the counterterms cancel, in the same
way are also cancelled all the contributions stemming from the changes of the
counterterms couplings with change of the renormalization scale\footnote{%
Note that within dimensional regularization, the coefficient at the $\ln
\left( \mu /\mu ^{\prime }\right) $ term in the formula for the running of
the\ renormalized one-loop coupling coincide with the coefficient at the $%
\Lambda $ in formula for the bare coupling.}.

\section{Summary}

In this paper we have studied the duality transformations of the general
Galileon theories in $d$ dimensions. First we have reviewed the
interpretation of the Galileon as a Goldstone boson of the spontaneous
symmetry breakdown according to the pattern $GAL(d,1)\rightarrow ISO(d-1,1)$
and the identification of its action as the generalized WZW term. Then we
have studied the most general coordinate transformations on the
corresponding coset space $GAL(d,1)/SO(d-1,1)$. The requirement that such a
general transformation acts linearly on the basic building blocks of the
Galileon Lagrangian (and therefore it represents a duality transformation)
constraints the form of the transformation uniquely up to four free
parameters. Under composition these duality transformations form a group
which can be identified with $GL(2,{\mathbf{R}})$. The explicit form of the
duality transformation for general $\alpha \equiv \{\alpha
_{ij}\}_{i,j=1}^{2}\in GL(2,{\mathbf{R}})$ reads%
\begin{eqnarray}
x_{\alpha } &=&\alpha _{11}x+\alpha _{12}\partial \phi (x)  \notag \\
\phi _{\alpha }(x_{\alpha }) &=&\det \left( \mathbf{\alpha }\right) \phi (x)
\notag \\
&&+\frac{1}{2}\left( \alpha _{12}\alpha _{22}\partial \phi (x)\cdot \partial
\phi (x)+2\alpha _{12}\alpha _{21}x\cdot \partial \phi (x)+\alpha
_{11}\alpha _{21}x^{2}\right) .  \notag
\end{eqnarray}%
All the up to now known Galileon dualities can be identified as special
elements (or one-parametric subgroups) of this duality group. We have also
studied its action on the space of the Galileon theories and found a basis
of the independent invariants of one of its most interesting one-parametric
subgroups
\begin{equation*}
\mathbf{\alpha }_{D}(\theta )=\left(
\begin{array}{cc}
1 & -2\theta \\
0 & 1%
\end{array}%
\right)
\end{equation*}%
This subgroup is represented in the space of fields as a field redefinition
which can be understood both as a simultaneous space-time coordinates and
field transformation or as a non-local change of the fields which includes
infinite number of derivative dependent terms.

We have then studied the applications of the duality group. We have shown
that we can relate the classical covariant phase spaces of dual theories and
enlarge the duality transformation to the classical observables. In order to
avoid apparent paradoxes, correct dual observables within the dual theory
have to be used when we want to get the results of the original theory.

The duality of phase spaces can be used to generate classical solution of
the interacting Galileon theory from the solution of the more simple one
even when the Galileon is coupled to the local external source. We have
studied two such sources, namely the point-like and string-like ones. Here
the duality appears to be an efficient tool because of the symmetries which
effectively reduce the dimensionality of the problems.

We have also discussed the fluctuations of the classical solutions in the
linearized approximation. We have found duality transformation of the
corresponding classical covariant phase spaces and corresponding observables
and discussed the geometrical aspects of the problem with superluminal
propagation of fluctuations. The general consideration has been illustrated
by two explicit examples, namely the fluctuations of the plane wave and
cylindrically symmetric classical solutions.

We have also established the dual formulation of the additional symmetries
of the Lagrangian. We have shown that these symmetries might be hidden
within the dual theory and found e.g. that the $Z_{2}$ symmetry and space
time translations are realized non-linearly and non-locally.

Then we have discussed the transformation properties of the $S$ matrix and
established its formal invariance within the dimensional regularization,
though only the tree-level (classical) part of the complete action
transforms nicely under the duality field redefinition. As a next issue we
have demonstrated the usefulness of the $S$ matrix duality for calculations
of the tree on-shell scattering amplitudes and for finding the relations
between the contributions of the apparently very different Feynman graphs
with completely different topologies.

As another example we have classified the equivalence classes (with respect
to the duality subgroup $\alpha _{D}(\theta )$ combined with scaling) of the
Galileon theories (and at the same time of the tree-level $S$ matrices) in
three and four dimensions. We found e.g. that there is up to the above
dualities only one nontrivial interacting theory in three dimensions which
exhibits the $Z_{2}$ symmetry. Then we have discussed the transformation
properties of the $S$ matrix on the loop level. As we have discussed on a
concrete example of the one-loop four-point on-shell amplitude, due to the
counterterms the duality is not completely straightforward. It rather holds
on the regularized level for the loop graphs with vertices from the basic
tree-level Lagrangian. We have also touched the problem of the counterterms
classification based on a generalization of the Weinberg formula and with
help of the latter we discussed the non-renormalization theorem.

\textbf{Note added}: After this work was completed two works \cite%
{Creminelli:2014zxa,deRham:2014lqa} closely connected with the topic studied
in this paper appeared. Both these papers concern the properties of the
one-parametric duality subgroup denoted as $\alpha_D(\theta)$ in our
notation and partially overlap with our results.


\section*{Acknowledgement}

We would like to thank Nima Arkani-Hamed and Jaroslav Trnka for turning our
attention to Galileon theory and Jaroslav Trnka and Ji\v{r}\'{\i} Ho\v{r}ej%
\v{s}\'{\i} for useful discussions and comments on the manuscript. This work
is supported by Charles University in Prague, project PRVOUK P45 and by
Ministry of Education of the Czech Republic, grants LG 13031 and LH 14035.

\appendix

\section{Bottom up construction of the duality subgroup $\mathbf{\protect%
\alpha }_{D}\left( \protect\theta \right)$}

\label{bu}

In this Appendix we get more elementary treatment of the Galileon duality
corresponding to the subgroup $\mathbf{\alpha }_{D}\left( \theta \right) $.
In fact this was the way we had started to think about the Galileon duality.

Let us assume an infinitesimal field transformation
\begin{equation}
\phi \rightarrow \phi +\theta \partial \phi \cdot \partial \phi ,
\label{basic_transformation}
\end{equation}%
where $\theta $ infinitesimal parameter. The infinitesimal change of $\phi $
can be also understood as an action of the following operator (which is
defined on the space of the functionals $F[\phi ]$ of the field $\phi $)
\begin{equation}
\delta _{\theta }\equiv \theta \left\langle \delta \phi \frac{\delta }{%
\delta \phi }\right\rangle =\theta \left\langle \partial \phi \cdot \partial
\phi \frac{\delta }{\delta \phi }\right\rangle
\end{equation}%
on a special functional $F[\phi ](x)=\phi (x).$ Here and in what follows we
again abbreviate $\left\langle \cdot \right\rangle \equiv \int \mathrm{d}%
^{d}x\left( \cdot \right) $.

Acting by the operator $\delta _{\theta }$ on the Galileon action (cf. (\ref%
{galileon_Lagrangian}))
\begin{equation}
S[\phi ]=\sum_{n=1}^{d+1}d_{n}\left\langle \phi \mathcal{L}_{n-1}^{\mathrm{%
der}}\right\rangle
\end{equation}
we get
\begin{equation*}
\delta _{\theta }S[\phi ]=\theta \left\langle \partial \phi \cdot \partial
\phi \frac{\delta S[\phi ]}{\delta \phi }\right\rangle
=\sum_{n=2}^{d+1}\theta nd_{n}\left\langle \partial \phi \cdot \partial \phi
\mathcal{L}_{n-1}^{\mathrm{der}}\right\rangle
\end{equation*}%
which can be rewritten with help of the formula (\ref{prevod}) to the form%
\begin{equation}
\delta _{\theta }S[\phi ]=-2\sum_{n=3}^{d+1}\theta \frac{n-1}{n}%
(d-n+2)d_{n-1}\left\langle \phi \mathcal{L}_{n-1}^{\mathrm{der}%
}\right\rangle =\sum_{n=3}^{d+1}\theta \delta d_{n}\left\langle \phi
\mathcal{L}_{n-1}^{\mathrm{der}}\right\rangle
\end{equation}%
with%
\begin{equation}
\delta d_{n}=-2\frac{n-1}{n}(d-n+2)d_{n-1}
\end{equation}%
Therefore to the first order in $\theta $ the transformation (\ref%
{basic_transformation}) conserves the Galileon structure of the Lagrangian
and merely shifts the coupling constants $d_{n}$ by $\delta d_{n}$. Note
that the transformation (\ref{basic_transformation}) with finite $\theta $
can be used to eliminate the cubic term from the interaction Lagrangian,
however, the Galileon structure of the Lagrangian is spoiled with additional
interaction terms which are generated by the transformation. The way how to
eliminate the cubic term consistently without leaving the space of the
Galileon theories is now clear. It suffices to construct the finite
transformation by means of iteration of the infinitesimal one, i.e. to
exponentialize it according to
\begin{equation}
\phi _{\theta }=\exp \left( \delta _{\theta }\right) \phi =\exp \left\langle
\theta \partial \phi \cdot \partial \phi \frac{\delta }{\delta \phi }%
\right\rangle \phi =\phi +\theta \partial \phi \cdot \partial \phi +2\theta
^{2}\partial \phi \cdot \partial \partial \phi \cdot \partial \phi +\ldots
\label{dualita}
\end{equation}%
Applying this finite transformation to the Galileon action results in a dual
action $S_{\theta }[\phi ]$ defined as
\begin{equation}
S_{\theta }[\phi ]\equiv S[\phi _{\theta }]=\exp \delta _{\theta }S[\phi
]=\sum_{n=2}^{d+1}d_{n}(\theta )\left\langle \phi \mathcal{L}_{n-1}^{\mathrm{%
der}}\right\rangle .
\end{equation}%
It is not difficult to show that $d_{n}(\theta )=d_{n}\left( \mathbf{\alpha }%
_{D}\left( \theta \right) \right) $ where the right hand side is given by (%
\ref{d_theta}). From this construction it is clear that the transformations $%
\phi _{\theta }$ form a one parametric group.

In what follows we will give an alternative elementary derivation of the
explicit form of $\phi _{\theta }$. Let us denote%
\begin{equation}
\phi (\theta ,x)=\exp \left\langle \theta \partial \phi \cdot \partial \phi
\frac{\delta }{\delta \phi }\right\rangle \phi (x),
\end{equation}%
then we get by derivative with respect to $\theta $%
\begin{eqnarray}
\frac{\partial \phi (\theta ,x)}{\partial \theta } &=&\exp \left\langle
\theta \partial \phi \cdot \partial \phi \frac{\delta }{\delta \phi }%
\right\rangle \left\langle \partial \phi \cdot \partial \phi \frac{\delta }{%
\delta \phi }\right\rangle \phi (x)  \notag \\
&=&\exp \left\langle \theta \partial \phi \cdot \partial \phi \frac{\delta }{%
\delta \phi }\right\rangle \partial _{\mu }\phi (\theta ,x)\partial ^{\mu
}\phi (\theta ,x)=\partial _{\mu }\phi (\theta ,x)\partial ^{\mu }\phi
(\theta ,x).
\end{eqnarray}%
Therefore the function $\phi (\theta ,x)$ is a solution of the following
Cauchy problem for the partial differential equation of the first order%
\begin{equation}
\frac{\partial \phi (\theta ,x)}{\partial \theta }=\partial _{\mu }\phi
(\theta ,x)\partial ^{\mu }\phi (\theta ,x),~~~\phi (0,x)=\phi (x).~
\label{cauchy}
\end{equation}%
This problem can be solved by standard method of characteristics which are
the solutions of a set of ordinary differential equations%
\begin{equation}
\frac{\mathrm{d}\theta }{\mathrm{d}t}=1,~~~~~\frac{\mathrm{d}P}{\mathrm{d}t}%
=0~,~~~~\frac{\mathrm{d}x}{\mathrm{d}t}=-2p,~~\frac{\mathrm{d}p}{\mathrm{d}t}%
=0,~~~~\frac{\mathrm{d}\phi }{\mathrm{d}t}=-2p\cdot p+P^{2}.
\end{equation}%
The solution of these equation is%
\begin{equation}
\theta =\theta _{0}+t,~~~P=P_{0},~~~p=p_{0},~~x=x_{0}-2tp_{0},~~~~\phi =\phi
_{0}+t\left( P_{0}^{2}-2p_{0}\cdot p_{0}\right) .~~~~
\end{equation}%
The general recipe how to get the $d+1$-dimensional integral surface
corresponding to the equation (\ref{cauchy}) consist of two steps. First we
replace the integrations constants $\theta _{0},\ldots ,\phi _{0}$ with
functions of $d$ parameters $a_{i}$, $i=1,\ldots ,d$ in such a way that the
following conditions are satisfied%
\begin{equation}
P_{0}(a_{i})-p_{0}(a_{i})\cdot p_{0}(a_{i})=0,~~~~~~\mathrm{d}\phi
_{0}(a_{i})=p_{0}(a_{i})\cdot \mathrm{d}x_{0}(a_{i})+P_{0}(a_{i})\mathrm{d}%
\theta _{0}(a_{i})  \notag
\end{equation}%
and subsequently we eliminate the parameters $a_{i}$ and $t$ from the
equations
\begin{eqnarray}
\theta &=&\theta _{0}(a_{i})+t,~~~~x=x_{0}(a_{i})-2tp_{0}(a_{i})  \notag \\
\phi &=&\phi _{0}(a_{i})+t\left( P_{0}^{2}(a_{i})-2p_{0}(a_{i})\cdot
p_{0}(a_{i})\right) .
\end{eqnarray}%
Let us choose the parameters $a_{i}$ to be just $x_{0}$, we get then%
\begin{equation}
t=\theta -\theta _{0}(x_{0}),~~~~~P_{0}(x_{0})=p_{0}(x_{0})\cdot
p_{0}(x_{0}),~~~~p_{0}(x_{0})=\partial \phi _{0}(x_{0})-P_{0}(x_{0})\partial
\theta _{0}(x_{0}),  \notag
\end{equation}%
and
\begin{eqnarray}
x &=&x_{0}-2tp_{0}(x_{0})  \notag \\
&=&x_{0}-2\left( \theta -\theta _{0}(x_{0})\right) \left( \partial \phi
_{0}(x_{0})-p_{0}(x_{0})\cdot p_{0}(x_{0})\partial \theta _{0}(x_{0})\right)
\notag \\
\phi &=&\phi _{0}(x_{0})+t\left( P_{0}^{2}(x_{0})-2p_{0}(x_{0})\cdot
p_{0}(x_{0})\right)  \notag \\
&=&\phi _{0}(x_{0})-\left( \theta -\theta _{0}(x_{0})\right)
p_{0}(x_{0})\cdot p_{0}(x_{0}).
\end{eqnarray}%
A special choice $\theta _{0}(x_{0})=0$ gives
\begin{equation}
x=x_{0}-2\theta \partial \phi _{0}(x_{0}),~~~~~\phi (\theta ,x)=\phi
_{0}(x_{0})-\theta \partial \phi _{0}(x_{0})\cdot \partial \phi _{0}(x_{0}).
\end{equation}%
The initial condition of the Cauchy problem is $\phi (0,x)=\phi (x)$ and
therefore $\phi _{0}(x_{0})=\phi (x_{0})$. Thus the final solution of the
Cauchy problem is
\begin{equation}
x=x_{0}-2\theta \partial \phi (x_{0}),~~~~~~~\phi (\theta ,x)=\phi
(x_{0})-\theta \partial \phi (x_{0})\cdot \partial \phi (x_{0})
\end{equation}%
which is nothing else but the duality transformation (\ref%
{one_parametric_duality}).

\section{Compatibility of duality and IHC constraint}

In this Appendix we demonstrate by explicit calculation the consistency of
the duality transformation with the IHC constraint. \ For the derivative of
the field $\phi $ with respect to the unprimed coordinates we get with help
of the second row of (\ref{generalized_duality_alpha})
\begin{eqnarray}
\partial \phi &=&\partial x^{\prime }\cdot \partial ^{\prime }\phi  \notag \\
&=&\partial x^{\prime }\cdot \left[ \det \left( \alpha _{IJ}\right) \partial
^{\prime }\phi ^{\prime }\phantom{\frac{1}{2}}\right.  \notag \\
&+&\left. \frac{1}{2}\left( 2\alpha _{PB}\alpha _{BB}\partial ^{\prime
}\partial ^{\prime }\phi ^{\prime }\cdot \partial ^{\prime }\phi ^{\prime
}+2\alpha _{PB}\alpha _{BP}\partial ^{\prime }\phi ^{\prime }+2\alpha
_{PB}\alpha _{BP}x^{\prime }\cdot \partial ^{\prime }\partial ^{\prime }\phi
^{\prime }+2\alpha _{PP}\alpha _{BP}x^{\prime }\right) \right] .  \notag \\
&&
\end{eqnarray}%
Differentiation of the first row of (\ref{generalized_duality_alpha}) we get
\begin{equation}
\eta =\partial x^{\prime }\cdot \left( \alpha _{PP}\eta +\alpha
_{PB}\partial ^{\prime }\partial ^{\prime }\phi ^{\prime }\right)
\end{equation}%
thus%
\begin{eqnarray}
\partial \phi &=&\det \left( \alpha _{IJ}\right) \partial x^{\prime }\cdot
\partial ^{\prime }\phi ^{\prime }+\alpha _{BB}\alpha _{PB}\partial
x^{\prime }\cdot \partial ^{\prime }\partial ^{\prime }\phi ^{\prime }\cdot
\partial ^{\prime }\phi ^{\prime }  \notag \\
&&+\alpha _{PB}\alpha _{BP}\partial x^{\prime }\cdot \partial ^{\prime }\phi
^{\prime }+\alpha _{PB}\alpha _{BP}\partial x^{\prime }\cdot \partial
^{\prime }\partial ^{\prime }\phi ^{\prime }\cdot x^{\prime }+\alpha
_{PP}\alpha _{BP}\partial x^{\prime }\cdot x^{\prime }  \notag \\
&=&\det \left( \alpha _{IJ}\right) \partial x^{\prime }\cdot \partial
^{\prime }\phi ^{\prime }+\alpha _{BB}\left( \eta -\alpha _{PP}\partial
x^{\prime }\right) \cdot \partial ^{\prime }\phi ^{\prime }+\alpha
_{PB}\alpha _{BP}\partial x^{\prime }\cdot \partial ^{\prime }\phi ^{\prime }
\notag \\
&&+\alpha _{BP}\left( \eta -\alpha _{PP}\partial x^{\prime }\right) \cdot
x^{\prime }+\alpha _{PP}\alpha _{BP}\partial x^{\prime }\cdot x^{\prime }
\notag \\
&=&\det \left( \alpha _{IJ}\right) \partial x^{\prime }\cdot \partial
^{\prime }\phi ^{\prime }+\alpha _{BB}\partial ^{\prime }\phi ^{\prime
}-\left( \alpha _{BB}\alpha _{PP}-\alpha _{PB}\alpha _{BP}\right) \partial
x^{\prime }\cdot \partial ^{\prime }\phi ^{\prime }+\alpha _{BP}x^{\prime }
\notag \\
&=&\alpha _{BB}\partial ^{\prime }\phi ^{\prime }+\alpha _{BP}x^{\prime }
\end{eqnarray}%
where we have used the integrability constraint (\ref{integrability_alpha})
in the last line.

\section{Remark on dual observables in the linearized fluctuation theories
\label{appendix_fluctuation_duality}}

In section \ref{section_fluctuations} we have shown that linearized actions
for fluctuations satisfy%
\begin{equation}
S_{\theta }[\phi _{\ast },\chi ]=S[\left( \phi _{\ast }\right) _{\theta
},\chi _{\theta }]
\end{equation}%
where%
\begin{eqnarray}
S_{\theta }[\phi _{\ast },\chi ] &=&\frac{1}{2}\int \mathrm{d}^{d}x\mathrm{d}%
^{d}y\chi (x)\frac{\delta ^{2}S_{\theta }[\phi ]}{\delta \phi (x)\delta \phi
(y)}|_{\phi _{\ast }}\chi (y)  \notag \\
S[\left( \phi _{\ast }\right) _{\theta },\chi ] &=&\frac{1}{2}\int \mathrm{d}%
^{d}x\mathrm{d}^{d}y\chi (x)\frac{\delta ^{2}S[\phi ]}{\delta \phi (x)\delta
\phi (y)}|_{\left( \phi _{\ast }\right) _{\theta }}\chi (y)
\end{eqnarray}%
This relation enabled us to relate the solutions for the fluctuations in
both theories and e.g. to prove, that the general theory might show
superluminal propagation of the fluctuations even though it corresponds to a
dual theory which is healthy. Here we will show opposite, i.e. that
apparently healthy theory might show superluminal propagation of the
fluctuations of some properly chosen operators. Let us add to $S[\left( \phi
_{\ast }\right) _{\theta },\chi _{\theta }]$ a source term%
\begin{equation}
S_{J}[\chi _{\theta }]=\int \mathrm{d}^{d}xJ(x)\chi _{\theta }(x).
\end{equation}%
After some manipulations we get%
\begin{equation}
S_{J}[\chi _{\theta }]=\int \mathrm{d}^{d}xJ(x)\chi (X[\phi _{\ast }](x))
\end{equation}%
Therefore we have%
\begin{equation}
S_{\theta }[\phi _{\ast },\chi ]+\int \mathrm{d}^{d}xJ(x)\chi (X[\phi _{\ast
}](x))=S[\left( \phi _{\ast }\right) _{\theta },\chi _{\theta }]+\int
\mathrm{d}^{d}xJ(x)\chi _{\theta }(x).
\end{equation}%
Suppose, that we have chosen $\theta $ in such a way that the $S_{\theta
}[\phi _{\ast },\chi ]$ is healthy (e.g. $d_{3}(\theta )=0$ and the
background is a plane wave). Therefore, in the framework of such a healthy
theory the nonlocal operator $\chi (X[\phi _{\ast }](x))$ has the same
superluminal propagation as the perturbations $\chi (x)$ in the theory with
action $S[\left( \phi _{\ast }\right) _{\theta },\chi ]$. Indeed, the
generating functional of the correlators of the operators $\chi (X[\phi
_{\ast }](x))$ in the healthy theory
\begin{equation}
Z_{\theta }[J,\phi _{\ast }]=\int \mathcal{D}\chi \exp \left( \frac{\mathrm{i%
}}{\hbar }S_{\theta }[\phi _{\ast },\chi ]+\frac{\mathrm{i}}{\hbar }\int
\mathrm{d}^{d}xJ(x)\chi (X[\phi _{\ast }](x))\right)
\end{equation}%
can be obtained form the generating functional for the perturbations $\chi $
in the pathological theory by means of change of variables,
\begin{eqnarray}
Z[J,\left( \phi _{\ast }\right) _{\theta }] &=&\int \mathcal{D}\chi \exp
\left( \frac{\mathrm{i}}{\hbar }S[\left( \phi _{\ast }\right) _{\theta
},\chi ]+\frac{\mathrm{i}}{\hbar }\int \mathrm{d}^{d}xJ(x)\chi (x)\right)
\notag \\
&=&\int \mathcal{D}\chi \det \left( \frac{\delta \chi _{\theta }}{\delta
\chi }\right) \exp \left( \frac{\mathrm{i}}{\hbar }S_{\theta }[\phi _{\ast
},\chi ]+\frac{\mathrm{i}}{\hbar }\int \mathrm{d}^{d}xJ(x)\chi (X[\phi
_{\ast }](x))\right)  \notag \\
&=&Z_{\theta }[J,\phi _{\ast }].  \label{fluctuation_correlators_generators}
\end{eqnarray}%
Here we have used
\begin{equation}
\frac{\delta \chi _{\theta }(x)}{\delta \chi (y)}=\frac{\delta \chi (X[\phi
_{\ast }](x))}{\delta \chi (y)}=\delta ^{(d)}\left( X[\phi _{\ast
}](x)-y\right)
\end{equation}%
and thus $\det \left( \delta \chi _{\theta }/\delta \chi \right) =1$ within
dimensional regularization as we have shown in Section \ref{S_duality}.

Note however, that the operator$~$ $\chi (X[\phi _{\ast }](x))$ is $\phi
_{\ast }$ dependent and nonlocal, namely, because%
\begin{equation}
X[\phi _{\ast }](x)=x+2\theta \left( \partial \phi _{\ast }\right) _{\theta
}(x)
\end{equation}%
and thus%
\begin{equation}
\chi (X[\phi _{\ast }](x))=\sum\limits_{n=0}^{\infty }\frac{(2\theta )^{n}}{%
n!}\left( \partial ^{\mu _{1}}\phi _{\ast }\right) _{\theta }(x)\ldots
\left( \partial ^{\mu _{n}}\phi _{\ast }\right) _{\theta }(x)\left( \partial
_{\mu _{1}}\ldots \partial _{\mu _{n}}\chi \right) (x)
\end{equation}%
Therefore $\chi (X[\phi _{\ast }](x))$ is an infinite linear combinations of
local operators $\left( \partial _{\mu _{1}}\ldots \partial _{\mu _{n}}\chi
\right) (x)$ with $x-$dependent coefficients i.e. it is not translation
invariant. The latter fact is the reason why we cannot use (\ref%
{fluctuation_correlators_generators}) to argue that also $S$ matrices for
fluctuations are the same in both theories (cf. Section \ \ref{S_duality}).

\section{The group velocity of the plane wave perturbation\label%
{v_group_appendix}}

The group velocity can be obtained by means of differentiation of condition
(\ref{wave_packet_center}) for the center of the wave packet (here $\widehat{%
\mathbf{k}}=\mathbf{k}/|\mathbf{k}|$)
\begin{equation}
\mathbf{X}[\phi _{\ast }]-\widehat{\mathbf{k}}X^{0}[\phi _{\ast }]=\mathbf{%
const}.
\end{equation}%
with respect to $t$, the group velocity is then $\mathbf{v}_{group}(x)=%
\mathrm{d}\mathbf{x}/\mathrm{d}t$. Explicitly we get%
\begin{equation}
X[\phi _{\ast }](x)=x+2\theta nF^{\prime }(n\cdot x),
\end{equation}%
and thus writing $n=(1,\mathbf{n})$%
\begin{equation}
\mathbf{v}_{group}+2\theta \mathbf{n}F^{\prime \prime }(n\cdot x)\left( 1-%
\mathbf{n\cdot v}_{group}\right) -\widehat{\mathbf{k}}\left[ 1+2\theta
F^{\prime \prime }(n\cdot x)\left( 1-\mathbf{n\cdot v}_{group}\right) \right]
=0  \label{v_group}
\end{equation}%
and therefore denoting $v_{\parallel }\equiv \left( \mathbf{n\cdot v}%
_{group}\right) $ the component of $\mathbf{v}_{group}$ parallel to $\mathbf{%
n}$
\begin{equation}
v_{\parallel }+2\theta F^{\prime \prime }(n\cdot x)\left( 1-v_{\parallel
}\right) -\mathbf{n}\cdot \widehat{\mathbf{k}}\left[ 1+2\theta F^{\prime
\prime }(n\cdot x)\left( 1-v_{\parallel }\right) \right] =0.
\end{equation}%
As a result%
\begin{equation}
v_{\parallel }=\frac{\mathbf{n}\cdot \widehat{\mathbf{k}}\mathbf{-}2\theta
F^{\prime \prime }(n\cdot x)\left( 1-\mathbf{n}\cdot \widehat{\mathbf{k}}%
\right) }{1-2\theta F^{\prime \prime }(n\cdot x)\left( 1-\mathbf{n}\cdot
\widehat{\mathbf{k}}\right) }
\end{equation}%
Inserting this to (\ref{v_group}) we get finally
\begin{eqnarray}
\mathbf{v}_{group}\mathbf{\ } &\mathbf{=}&\widehat{\mathbf{k}}\left[
1+2\theta F^{\prime \prime }(n\cdot x)\left( 1-v_{\parallel }\right) \right]
-2\theta \mathbf{n}F^{\prime \prime }(n\cdot x)\left( 1-v_{\parallel }\right)
\notag \\
&=&\frac{\widehat{\mathbf{k}}-2\theta \mathbf{n}F^{\prime \prime }(n\cdot
x)\left( 1-\mathbf{n}\cdot \widehat{\mathbf{k}}\right) }{1-2\theta F^{\prime
\prime }(n\cdot x)\left( 1-\mathbf{n}\cdot \widehat{\mathbf{k}}\right) }.
\end{eqnarray}

\section{Perturbative calculation of the two-point amplitude\label{appendix_amplitude}}

Here we explicitly calculate the first two perturbative contributions to the
amplitude $\mathcal{M}(\mathbf{k},\mathbf{k}^{\prime })$. The first order
contribution corresponds to the first graph on the right hand side of the
graphical equation in the last row of the Figure \ref{fig:F_rules} and we
get it simply by setting $p\rightarrow k$, $q\rightarrow -k^{\prime }$ in the Feynman rule for interaction vertex and
putting the external lines on shell. That means, we get for the first order
contribution to the $S$ matrix
\begin{equation}
S_{fi}^{(1)}(\mathbf{k},\mathbf{k}^{\prime })=-4\mathrm{i}\theta _{\ast
}\left( k^{+}\right) ^{2}\left( 2\pi \right) ^{3}\delta (k^{+}-k^{+\prime
})\delta ^{(2)}(\mathbf{k}_{\bot }-\mathbf{k}_{\bot }^{\prime })\int \mathrm{%
d}x^{+}\mathrm{e}^{-\frac{\mathrm{i}}{2}(k^{-}-k^{-\prime })x^{+}}F^{\prime
\prime }(x^{+}).
\end{equation}%
Provided $k$ and $k^{\prime }$ are on shell, i.e. $k^{-}=\mathbf{k}_{\bot
}^{2}/k^{+}$ and similarly for $k^{\prime }$, the exponential factor in the
integrand is just one and we get (let us remind $\psi ^{\pm
}=\lim_{x^{+}\rightarrow \pm \infty }F^{\prime }(x^{+})$)
\begin{equation}
S_{fi}^{(1)}(\mathbf{k},\mathbf{k}^{\prime })=\left[ 2\mathrm{i}\theta
_{\ast }k^{+}\left( \psi ^{-}-\psi ^{+}\right) \right] 2k^{+}\left( 2\pi
\right) ^{3}\delta (k^{+}-k^{+\prime })\delta ^{(2)}(\mathbf{k}_{\bot }-%
\mathbf{k}_{\bot }^{\prime })\left( \psi ^{-}-\psi ^{+}\right) .
\end{equation}%
Using the identity valid for on-shell $k$%
\begin{equation}
\delta (k^{+}-k^{+\prime })\delta ^{(2)}(\mathbf{k}_{\bot }-\mathbf{k}_{\bot
}^{\prime })=\frac{|\mathbf{k|}}{k^{+}}\delta ^{(3)}(\mathbf{k}-\mathbf{k}%
^{\prime })=\frac{1}{k^{+}|\mathbf{k|}}\delta (|\mathbf{k|}-|\mathbf{k}%
^{\prime }|)\delta ^{(2)}(\widehat{\mathbf{k}}-\widehat{\mathbf{k}}^{\prime
})
\end{equation}%
we get finally%
\begin{equation}
\mathcal{M}^{(1)}(\mathbf{k},\mathbf{k}^{\prime })=(4\pi )^{2}\frac{2\mathrm{%
i}\theta _{\ast }k^{+}\left( \psi ^{-}-\psi ^{+}\right) }{2\mathrm{i}|%
\mathbf{k|}}\delta ^{(2)}(\widehat{\mathbf{k}}-\widehat{\mathbf{k}}^{\prime
}).
\end{equation}%
The next term corresponding to the graph with two vertices and one
propagator (see the second graph on the right hand side in the last row of
Figure \ref{fig:F_rules}) gives according the standard Feynman rules%
\begin{eqnarray}
&&S_{fi}^{(2)}(\mathbf{k},\mathbf{k}^{\prime })=  \notag \\
&=&\int \frac{\mathrm{d}^{4}p}{(2\pi )^{4}}\frac{\mathrm{i}}{p^{+}p^{-}-%
\mathbf{p}_{\bot }^{2}+\mathrm{i}0}  \notag \\
&&\times \left[ -4\mathrm{i}\theta _{\ast }\left( k^{+}p^{+}\right) \left(
2\pi \right) ^{3}\delta (k^{+}-p^{+})\delta ^{(2)}(\mathbf{k}_{\bot }-%
\mathbf{p}_{\bot })\int \mathrm{d}x^{+}\mathrm{e}^{-\frac{\mathrm{i}}{2}%
(p^{-}-k^{-})x^{+}}F^{\prime \prime }(x^{+})\right]   \notag \\
&&\times \left[ -4\mathrm{i}\theta _{\ast }\left( k^{+\prime }p^{+}\right)
\left( 2\pi \right) ^{3}\delta (k^{+\prime }-p^{+})\delta ^{(2)}(\mathbf{k}%
_{\bot }^{\prime }-\mathbf{p}_{\bot })\int \mathrm{d}y^{+}\mathrm{e}^{+\frac{%
\mathrm{i}}{2}(p^{-}-k^{-\prime })x^{+}}F^{\prime \prime }(y^{+})\right].\notag \\
&&
\end{eqnarray}%
Writing $\mathrm{d}^{4}p=\left( 1/2\right) \mathrm{d}p^{-}\mathrm{d}p^{+}%
\mathrm{d}^{2}\mathbf{p}_{\bot }$ and integrating out the delta functions we
get%
\begin{eqnarray}
S_{fi}^{(2)}(\mathbf{k},\mathbf{k}^{\prime }) &=&8\left( \mathrm{i}\theta
_{\ast }\right) ^{2}\left( k^{+}\right) ^{4}\left( 2\pi \right) ^{3}\delta
(k^{+}-k^{+\prime })\delta ^{(2)}(\mathbf{k}_{\bot }-\mathbf{k}_{\bot
}^{\prime })  \notag \\
&&\times \int \mathrm{d}x^{+}\mathrm{d}y^{+}F^{\prime \prime
}(x^{+})F^{\prime \prime }(y^{+})  \notag \\
&&\times \int \frac{\mathrm{d}p^{-}}{2\pi }\mathrm{e}^{-\frac{\mathrm{i}}{2}%
(p^{-}-k^{-})(x^{+}-y^{+})}\frac{\mathrm{i}}{k^{+}p^{-}-\mathbf{k}_{\bot
}^{2}+\mathrm{i}0}.
\end{eqnarray}%
Using the on-shell condition $\mathbf{k}_{\bot }^{2}=k^{-}k^{+}$ and
shifting the integration variable $p^{-}\rightarrow p^{-}+k^{-}$ we can
rewrite this as (note that $k^{+}>0$)%
\begin{eqnarray}
S_{fi}^{(2)}(\mathbf{k},\mathbf{k}^{\prime }) &=&8\left( \mathrm{i}\theta
_{\ast }\right) ^{2}\left( k^{+}\right) ^{3}\left( 2\pi \right) ^{3}\delta
(k^{+}-k^{+\prime })\delta ^{(2)}(\mathbf{k}_{\bot }-\mathbf{k}_{\bot
}^{\prime })  \notag \\
&&\times \int \mathrm{d}x^{+}\mathrm{d}y^{+}F^{\prime \prime
}(x^{+})F^{\prime \prime }(y^{+})\int \frac{\mathrm{d}p^{-}}{2\pi }\mathrm{e}%
^{-\frac{\mathrm{i}}{2}p^{-}(x^{+}-y^{+})}\frac{\mathrm{i}}{p^{-}+\mathrm{i}0%
} 
\end{eqnarray}%
We can recognize the integral representation of the Heaviside theta function
in the last line, i.e.%
\begin{eqnarray}
S_{fi}^{(2)}(\mathbf{k},\mathbf{k}^{\prime }) &=&\left( 2\mathrm{i}\theta
_{\ast }k^{+}\right) ^{2}2k^{+}\left( 2\pi \right) ^{3}\delta
(k^{+}-k^{+\prime })\delta ^{(2)}(\mathbf{k}_{\bot }-\mathbf{k}_{\bot
}^{\prime })  \notag \\
&&\times \int \mathrm{d}x^{+}\mathrm{d}y^{+}F^{\prime \prime
}(x^{+})F^{\prime \prime }(y^{+})\theta (x^{+}-y^{+}).
\end{eqnarray}%
The remaining integral is elementary and we get finally%
\begin{equation}
S_{fi}^{(2)}(\mathbf{k},\mathbf{k}^{\prime })=\frac{1}{2!}\left[ 2\mathrm{i}%
\theta _{\ast }k^{+}\left( \psi ^{-}-\psi ^{+}\right) \right]
^{2}2k^{+}\left( 2\pi \right) ^{3}\delta (k^{+}-k^{+\prime })\delta ^{(2)}(%
\mathbf{k}_{\bot }-\mathbf{k}_{\bot }^{\prime })
\end{equation}%
and%
\begin{equation*}
\mathcal{M}^{(2)}(\mathbf{k},\mathbf{k}^{\prime })=(4\pi )^{2}\frac{\left( 2%
\mathrm{i}\theta _{\ast }k^{+}\left( \psi ^{-}-\psi ^{+}\right) \right)
^{2}/2!}{2\mathrm{i}|\mathbf{k|}}\delta ^{(2)}(\widehat{\mathbf{k}}-\widehat{%
\mathbf{k}}^{\prime }).
\end{equation*}%
The sum $\mathcal{M}^{(1)}+\mathcal{M}^{(2)}$ thus reproduces the first two
terms of the expansion of the complete amplitude (\ref{2pt_amplitude}) in
powers of the phase shift.

\section{The fluctuation operator of the static cylindrically symmetric
solution\label{cylindric_g_mu_nu}}

The fluctuation operator is%
\begin{equation}
\frac{\delta ^{2}S_{\theta }[\phi ]}{\delta \phi (x)\delta \phi (y)}%
=\sum\limits_{n=2}^{d}nd_{n}(\theta )\frac{\partial \mathcal{L}_{n-1}^{%
\mathrm{der}}(\partial \partial \phi (x))}{\partial \left( \partial _{\mu
}\partial _{\nu }\phi (x)\right) }\partial _{\mu }\partial _{\nu }\delta
^{(d)}(x-y)
\end{equation}%
From the generating function of $\mathcal{L}_{k}^{\mathrm{der}}(\partial
\partial \phi (x))$%
\begin{equation}
4!\det \left( \eta +w\partial \partial \phi \right)
=\sum\limits_{n=0}^{d}w^{k}\left(
\begin{array}{c}
4 \\
k%
\end{array}%
\right) \mathcal{L}_{k}^{\mathrm{der}}(\partial \partial \phi )
\end{equation}%
we get%
\begin{equation}
\frac{\partial }{\partial \left( \partial _{\mu }\partial _{\nu }\phi
\right) }\det \left( \eta +w\partial \partial \phi \right) =\frac{1}{4!}%
\sum\limits_{n=0}^{d}w^{k}\left(
\begin{array}{c}
4 \\
k%
\end{array}%
\right) \frac{\partial \mathcal{L}_{k}^{\mathrm{der}}(\partial \partial \phi
)}{\partial \left( \partial _{\mu }\partial _{\nu }\phi \right) }
\end{equation}%
Left hand side gives%
\begin{eqnarray}
\frac{\partial }{\partial \left( \partial _{\mu }\partial _{\nu }\phi
\right) }\det \left( \eta +w\partial \partial \phi \right) &=&-\frac{%
\partial }{\partial \left( \partial _{\mu }\partial _{\nu }\phi \right) }%
\det \left( \delta +w\eta \cdot \partial \partial \phi \right)  \notag \\
&=&-\frac{\partial }{\partial \left( \partial _{\mu }\partial _{\nu }\phi
\right) }\exp \mathrm{Tr}\ln \left( \delta +w\eta \cdot \partial \partial
\phi \right)  \notag \\
&=&\det \left( \eta +w\partial \partial \phi \right) \frac{\partial }{%
\partial \left( \partial _{\mu }\partial _{\nu }\phi \right) }\mathrm{Tr}\ln
\left( \delta +w\eta \cdot \partial \partial \phi \right)  \notag \\
&=&w\det \left( \eta +w\partial \partial \phi \right) \left[ \left( \delta
+w\eta \cdot \partial \partial \phi \right) ^{-1}\right] _{\beta }^{\mu
}\eta ^{\nu \beta }  \notag \\
&=&w\det \left( \eta +w\partial \partial \phi \right) \left[ \left( \eta
+w\partial \partial \phi \right) ^{-1}\right] ^{\mu \nu }
\end{eqnarray}%
For cylindrically symmetric static solution $\phi \equiv \phi (z\overline{z}%
) $ we get
\begin{equation}
\eta +w\partial \partial \phi =\left(
\begin{array}{cccc}
1 & 0 & 0 & 0 \\
0 & -1 & 0 & 0 \\
0 & 0 & -1+w\left( \partial +\overline{\partial }\right) ^{2}\phi & w\mathrm{%
i}\left( \partial ^{2}-\overline{\partial }^{2}\right) \phi \\
0 & 0 & w\mathrm{i}\left( \partial ^{2}-\overline{\partial }^{2}\right) \phi
& -1-w\left( \partial -\overline{\partial }\right) ^{2}\phi%
\end{array}%
\right)
\end{equation}%
and thus%
\begin{equation}
\det \left( \eta +w\partial \partial \phi \right) =-1+4w\partial \overline{%
\partial }\phi +4w^{2}\left[ \partial ^{2}\phi \overline{\partial }^{2}\phi
-\left( \partial \overline{\partial }\phi \right) ^{2}\right]
\end{equation}%
and
\begin{eqnarray}
&&\det \left( \eta +w\partial \partial \phi \right) \left( \eta +w\partial
\partial \phi \right) ^{-1}  \notag \\
&=&\left(
\begin{array}{cccc}
\det \left( \eta +w\partial \partial \phi \right) & 0 & 0 & 0 \\
0 & -\det \left( \eta +w\partial \partial \phi \right) & 0 & 0 \\
0 & 0 & 1+w\left( \partial -\overline{\partial }\right) ^{2}\phi & w\mathrm{i%
}\left( \partial ^{2}-\overline{\partial }^{2}\right) \phi \\
0 & 0 & w\mathrm{i}\left( \partial ^{2}-\overline{\partial }^{2}\right) \phi
& 1-w\left( \partial +\overline{\partial }\right) ^{2}\phi%
\end{array}%
\right)
\end{eqnarray}%
As a result%
\begin{eqnarray}
&&\frac{1}{4!}\sum\limits_{n=1}^{d}w^{k}\left(
\begin{array}{c}
4 \\
k%
\end{array}%
\right) \frac{\partial \mathcal{L}_{k}^{\mathrm{der}}(\partial \partial \phi
)}{\partial \left( \partial \partial \phi \right) }  \notag \\
&=&w\left(
\begin{array}{cccc}
\det \left( \eta +w\partial \partial \phi \right) & 0 & 0 & 0 \\
0 & -\det \left( \eta +w\partial \partial \phi \right) & 0 & 0 \\
0 & 0 & 1+w\left( \partial -\overline{\partial }\right) ^{2}\phi & w\mathrm{i%
}\left( \partial ^{2}-\overline{\partial }^{2}\right) \phi \\
0 & 0 & w\mathrm{i}\left( \partial ^{2}-\overline{\partial }^{2}\right) \phi
& 1-w\left( \partial +\overline{\partial }\right) ^{2}\phi%
\end{array}%
\right)
\end{eqnarray}%
and thus%
\begin{eqnarray}
\frac{\partial \mathcal{L}_{1}^{\mathrm{der}}(\partial \partial \phi )}{%
\partial \left( \partial _{\mu }\partial _{\nu }\phi \right) } &=&-6\eta
\notag \\
\frac{\partial \mathcal{L}_{2}^{\mathrm{der}}(\partial \partial \phi )}{%
\partial \left( \partial _{\mu }\partial _{\nu }\phi \right) } &=&4\left(
\begin{array}{cccc}
4\partial \overline{\partial }\phi & 0 & 0 & 0 \\
0 & -4\partial \overline{\partial }\phi & 0 & 0 \\
0 & 0 & \left( \partial -\overline{\partial }\right) ^{2}\phi & \mathrm{i}%
\left( \partial ^{2}-\overline{\partial }^{2}\right) \phi \\
0 & 0 & \mathrm{i}\left( \partial ^{2}-\overline{\partial }^{2}\right) \phi
& -\left( \partial +\overline{\partial }\right) ^{2}\phi%
\end{array}%
\right)  \notag \\
\frac{\partial \mathcal{L}_{3}^{\mathrm{der}}(\partial \partial \phi )}{%
\partial \left( \partial _{\mu }\partial _{\nu }\phi \right) } &=&6\left(
\begin{array}{cccc}
4\left[ \partial ^{2}\phi \overline{\partial }^{2}\phi -\left( \partial
\overline{\partial }\phi \right) ^{2}\right] & 0 & 0 & 0 \\
0 & -4\left[ \partial ^{2}\phi \overline{\partial }^{2}\phi -\left( \partial
\overline{\partial }\phi \right) ^{2}\right] & 0 & 0 \\
0 & 0 & 0 & 0 \\
0 & 0 & 0 & 0%
\end{array}%
\right)
\end{eqnarray}%
Inserting this to the formula%
\begin{equation}
g[\phi ]^{\mu \nu }=\sum\limits_{n=2}^{d}nd_{n}(\theta )\frac{\partial
\mathcal{L}_{n-1}^{\mathrm{der}}(\partial \partial \phi (x))}{\partial
\left( \partial _{\mu }\partial _{\nu }\phi (x)\right) }
\end{equation}%
gives (\ref{general_g[phi]}).

\section{Full form of 4-pt scattering amplitude and self-energy\label%
{A_Sigma}}

The corresponding contributions are
\begin{eqnarray}
A_{1} &=&d_{3}^{4}\frac{B(s)}{(4\pi )^{2}}\frac{%
81s^{4}[(d^{2}+6d+32)s^{2}-72tu]}{32(d^{2}-1)}+\text{cycl} \\
A_{2} &=&-d_{3}^{4}\frac{B(s)}{(4\pi )^{2}}\frac{81(d+2)s^{6}}{4(d-1)}+\text{%
cycl} \\
A_{3} &=&d_{3}^{4}\frac{B(s)}{(4\pi )^{2}}\frac{81s^{6}}{2}+\text{cycl} \\
A_{4} &=&d_{4}^{2}\frac{B(s)}{(4\pi )^{2}}\frac{9s^{4}[(d^{2}-2d)s^{2}-8tu]}{%
8(d^{2}-1)}+\text{cycl}  \label{fullA4} \\
A_{5} &=&d_{3}^{2}d_{4}\frac{B(s)}{(4\pi )^{2}}\frac{%
27s^{4}[(d+4)(d-2)s^{2}+24tu]}{8(d^{2}-1)}+\text{cycl} \\
A_{6} &=&-d_{3}^{2}d_{4}\frac{B(s)}{(4\pi )^{2}}\frac{27s^{6}(d-2)}{2(d-1)}+%
\text{cycl}
\end{eqnarray}%
We have used the cyclic summation over all Mandelstam variables (e.g. $%
(s^{2}+\text{cycl})=s^{2}+t^{2}+u^{2}$). The loop function is given by
\begin{equation}
B(s)=\frac{1}{(4\pi )^{d/2-2}}\frac{1}{d-3}\Gamma (2-d/2)s^{d/2-2}
\end{equation}%
Summing up all diagrams leads to
\begin{equation}
A=(d_{4}-\tfrac{9}{2}d_{3}^{2})^{2}\frac{B(s)}{(4\pi )^{2}}\frac{%
9s^{4}[d(d-2)s^{2}-8tu]}{8(d^{2}-1)}+\text{cycl}.  \label{fullA}
\end{equation}%
The full result for the one-loop self-energy reads%
\begin{equation}
\Sigma (p)=-\frac{1}{(4\pi )^{d/2}}\frac{9}{2}d_{3}^{2}\left( p^{2}\right)
^{4}B\left( p^{2}\right) .
\end{equation}

\end{document}